# QUANTUM CONTROL AND QUANTUM TOMOGRAPHY ON NEUTRAL ATOM QUDITS

by

Hector Sosa Martinez

---

A Dissertation Submitted to the Faculty of the

DEPARTMENT OF PHYSICS

In Partial Fulfillment of the Requirements

For the Degree of

DOCTOR OF PHILOSOPHY

In the Graduate College

THE UNIVERSITY OF ARIZONA

2 0 1 6

# ACKNOWLEDGEMENTS

The work shown in this dissertation was made possible thanks to the knowledge, effort, and sacrifice of several wonderful people. I am grateful to all of them.

First and foremost, my advisor, Poul S. Jessen. He is a brilliant scientist, with impressive physical intuition. He always provided me with clear guidance along with complete freedom to explore. Always willing to make time to discuss any aspect of the experiment and offer support when necessary. I could not imagine a better doctoral advisor over the past six years and I am extremely grateful to him.

Brian E. Anderson and Aaron Smith, into whose research project I stepped in immediately after joining Poul's group. In a few months, on top of everything else going on in the laboratory, they provided me with plenty of knowledge and excitement about our research which gave me an excellent start.

I owe big thanks to Nathan Lysne, my friend and laboratory partner for the past few years. Nathan is a very intelligent technical thinker who helped me perform many of the experiments for this dissertation. Despite the fact that we worked on different projects, Enrique Montaño, Daniel Hemmer, David Melchior, and Jae Hoon Lee were also contributors to my research through constructive discussions. Accomplishing many of the objectives in the experiment would have proved very difficult without being around such talented people.

Much of the work in this dissertation emerged from a collaboration with Ivan Deutsch and his theory group at the University of New Mexico. Ivan is an extraordinary person and I have great admiration for his passion for quantum physics. I thank Charles Baldwin, one of Ivan's students, for tirelessly working alongside from the theory front and for answering my countless theoretical questions.

I especially thank, with all my heart, my beloved mother Amanda and my father Martiniano. After all these years away from home, I am absolutely convinced that they are the greatest educators I will ever have. Thank you for your love, unconditional support, and for inculcating me all the values that have sculpted my character. My sisters, Eva and Elizabeth, I thank you for your support and for being such amazing individuals.

Finally, I would like to thank my lovely wife, Margarita. Since I met her, fifteen years ago, she has been my day to day inspiration. Having her by my side and witnessing all her demonstrations of unwavering love and support makes me feel the happiest person in the world.

# DEDICATION

*To my family*

# TABLE OF CONTENTS

TABLE OF CONTENTS – *Continued*

# LIST OF FIGURES

LIST OF FIGURES – *Continued*

# LIST OF TABLES

# ABSTRACT

Neutral atom systems are an appealing platform for the development and testing of quantum control and measurement techniques. This dissertation presents experimental investigations of control and measurement tools using as a testbed the 16-dimensional hyperfine manifold associated with the electronic ground state of $^{133}$Cs atoms. On the control side, we present an experimental realization of a protocol to implement robust unitary transformations in the presence of static and dynamic perturbations. We also present an experimental realization of inhomogeneous quantum control. Specifically, we demonstrate our ability to perform two different unitary transformations on atoms that see different light shifts from an optical addressing field. On the measurement side, we present experimental realizations of quantum state and process tomography. The state tomography project encompasses a comprehensive evaluation of several measurement strategies and state estimation algorithms. Our experimental results show that in the presence of experimental imperfections, there is a clear tradeoff between accuracy, efficiency and robustness in the reconstruction. The process tomography project involves an experimental demonstration of efficient reconstruction by using a set of intelligent probe states. Experimental results show that we are able to reconstruct unitary maps in Hilbert spaces with dimension ranging from $d = 4$ to $d = 16$. To the best of our knowledge, this is the first time that a unitary process in $d = 16$ is successfully reconstructed in the laboratory.

# CHAPTER 1

# INTRODUCTION

The theory of classical computation was laid down in the 1930s [1]. Within a decade, the first digital computers started to appear, using vacuum tubes as their building blocks. These rudimentary computers were typically the size of a large room. The introduction of the transistor in the late 1940s launched a race towards smaller and faster processors. Today, more than half a century later, nearly all the information we digitally process is encoded in powerful and often portable computers.

In spite of the great success of classical computation, as the size of the fundamental components in emerging technologies becomes smaller, there will be a point where quantum mechanical effects govern their principal behavior. Quantum information science (QIS) is an expanding field with roots that go back almost twenty years, when pioneers such as R. Feynman, C. Bennett, P. Benioff, and others began thinking about the implications of combining quantum mechanics with classical computing.

Nowadays, quantum information science provides a framework in which unique quantum mechanical phenomena such as *superposition* and *entanglement* can be utilized to substantially improve the acquisition and processing of information. As with any revolutionary scientific insight, the ultimate impact of the development of quantum information technologies remains an open question. Nonetheless, QIS has already provided us with new ways to describe how nature works, and with novel approaches for a wide variety of scientific and technical questions.

Information technologies based on quantum mechanics perform calculations on fundamental pieces of information called quantum bits, or *qubits*. Qubits are 2-dimensional quantum systems that can be encoded in a variety of physical systems. A few examples include the electronic states of an atom, spin states of an atomic nucleus, flux (direction of a current) states of a superconducting circuit, etc. A fundamental challenge shared among all these systems consist in the development of tools to accurately and robustly control the qubit system in the presence of real-world imperfections. During recent years, a large amount of theoretical studies on this subject have provided answers as to when and how a quantum system can be fully controlled. On the experimental side, developments have been made on several physical systems including trapped ions [2], nuclear magnetic resonance (NMR) [3], neutral atoms [4], cavity quantum electrodynamics [5], solid state devices [6, 7], and superconducting circuits [8].

Neutral atom systems possess several attributes that make them an attractive platform for the development of quantum technologies. They have a simple quantum-level structure, excellent isolation from the decohering influence of the environment, and can be trapped and manipulated in a large ensemble of atoms. Quantum information techniques using neutral atoms have been studied in several experimental settings such as optical lattices [9, 10], microscopic optical traps [11], Rydberg atoms [12, 13], and single-atom traps [14, 15].

Most of the physical platforms listed thus far, including neutral atoms, possess a total Hilbert space with dimension larger than two. Quantum control tools for $d$-dimensional Hilbert space systems, known as *qudits*, remain as an unexplored field of study. Development of control tool for qudits represents an important challenge that may open up interesting lines of research in quantum information processing. Similar to qubits, qudits can be used as fundamental quantum information processing elements [16]. Alternatively, single qubits can be embedded in a qudit space system allowing for robust qubit manipulation [17]. Qudit control techniques can also enable fundamental studies in problems such as quantum chaos [18]. In the case of collective spin system, qudit control can be used to enhance collective spin squeezing [19].

As quantum control tools improve and the complexity of quantum information processors grows, it becomes increasingly difficult to implement measurements to determine if the quantum systems are performing as expected. As a result, sources of errors in the laboratory are harder to identify. One aspect of quantum measurement centers on the development of accurate, efficient, and robust tools to characterize a given quantum device.

The procedure by which a quantum information processor is fully characterize, is known as quantum tomography. Quantum tomography is divided into three techniques: quantum state tomography [20], quantum process tomography [21], and quantum detector tomography [22]. Each of these techniques is used to estimate one of the three components that describe a quantum information processor: state preparation, evolution, and readout. Quantum state tomography has been demonstrated in several experimental platforms, for instance, trapped ions [23], neutral atoms [24, 25], and superconducting qubits [26]. Quantum process tomography has been studied in trapped ions [27], optical systems [28], and NMR [29]. Quantum detector tomography has been used to characterize detectors in optical systems [30].

In spite of the numerous experimental demonstrations, quantum tomography remains as an impractical tool to fully characterize quantum devices. This is mainly because current quantum tomography protocols are subject to state preparation and measurement errors when implemented in the laboratory. In addition, quantum tomography for systems with large Hilbert spaces is a demanding task that requires a large amount of measurements to produce accurate estimates.

In the present dissertation, we survey our effort toward the experimental demonstration of new control and measurement tools for neutral atom qudits. On the control side, we expand the available toolbox by implementing inhomogeneous quantum control designed using optimal control ideas. On the measurement side, we explore quantum state and process tomography using several measurement strategies to find the tradeoffs between accuracy, efficiency and robustness in the presence of experimental imperfections.

The body of this dissertation is structured as follows. In chapter 2, we present the theoretical foundations necessary to understand the effects of magnetic and light fields on the ground state of cesium atoms. We also describe the basic concepts to understand how a quantum measurement is carried out, as well as a brief description of quantum tomography. In chapter 3, we describe our experimental apparatus along with our control and measurement toolbox. In chapter 4, we present experimental results that demonstrate our ability to perform high-accuracy and robust unitary transformation in the presence of, static and dynamical errors and perturbations. We also present results to demonstrate that inhomogeneous quantum control can be achieved using the tools of optimal control. In chapter 5, we present experimental results demonstrating our ability to perform quantum state tomography in a 16-dimensional Hilbert space. In this study we implemented several measurement strategies to determine which is the most accurate, efficient and robust in the presence of real-world experimental imperfections. In chapter 6, we present experimental results demonstrating our ability to perform efficient and robust quantum process tomography in a 16-dimensional Hilbert space. Chapter 7 summarizes the results and accomplishments of this work.

# CHAPTER 2

# THEORETICAL BACKGROUND

This chapter covers the theoretical background required for our work on quantum control and tomography of systems with large Hilbert spaces (qudits). The aim throughout is to introduce relevant theoretical formalism needed for the work presented in this dissertation; the discussion is not intended to be comprehensive or self-contained, but rather to provide an overview of ideas and concepts, with references to existing literature provided in the appropriate places. Much of the formalism was developed in collaboration with Ivan Deutsch and his research group at the University of New Mexico, and some of what follows are excerpts from Refs. [31, 32].

## 2.1 The Cesium Atom in a Magnetic Field

Alkali atoms are commonly used in experiments where trapping and cooling of neutral atoms is necessary. This is largely due to their simple level structure and the ease with which they can be manipulated with optical and magnetic fields. These same properties make individual alkali atoms an excellent physical platform to perform quantum control and measurement experiments. For examples of such work see, e. g., Refs [31, 32].

Fig. 2.1 shows an energy level diagram of the relevant hyperfine structure of cesium. The absence of spontaneous decay in the $6S_{1/2}$ electronic ground state allows long-lived populations and coherences in the associated hyperfine manifold and makes it an excellent candidate to perform quantum control tasks on it. Preparing an ensemble of cold, trapped atoms requires optical transitions on the $D_2$ line, coupling $6S_{1/2}$ to $6P_{3/2}$. Implementing inhomogeneous quantum control using the light shift from an addressing optical field requires optical transitions on the $D_1$ or $D_2$ line, coupling $6S_{1/2}$ to $6P_{1/2}$ or $6S_{1/2}$ to $6P_{3/2}$ respectively, and will be discussed in chapter 4.

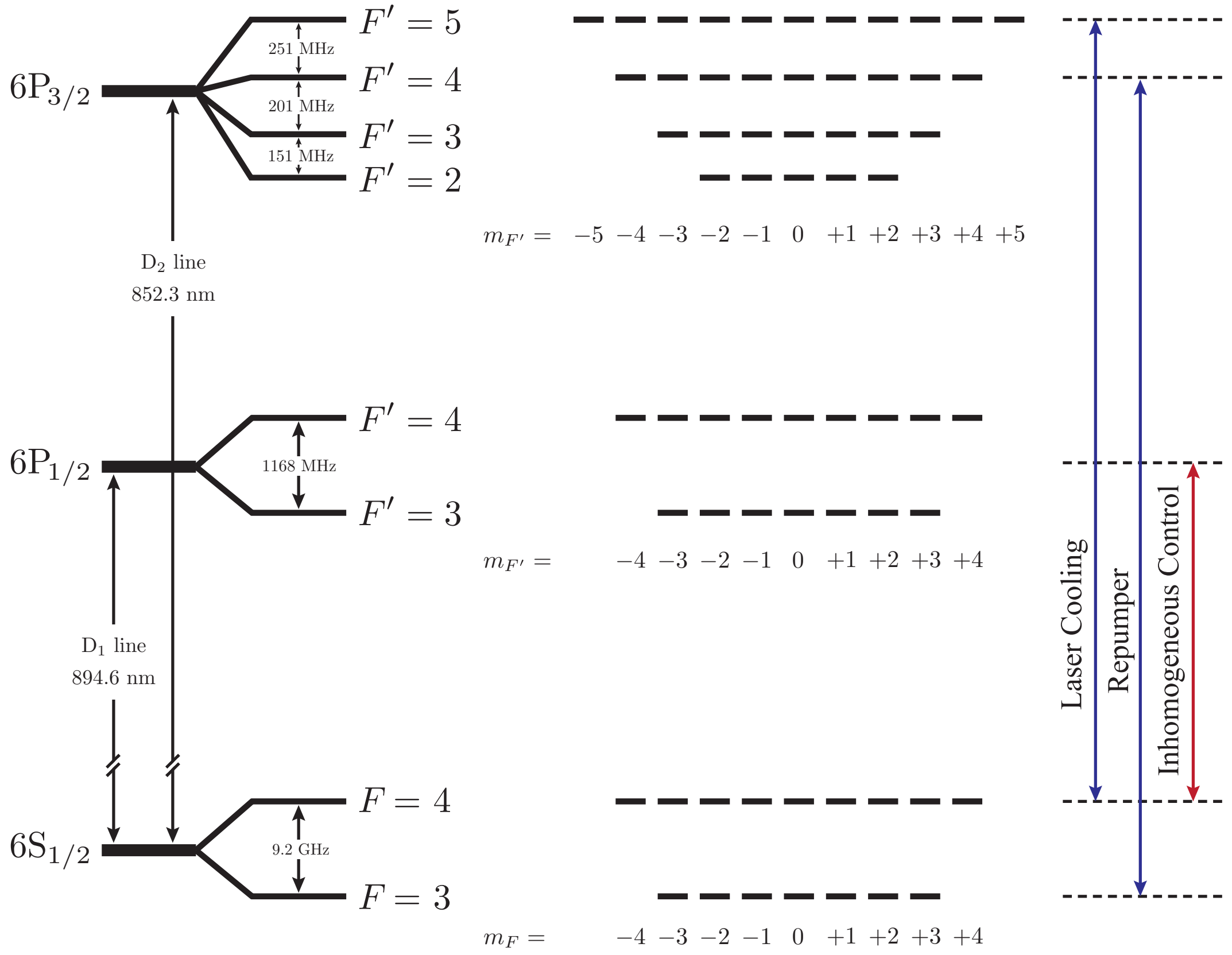

Figure 2.1: Cesium energy level structure. Magnetic sublevels ($m_F$) are shown for each hyperfine level ($F$). Laser cooling and trapping requires optical transitions on the $D_2$ line (blue arrows). Inhomogeneous quantum control is implemented using an optical transition on the $D_1$ line (red arrow). Energy level separations are not drawn to scale.

The rest of this section focuses on the physics occurring in the ground manifold $6S_{1/2}$, which encodes quantum information on its total atomic spin state. The total atomic spin consists of the sum of the single valence electron spin and nuclear spin, $\hat{\mathbf{F}} = \hat{\mathbf{S}} + \hat{\mathbf{I}}$, with quantum numbers $S = 1/2$, $I = 7/2$, and $F^{(\pm)} = 3, 4$. The set of magnetic sublevels $\{|F, m_F\rangle\}$ form a basis (the "logical basis"), with states that are simultaneous eigenstates of $\hat{\mathbf{F}}^2$ and $\hat{F}_z$, providing for a total of $d = (2S+1)(2I+1) = 16$ Hilbert space dimensions. More details of the physical and optical properties of cesium can be found in [33, 34].

In order to implement quantum control on a cesium atom, we apply a well-chosen magnetic field. The corresponding interaction between the cesium atom and the magnetic field can be described with the control Hamiltonian,

$$\hat{H}_{\mathrm{C}}(t) = A\hat{\mathbf{I}} \cdot \hat{\mathbf{S}} + \left(\frac{g_{\mathrm{S}}\mu_{\mathrm{B}}}{\hbar}\hat{\mathbf{S}} + \frac{g_{\mathrm{I}}\mu_{\mathrm{B}}}{\hbar}\hat{\mathbf{I}}\right) \cdot \mathbf{B}(t), \tag{2.1}$$

where $g_{\mathrm{S}}$ and $g_{\mathrm{I}}$ are the electron and nuclear $g$-factor respectively. When the magnetic interaction is negligible compared to the hyperfine interaction, $\mu_{\mathrm{B}}|\mathbf{B}| \ll A$, Eq. 2.1 can be rewritten in terms of operators that act separately in the $F^{(\pm)} = 3, 4$

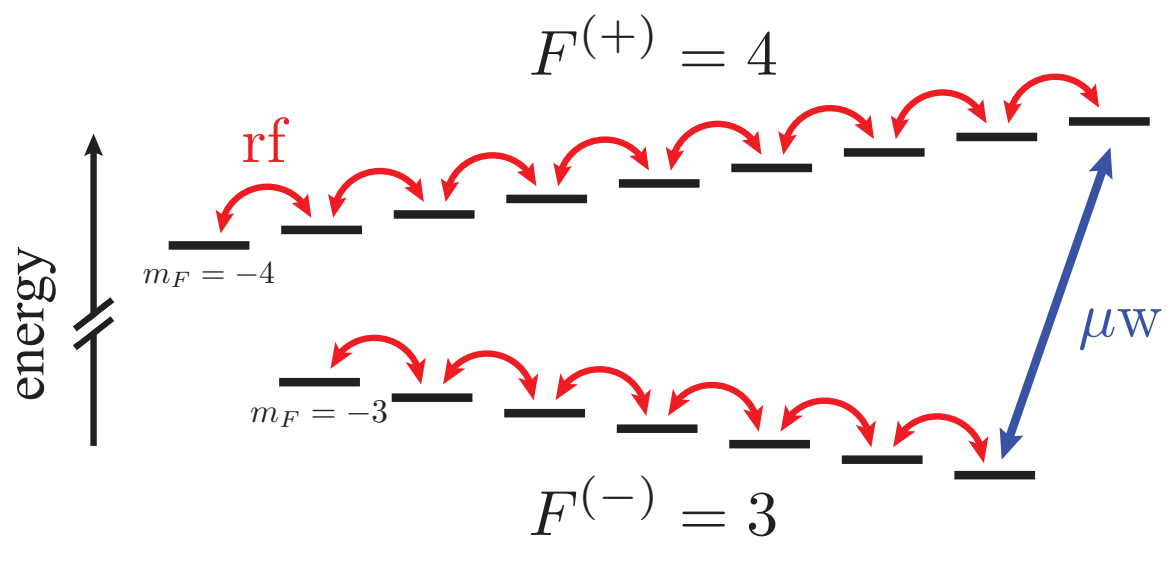

Figure 2.2: Ground hyperfine manifold of $^{133}$Cesium. Rf magnetic fields couple individual magnetic sublevels and generate independent rotations in each manifold (red arrows). Microwave magnetic fields couple states $|F^{(\pm)}, m_F = F^{(\pm)}\rangle$, and generate rotations of the corresponding pseudo-spin (blue arrow). A static bias magnetic field is applied to lift energy degeneracies.

manifolds,

$$\hat{H}_C(t) = \frac{\Delta E_{\text{HF}}}{2}\left(P^{(+)} - P^{(-)}\right) + g_+\mu_B\hat{\mathbf{F}}^{(+)}\cdot\mathbf{B}(t) + g_-\mu_B\hat{\mathbf{F}}^{(-)}\cdot\mathbf{B}(t). \tag{2.2}$$

Here $\Delta E_{\text{HF}}$ is the hyperfine splitting, and $P^{(\pm)}$, $\hat{\mathbf{F}}^{(\pm)}$ and $g_\pm$ are the projectors, angular momenta, and Landé $g$-factors associated with the $F^{(\pm)}$ manifolds, respectively.

In our experiment, $\mathbf{B}(t)$ is an external magnetic field given by

$$\mathbf{B}(t) = B_0\mathbf{z} + \text{Re}[B_{\text{rf}}\left(\mathbf{x}e^{-i\phi_x(t)} + \mathbf{y}e^{-i\phi_y(t)}\right)e^{-i\omega_{\text{rf}}t}] + \text{Re}[\boldsymbol{B}_{\mu\text{w}}e^{-i\phi_{\mu\text{w}}(t)}e^{-i\omega_{\mu\text{w}}t}]. \tag{2.3}$$

As shown in [35, 36], our system is fully controllable through the use of a static bias magnetic field along $\mathbf{z}$, a pair of phase modulated rf magnetic fields along $\mathbf{x}$ and $\mathbf{y}$, and a phase modulated $\mu$w magnetic field coupling states $|F^{(\pm)}, m_F = F^{(\pm)}\rangle$.

Using standard rotating wave approximations for the interaction between the bias magnetic field with rf and $\mu$w fields to model the dynamics, the control

Hamiltonian has the form

$$\hat{H}_{\mathrm{C}}(t) = \hat{H}_0 + \hat{H}_{\mathrm{rf}}^{(+)}[\phi_x(t), \phi_y(t)] + \hat{H}_{\mathrm{rf}}^{(-)}[\phi_x(t), \phi_y(t)] + \hat{H}_{\mu\mathrm{w}}[\phi_{\mu\mathrm{w}}(t)]. \tag{2.4}$$

Here $\hat{H}_0$ is a static term including the hyperfine interaction and Zeeman shift from the bias field, the $\hat{H}_{\mathrm{rf}}^{(\pm)}$ generate $SU(2)$ rotations of the $F^{(\pm)}$ hyperfine spins depending on the rf phases, and $\hat{H}_{\mu\mathrm{w}}$ generates $SU(2)$ rotations of the $|F^{(\pm)}, m_F = F^{(\pm)}\rangle$ pseudospin depending on the $\mu$w phases (see Fig. 2.2). Derivation and further details regarding the explicit form of Eq. 2.4 can be found in [37, 38], as well as in the supplemental material of publications presented in App. **??**.

## 2.2 Optimal Control

In order to implement a desired quantum control task, we employ the tools of optimal control [39] to design rf and $\mu$w control waveforms.

We define a control waveform as a vector of phases, $\vec{\phi} = \{\phi_x(t), \phi_y(t), \phi_{\mu\mathrm{w}}(t)\}$. First, the control waveform is coarse-grained in time to yield a discrete vector of control parameters such that

$$\vec{\phi} = \{\phi_x(t_j), \phi_y(t_j), \phi_{\mu\mathrm{w}}(t_j)\}\,, \tag{2.5}$$

where $j = 1, \ldots, N$. $N$ is the number of discrete phase steps given by $N = T/\Delta t$, where $\Delta t$ is the phase step duration and $T$ is the total control time. Because there are three sets of control fields, there are $3N$ independent control phases in the control waveform. We then feed an initial random guess for $\vec{\phi}$ to a gradient ascent

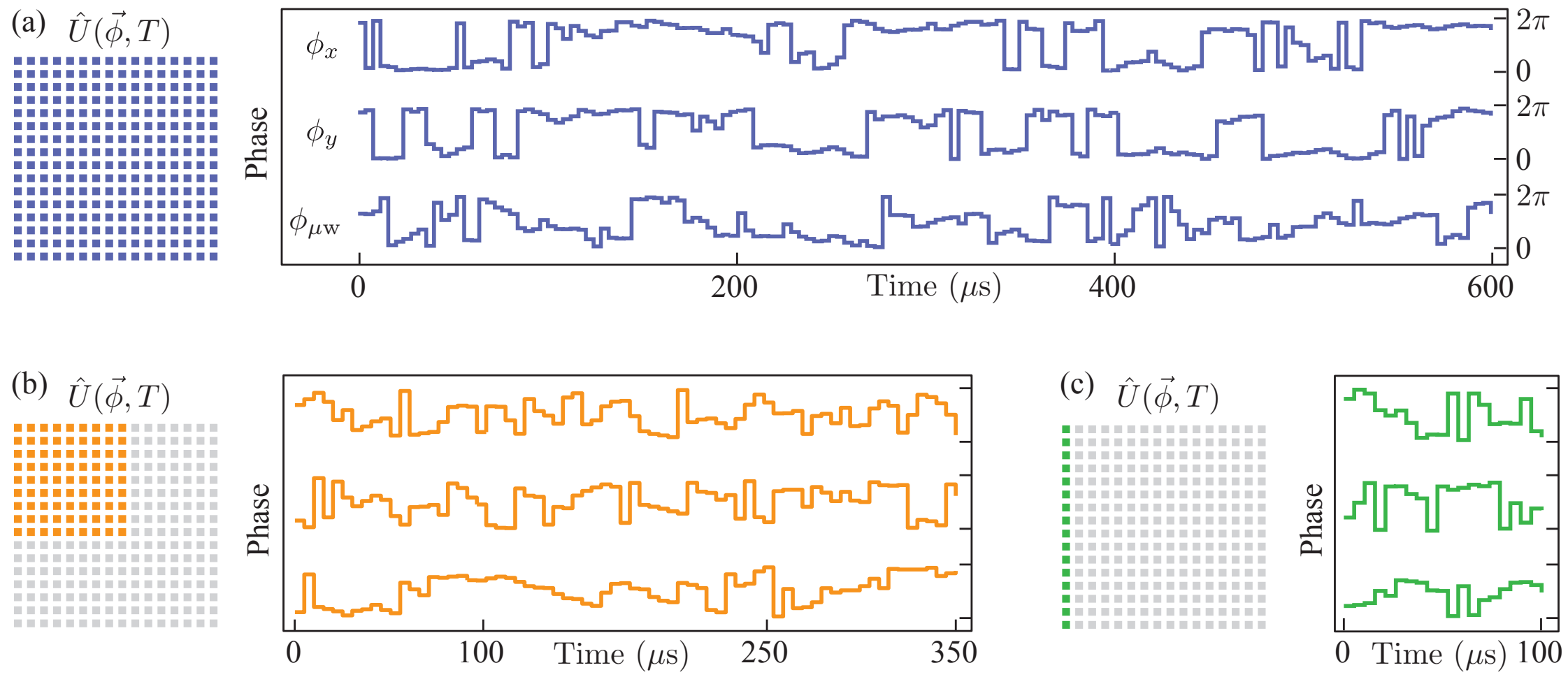

Figure 2.3: From [32]. Examples of phase modulation waveforms for different control tasks. (a) Unitary map on the entire Hilbert space, $k = d = 16$. (b) Unitary map in the $k = 9$ dimensional $F^{(+)}$ manifold. (c) A state-to-state map, which constrains a single column of $\hat{U}(\vec{\phi}, T)$.

algorithm. The algorithm searches for $\vec{\phi}$ that maximizes the fidelity

$$\mathcal{F} = \frac{1}{k^2} \left| \mathrm{Tr} \left[ \hat{W}^\dagger P_f \hat{U}(\vec{\phi}, T) P_i \right] \right|^2, \tag{2.6}$$

where $\hat{W}$ is the target map, $\hat{U}(\vec{\phi}, T)$ is the map generated by the Schrodinger equation at time $T$, $k$ is the dimension of the space and $P_i, P_f$ are the projectors onto the initial and final Hilbert spaces. If $k = 1$, this is a state-to-state map and Eq. 2.6 reduces to

$$\mathcal{F} = \left| \langle \psi_f | \hat{U}(\vec{\phi}, T) | \psi_i \rangle \right|^2. \tag{2.7}$$

If $k = d = 16$, the map is a unitary transformation on the entire space and Eq. 2.6 becomes

$$\mathcal{F} = \frac{1}{d^2} \left| \mathrm{Tr} \left[ \hat{W}^\dagger \hat{U}(\vec{\phi}, T) \right] \right|^2. \tag{2.8}$$

The length of a control waveform reflects the complexity of the corresponding control tasks. Fig. 2.3a shows a waveform designed for a unitary map on the 16-dimensional Hilbert space. In this case, every element of the matrix $\hat{U}(\vec{\phi}, T)$ and $\hat{W}$ are constrained to be identical. A $d$-dimensional unitary matrix $\hat{W}$ requires $d^2 - 1$ real numbers to specify, and thus the waveforms must have at least $d^2 - 1 = 255$ independent control phases. In our setup, the optimal control time and phase step duration for a 16-dimensional unitary map correspond to a total of 450 control phases. Fig. 2.3b shows a waveform for a unitary map on the 9-dimensional subspace of the $F^{(+)}$ manifold. In this case, only the upper left block of $\hat{U}(\vec{\phi}, T)$ must be specified, while the lower right block is an unspecified transformation on the complementary subspace $F^{(-)}$ which can take any form as long as is unitary. The waveform must contain at least $k^2 - 1 = 80$ control phases, and we have successfully used a total of 210. Finally, Fig. 2.3c shows a state-to-state map. In this case, we can choose a basis representation in which the initial state is the first basis state $|\psi_i\rangle = (1; 0; \ldots; 0)$. By doing so, only the first column of $\hat{U}(\vec{\phi}, T)$ is required to calculate the final target state $|\psi_f\rangle$. The waveform must contain at least $2d - 2 = 30$ control phases, and we have successfully used a total of 60. These examples illustrate how control waveforms require fewer independent control phases and become shorter as the constraints on $\hat{U}(\vec{\phi}, T)$ are relaxed.

Besides the control phases, the control Hamiltonian given by Eq. 2.4, is fully determined by an additional set of 6 parameters, $\Lambda = \{\Omega_0, \Omega_{\mathrm{rf}}^x, \Omega_{\mathrm{rf}}^y, \Omega_{\mu\mathrm{w}}, \Delta_{\mathrm{rf}}, \Delta_{\mu\mathrm{w}}\}$. Here $\Omega_0$ is the Larmor frequency at which the spin $F^{(+)}$ precesses in the bias field, $\Omega_{\mathrm{rf}}^x$ and $\Omega_{\mathrm{rf}}^y$ are the rf Larmor frequencies in the rotating frame, $\Omega_{\mu\mathrm{w}}$ is the $\mu$w Rabi frequency and, $\Delta_{\mathrm{rf}}$ and $\Delta_{\mu\mathrm{w}}$ are the detunings of the rf and $\mu$w fields from resonance. Even though these parameters are experimentally set as close as possible to their nominal values, there will be experimental errors and technical limitations to make them imperfect. In this case, we can search for robust control waveform by maximizing the average fidelity

$$\bar{\mathcal{F}} = \int_{\Lambda} p_{\Lambda} \mathcal{F}(\Lambda)\, d\Lambda, \tag{2.9}$$

where $p_{\Lambda}$ is the probability that the parameters take on the values $\Lambda$, and $\mathcal{F}(\Lambda)$ is the corresponding fidelity out of Eq. 2.6. In practice, we have found sufficient to average over discrete values of the parameters such that Eq. 2.9 becomes

$$\bar{\mathcal{F}} = \sum_{\Lambda} p_{\Lambda} \mathcal{F}(\Lambda), \tag{2.10}$$

and for simplicity, we assume each contribution is equally probable. This relatively coarse sampling of the probability distribution speeds up optimization, and we have found that the resulting, optimized control waveform performs well when its fidelity is averaged using a finer sampling of the estimated Gaussian distributions.

Robust control waveforms designed using this approach allows us to perform high-fidelity unitary maps in the presence of small static and/or time varying errors in the parameters. As expected, additional robustness requires more control phases and thus longer total control time. In practice most systems such as ours will have an upper limit on $T$, beyond which added robustness to imperfect parameters is overwhelmed by other errors.

## 2.3 The Cesium Atom in Optical Fields

We now consider the interaction of a cesium atom and a monochromatic optical field. When the frequency of the field is far-off resonance from any of the optical transition frequencies, there is negligible absorption. However, this interaction induces energy shifts of the magnetic sublevels in the ground state hyperfine manifold, which are of particular importance for the experiments presented in this dissertation. A comprehensive treatment of tensor light shifts in alkali atoms can be found in Ref. [4], and a brief summary of the main results is given here.

The interaction between a classical light field and the magnetic sublevels in the electronic ground state manifold can be described by the Hamiltonian

$$\hat{H}_{LS} = -\hat{\mathbf{d}} \cdot \mathbf{E}, \tag{2.11}$$

where $\hat{\mathbf{d}}$ is the induced dipole moment from the atom-light interaction and $\mathbf{E}$ is the electric field. In general, the induced dipole moment is not parallel to the electric field and we write it as $\hat{\mathbf{d}} = \overleftrightarrow{\hat{\alpha}}\mathbf{E}$, where $\overleftrightarrow{\hat{\alpha}}$ is a $3 \times 3$ matrix (the atomic polarizability tensor) that depends on the spin degrees of freedom and therefore is an operator acting in the ground hyperfine manifold. Expressing the electric field into its positive and negative frequency components, $\mathbf{E}^{(\pm)}$, Eq. 2.11 can be rewritten as

$$\hat{H}_{LS} = -\sum_{i,j} \hat{\alpha}_{ij} E_i{}^{(-)} E_j{}^{(+)}. \tag{2.12}$$

Calculating the atomic polarizability tensor using second order perturbation theory, Eq. 2.12 can be put in terms of its irreducible tensor components, and the atom-light interaction Hamiltonian takes the form

$$\begin{aligned}\hat{H}_{LS} = &- \hat{\alpha}^{(0)}\mathbf{E}^{(-)} \cdot \mathbf{E}^{(+)} - \hat{\vec{\alpha}}^{(1)} \cdot (\mathbf{E}^{(-)} \times \mathbf{E}^{(+)}) \\ &- \sum_{i,j} \hat{\alpha}_{ij}^{(2)} \big( E_i^{(-)} E_j^{(+)} - \frac{1}{3}\mathbf{E}^{(-)} \cdot \mathbf{E}^{(+)} \delta_{ij} \big),\end{aligned} \tag{2.13}$$

where $\hat{\alpha}^{(0)}$, $\hat{\vec{\alpha}}^{(1)}$, and $\hat{\alpha}_{ij}^{(2)}$ are the rank 0, 1, and 2 spherical tensor components of $\hat{\alpha}_{ij}$ respectively. Each component of the atom-light interaction acts to change the quantum state of the atomic spins as well as the polarization of the optical field. The polarization changes of the optical field were of particular importance in a quantum state tomography scheme developed during my first years working in the laboratory [25]. For the experiments presented in the main body of this dissertation, we make use of the effects on the atomic spins due to the optical field and ignore the polarization changes.

To better understand light shifts in the ground hyperfine manifold it is useful to write down the light shift Hamiltonian in terms of the optical field polarization $\boldsymbol{\epsilon}$ and the angular momentum $\hat{\mathbf{F}}$, such that

$$\hat{H}_{LS}(F, F') = \sum_{F'} V_0 \left[ C_{FF'}^{(0)} |\boldsymbol{\epsilon}|^2 + C_{FF'}^{(1)} (\boldsymbol{\epsilon}^* \times \boldsymbol{\epsilon}) \cdot \hat{\mathbf{F}} + C_{FF'}^{(2)} \left( |\boldsymbol{\epsilon} \cdot \hat{\mathbf{F}}|^2 - \frac{1}{3} \hat{\mathbf{F}}^2 |\boldsymbol{\epsilon}|^2 \right) \right], \tag{2.14}$$

where

$$V_0 = -\frac{1}{4} \alpha_{J'F'F} |E_0|^2 = \left( \frac{\hbar}{8} \frac{I}{I_{sat}} \right) \frac{\Delta_{FF'}/\Gamma}{(\Delta_{FF'}/\Gamma)^2 + 1/4}. \tag{2.15}$$

Here, $V_0$ is the AC stark shift associated with a light field of intensity $I$ acting on a transition with unit oscillator strength and saturation intensity $I_{\text{sat}} = c\epsilon_0 \hbar^2 \Gamma^2 / 4 |\langle J' || d || J \rangle|^2$. The detuning of the light from the exited to ground transition frequency is $\Delta_{FF'}$ and $\Gamma$ is the natural linewidth of the excited states. Each of the tensor coefficients $C^{(K)}_{FF'}$ come from the Wigner-Eckart theorem and are given by

$$C^{(0)}_{FF'} = (-1)^{3F-F'+1} \frac{1}{\sqrt{3}} \frac{2F'+1}{\sqrt{2F+1}} \left\{ \begin{array}{ccc} F & 1 & F' \\ 1 & F & 0 \end{array} \right\} \left| \mathcal{K}^{J'F'}_{JF} \right|^2, \tag{2.16}$$

$$C^{(1)}_{FF'} = (-1)^{3F-F'} \sqrt{\frac{3}{2}} \frac{2F'+1}{\sqrt{F(F+1)(2F+1)}} \left\{ \begin{array}{ccc} F & 1 & F' \\ 1 & F & 1 \end{array} \right\} \left| \mathcal{K}^{J'F'}_{JF} \right|^2, \tag{2.17}$$

$$C^{(2)}_{FF'} = (-1)^{3F-F'} \frac{\sqrt{30}(2F'+1)}{\sqrt{F(F+1)(2F+1)(2F-1)(2F+3)}} \times \left\{ \begin{array}{ccc} F & 1 & F' \\ 1 & F & 2 \end{array} \right\} \left| \mathcal{K}^{J'F'}_{JF} \right|^2, \tag{2.18}$$

and the coefficient $\mathcal{K}^{J'F'}_{JF}$ is given in terms of a Wigner 6j symbol

$$\mathcal{K}^{J'F'}_{JF} = (-1)^{F'+I+J'+1} \sqrt{(2J'+1)(2F+1)} \left\{ \begin{array}{ccc} F' & I & J' \\ J & 1 & F \end{array} \right\}. \tag{2.19}$$

Casting Eq. 2.14 in this form provides a clear picture of the different effects on the magnetic sublevels of the ground state hyperfine spin manifold due to the atom-light interaction (see Fig. 2.4). The first term and second part of the third term represent an equal energy shift for all magnetic sublevels in a particular hyperfine manifold, which depends on the total intensity of the light field, and is independent of its polarization. If we are restricted to a single hyperfine manifold (e.g., $F^{(+)}$), this linear rank-0 scalar component does not produce any relevant effect. In the present work, however, quantum control typically involves both the $F^{(+)}$ and $F^{(-)}$ manifolds and is strongly affected by any differential light shift $V_{LS} = \Delta_{F(+)} - \Delta_{F(-)}$, induced by the rank-0 scalar term (Fig. 2.4a). The rank-1 vector component is analogous to the interaction with a magnetic field, $H^{(1)} = \mathbf{B}_{\text{fict}} \cdot \hat{\mathbf{F}}$. Here $\mathbf{B}_{\text{fict}}$ stands for a fictitious magnetic field proportional to $(\boldsymbol{\epsilon}^* \times \boldsymbol{\epsilon})$, and thus depends on the ellipticity of the incident laser light (Fig. 2.4b). The rank-2 component contains a non-linear term proportional to $|\boldsymbol{\epsilon} \cdot \hat{\mathbf{F}}|^2$ which generates quadratic energy shifts when the quantization axis is along the light polarization axis (Fig. 2.4c). This later term played a main role in experiments developed in our laboratory during previous years [18].

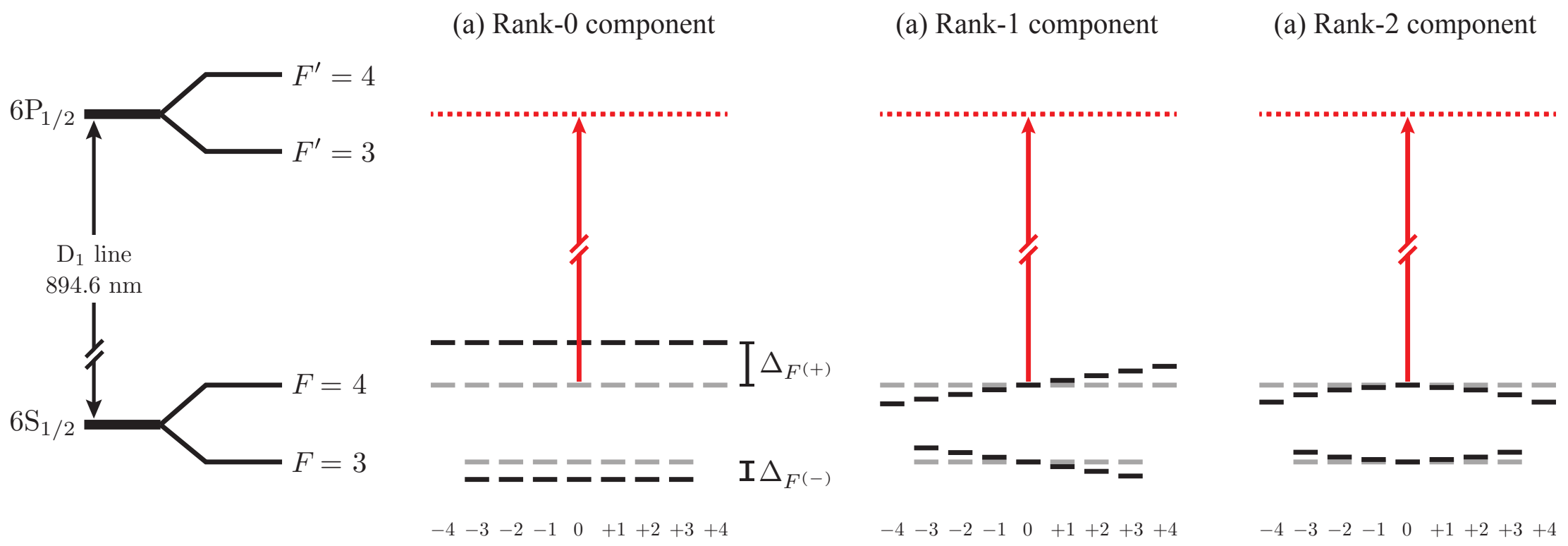

Figure 2.4: Energy diagram of light shifts on the $^{133}$Cs ground manifold from an optical field tuned near the $D_1$ transition. (a) Rank-0 component of light shift: independent of $m_F$. When both ground state manifolds are considered, a non-zero differential light shift $V_{LS} = \Delta_{F(+)} - \Delta_{F(-)}$ can be induced between them. (b) Rank-1 component of light shift: linear dependance on $m_F$. (c) Rank-2 component of light shift: quadratic dependance on $m_F$. Energy shifts are not drawn to scale.

The relative strength of these different contributions depends on the polarization of the light and the detuning from resonance. For linear polarization the rank-1 vector light shift term is always zero. Furthermore, when the detuning is much larger than the excited state hyperfine splitting the rank-0 scalar and rank-1 vector light shifts are much larger than the rank-2 tensor light shift.

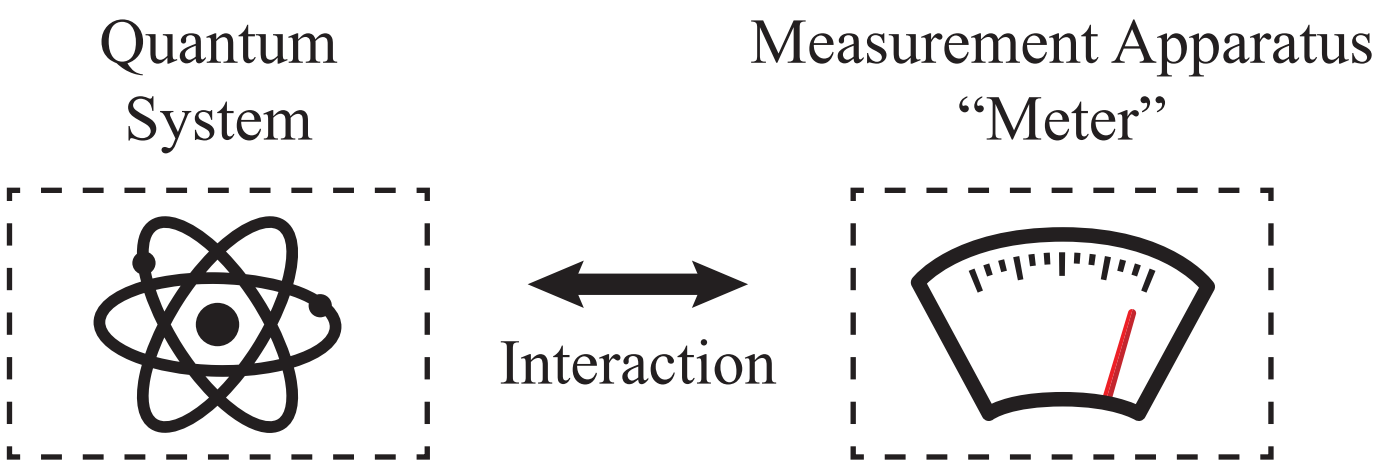

Figure 2.5: Measurement in quantum mechanics. The quantum system of interest interacts with the measurement apparatus or "meter". The interaction establishes a correlation between the quantum system and the meter.

## 2.4 Quantum Measurement

A quantum mechanical measurement is carried out by measuring an observable $\hat{E}$ or more generally implementing a Positive Operator-Valued Measurement (POVM) [40]. A POVM is a set of measurement operators or POVM elements $\{\hat{\mathcal{E}}_\alpha\}$, each of which satisfies the following properties:

1. Hermiticity: $\hat{\mathcal{E}}_\alpha = \hat{\mathcal{E}}_\alpha^\dagger$.
2. Positivity: $\mathrm{Tr}\left[\rho\hat{\mathcal{E}}_\alpha\right] \geq 0$ for any state $\rho$; typically, this property is written as $\hat{\mathcal{E}}_\alpha \geq 0$.
3. Completeness: $\sum_\alpha \hat{\mathcal{E}}_\alpha = \hat{I}$.

In a measurement, the probability of obtaining an outcome $\alpha$ is given by the Born rule

$$p_\alpha = \mathrm{Tr}\left[\rho\hat{\mathcal{E}}_\alpha\right], \tag{2.20}$$

where $\rho$ is the density matrix describing the state of the system. Notice that the properties mentioned above ensure that $p_\alpha \geq 0$ and $\sum_\alpha p_\alpha = 1$.

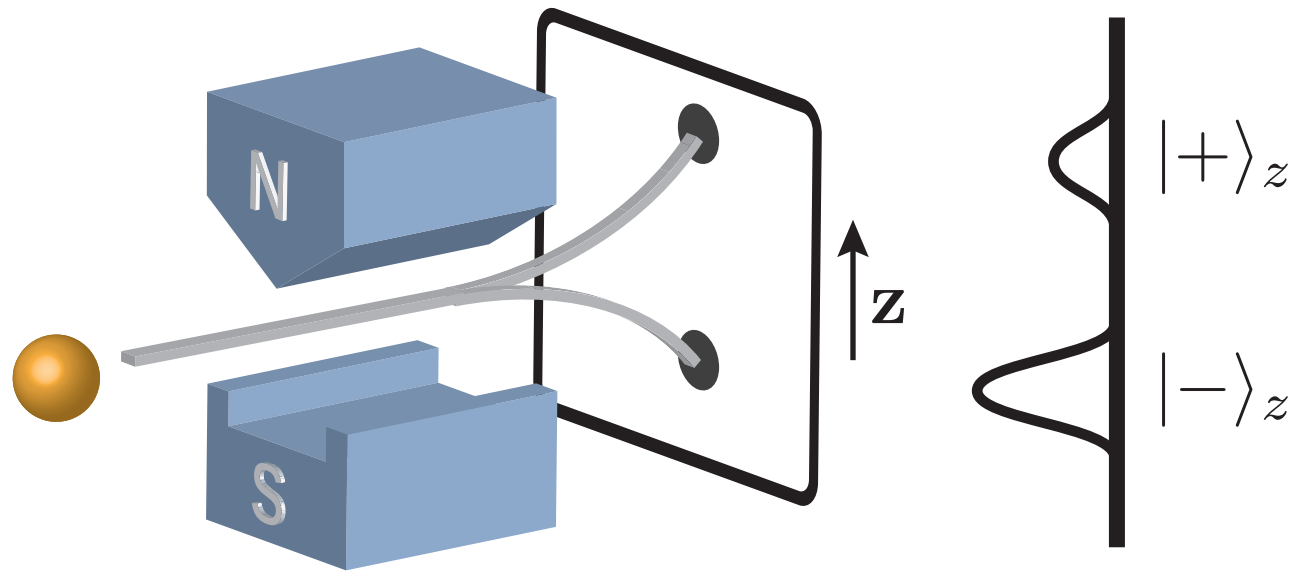

Figure 2.6: Stern-Gerlach analysis on a spin-1/2 particle. Particles pass through an inhomogeneous magnetic field which imparts an upwards or downwards momentum translation on them which depends on their **z**-component spin projection.

To understand the way in which measurement is carried out, we now review the basic concept behind the von Neumann model of projective measurements. To measure an observable $\hat{E}$ in a quantum system, we turn on an interaction between that system observable and another observable $\hat{M}$ that represents the measurement apparatus or "meter"(see Fig. 2.5). $\hat{M}$ might be a different degree of freedom of the same quantum system or a completely different physical object. The interaction establishes a correlation between the eigenstates of the observable (microscopic quantum property) and the distinguishable states of the meter (macroscopic classical property). The correlation allows us to observe the meter and infer the post-measurement state of the system (collapse of the system state), thus performing a measurement of the observable $\hat{E}$.

The model presented above can be applied to describe the measurement technique used in all the experiments presented in this dissertation: Stern-Gerlach analysis (SGA). In the simplest case, the objective of SGA is to measure $\hat{\sigma}_z$ of a spin-1/2 particle after it passes through an inhomogeneous magnetic field given by $\hat{\mathbf{B}} = |B|\hat{z}\mathbf{z}$ (see Fig. 2.6). The magnetic moment of the particle is $\hat{\boldsymbol{\mu}} = \mu\hat{\boldsymbol{\sigma}}$ and the interaction

induced by the magnetic field can be described with the interaction Hamiltonian

$$\begin{aligned}\hat{H} &= -\hat{\boldsymbol{\mu}} \cdot \hat{\mathbf{B}} \\ &= -|B|\mu\hat{\sigma}_z\hat{z}.\end{aligned} \tag{2.21}$$

In this case, $|B|\mu$ is a constant that determines the strength of the interaction and takes a non zero value during the time where the particle travels through the magnetic field, $\hat{\sigma}_z$ is the system observable to be measured which couples to the meter represented by the position degree of freedom $\hat{z}$ of the particle. Now, letting the interaction constant be on from time zero to time $T$, and expanding $\hat{\sigma}_z$ in its eigenbasis $\hat{\sigma}_z = \frac{1}{2}(|+_z\rangle\langle+_z| - |-_z\rangle\langle-_z|)$, the resulting time evolution operator is given by,

$$\begin{aligned}\hat{U}(T) &= e^{-i\hat{H}T} \\ &= |+_z\rangle\langle+_z|e^{i|B|\mu T\hat{z}/2} + |-_z\rangle\langle-_z|e^{-i|B|\mu T\hat{z}/2}.\end{aligned} \tag{2.22}$$

Let $\bar{\psi}(p)$ be the initial wave function of the particle in the momentum representation $\{|p\rangle\}$ and recall that the operator $\exp(ip_0\hat{z})$ generates a translation of the **z**-component of the momentum $\hat{P}$, such that

$$e^{ip_0\hat{z}}\bar{\psi}(p) = \bar{\psi}\left(p - p_0\right). \tag{2.23}$$

Now, if the initial spin state of the particle is a superposition of $\hat{\sigma}_z$ eigenstates given by $|\phi\rangle = a_+|+_z\rangle + a_-|-_z\rangle$, the time evolution operator given by Eq. 2.22 acts on

$|\Psi_0\rangle$ which is the initial product state of system and meter

$$\hat{U}(T)|\Psi_0\rangle = a_+\hat{U}(T)|+_z\rangle|\bar{\psi}(p)\rangle + a_-\hat{U}(T)|-_z\rangle|\bar{\psi}(p)\rangle$$
$$= a_+|+_z\rangle|\bar{\psi}(p - |B|\mu T/2)\rangle + a_-|-_z\rangle|\bar{\psi}(p + |B|\mu T/2)\rangle. \quad (2.24)$$

Eq. 2.24 shows how after the interaction the particle momentum is now correlated with the eigenvalues of the observable $\hat{\sigma}_z$. If the momentum distributions (position distributions at the detector) are distinguishable, then observing the position of the particle at the detector will project the system into the spin state $|+_z\rangle$ or $|-_z\rangle$ with probabilities $|\langle +_z|\phi\rangle|^2 = |a_+|^2$ and $|\langle -_z|\phi\rangle|^2 = |a_-|^2$, respectively. This special class of measurement is known as *projective measurement*. On the other hand, if the particle wavepacket is too wide to uniquely resolve the position distributions then a projective measurement can not be carried out.

In general, we can implement a measurement corresponding to an observable given by $\hat{\sigma}_{\mathbf{n}} = \mathbf{n} \cdot \hat{\boldsymbol{\sigma}}$ by rotating the magnets towards an arbitrary axis $\mathbf{n}$. Alternatively, one can measure the same observable $\hat{\sigma}_{\mathbf{n}}$ by keeping the magnets in their original position and applying a unitary transformation (Sec. 2.2) to the particle before it goes through the inhomogeneous magnetic field. The unitary transformation is designed to map the particle's **n**-axis spin projection into its **z**-axis spin projection. Fig. 2.7 illustrates the two equivalent ways to perform Stern-Gerlach Analysis for a particular example where the objective is to measure the **x**-component spin projection of the particle. Fig. 2.7a shows the rotated magnets into the **x** axis while Fig 2.7b shows a dashed line box representing a unitary transformation $\hat{U}$ which rotates the **x** spin projection into the **z** spin projection.

Stern-Gerlach analysis can be generalized to systems with spins larger than 1/2.

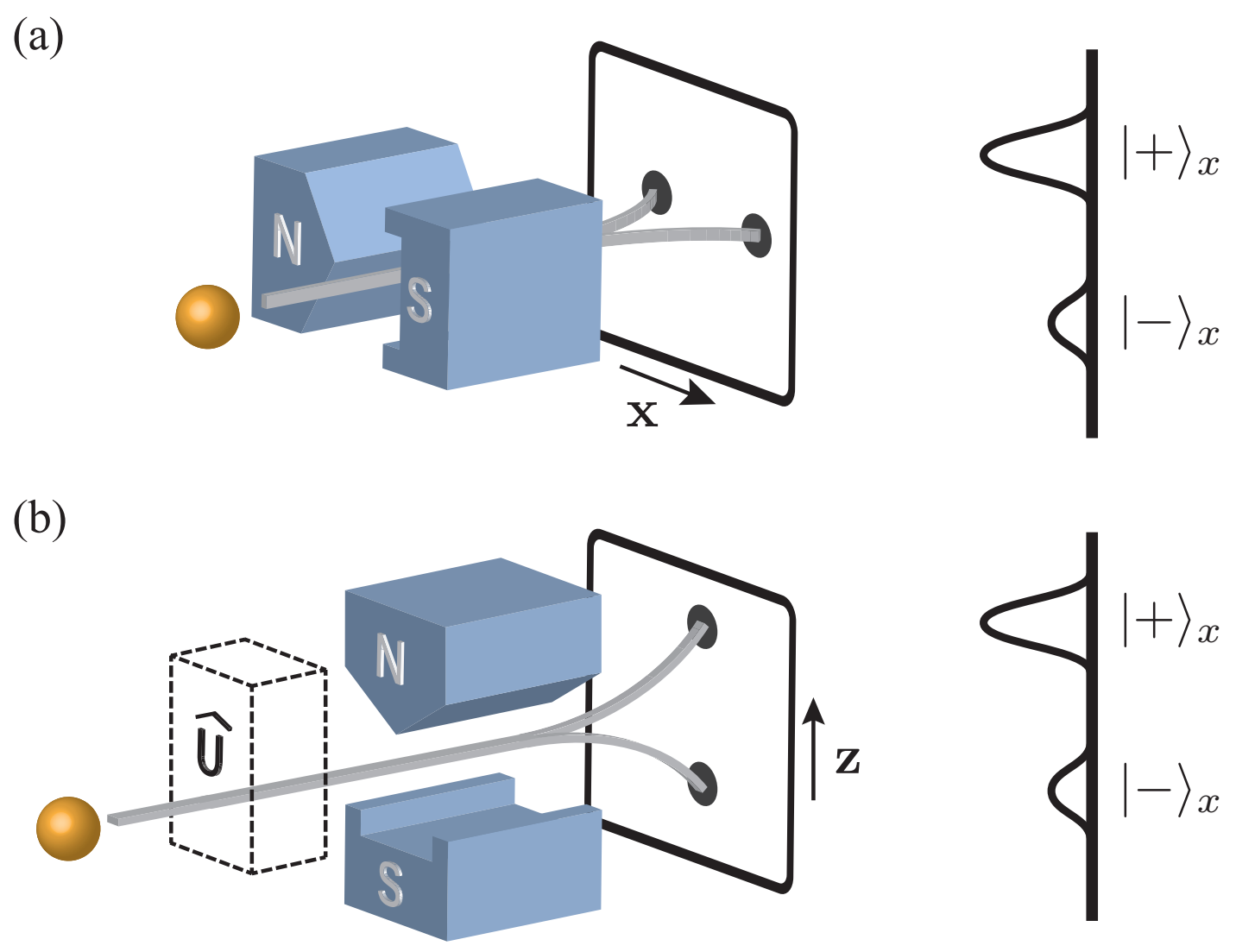

Figure 2.7: Equivalent ways to measure the **x**-component spin projection of a spin-1/2 particle using Stern-Gerlach analysis. (a) Measuring the **x**-component of the spin by rotating the magnets. (c) Measuring the **x**-component of the spin by applying a unitary transformation $\hat{U}$ to rotate the spin projection.

In Chapter 5 we will see how the combination of SGA and full unitary control can be used to experimentally implement any desired orthogonal measurement.

## 2.5 Quantum Tomography

In quantum mechanics, the behavior of an experiment can generally be broken down into three parts (see Fig. 2.8a). First, the system is initialized into a quantum state of interest, this is known as state preparation. For neutral atoms, this part can be implemented with standard techniques such as optical pumping and more advanced tools such as the ones described in [31]. Second, the system undergoes a dynamical evolution which transforms the prepared initial state into a final state. Generally, this evolution can be modeled with a time evolution map, which is linear, trace-preserving, and completely positive. Finally, one performs a measurement on the system to obtain information about it.

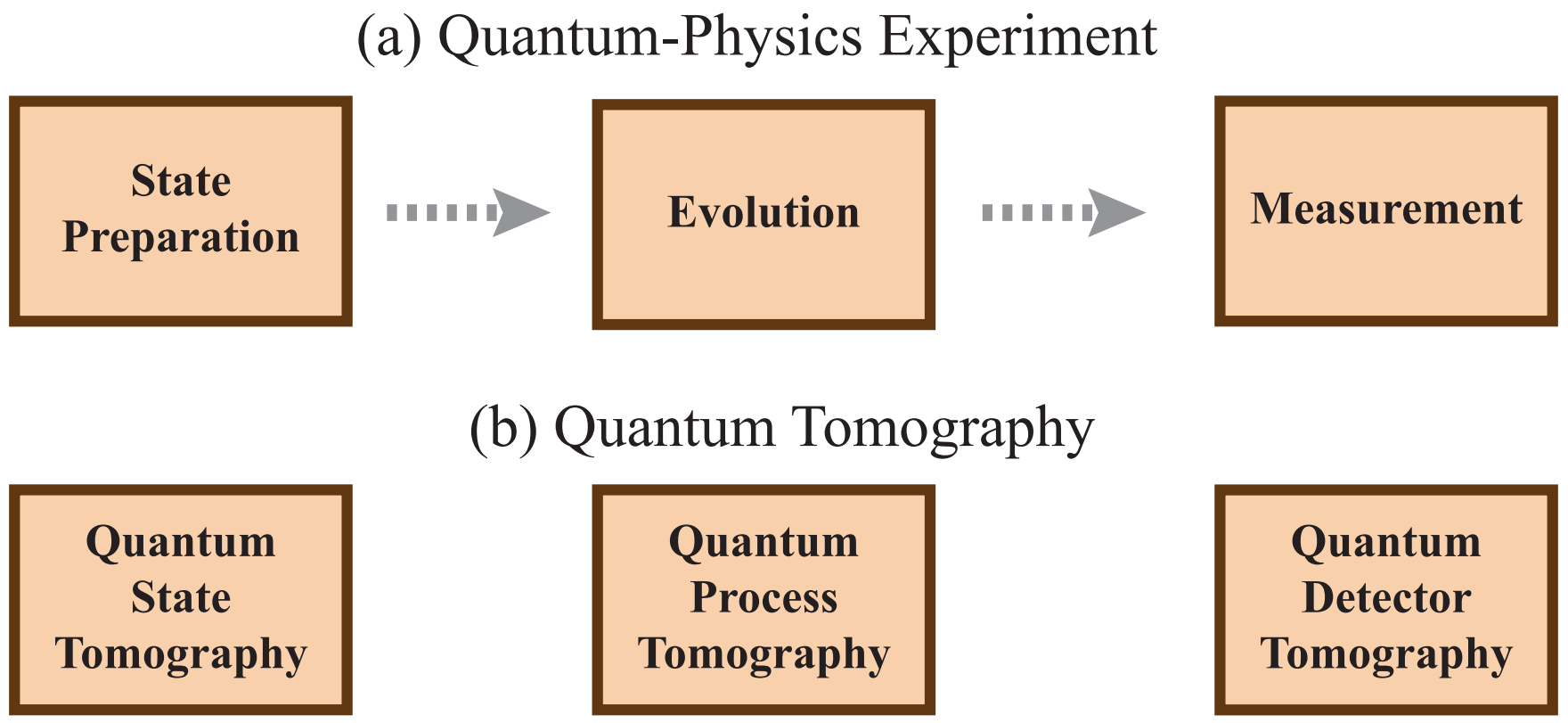

Figure 2.8: (a) A quantum mechanical experiment can be broken down into state preparation, evolution and measurement. (b) Quantum tomography techniques are used to fully characterize each part of the experiment.

In order to verify the performance and diagnose errors that occur in each part of the quantum-physics experiment described above, we can use quantum tomography (QT). Quantum tomography is a suite of techniques employed to fully characterize each part of a quantum-physics experiment (see Fig. 2.8b). Quantum state tomography (QST) is the procedure to experimentally determine an unknown quantum state. In quantum process tomography (QPT), known quantum states are used to probe an unknown quantum process to find out how the process can be described. Similarly, quantum detector tomography (QDT) make use of known states to estimate what measurement is being performed.

The general principle behind quantum state tomography is that, by performing a series of measurements of a set of POVM elements $\{\hat{\mathcal{E}}_\alpha\}$ acting on identically prepared copies of an unknown state $\rho$, one obtains the frequency of occurrences $f_\alpha$ to estimate the probability of outcomes $p_\alpha = \mathrm{Tr}[\rho\hat{\mathcal{E}}_\alpha]$. We then search for an estimated state $\rho_e$ such that the set of $\{p_\alpha^{(e)} = \mathrm{Tr}[\rho_e\hat{\mathcal{E}}_\alpha]\}$ matches the closest, according to some measure, with the set of frequencies of occurrences $\{f_\alpha\}$. When the series of measurements can uniquely identify any arbitrary state, the POVM $\{\hat{\mathcal{E}}_\alpha\}$ is said to be fully informationally complete.

As the dimension of the system of interest grows so does the number of POVM elements needed for an informationally complete tomographic reconstruction. Generally, for a $d$-dimensional Hilbert space, reconstruction is achieved by measuring $d$ observables $\hat{E}$, each with $d$ POVM elements, for a total of $d^2$ POVM elements. For this reason, quantum state tomography is generally a very time consuming procedure when applied to large dimensional systems. The amount of work gets even more demanding when performing quantum process tomography, in which the process is reconstructed by inputing a sequence of $d^2$ pure states into the unknown process and performing full quantum state tomography on each of the resulting output states, requiring $O(d^4)$ total POVM elements.

In order to reduce the resources needed to perform quantum tomography, the experiments presented in this dissertation make use of optimized measurement strategies to perform quantum state tomography. A measurement strategy is specified by a set of POVM elements tailored to be informationally complete for the class of quantum state of interest. For example, if the state we are trying to reconstruct is known to be pure, then there exist measurement strategies designed to take advantage of that information such that we get a reduction in the number of POVM elements required to estimate the unknown state.

# CHAPTER 3

# EXPERIMENTAL APPARATUS AND TECHNIQUES

In order to execute quantum control and measurement tasks in our physical system, a large variety of experimental techniques have to be implemented. This chapter describes the experimental apparatus and techniques that allows us to prepare, manipulate and measure the atomic spin state of an ensemble of cold cesium atoms. We begin with a brief discussion of laser cooling and trapping as well as spin polarization via optical pumping. The next section discusses how we produce static, rf, and $\mu$w fields for control of the atomic spin state. The final section describes the implementation of Stern-Gerlach analysis which allow us to measure the population of atoms in each magnetic sublevel in the ground state manifold.

## 3.1 Experimental Apparatus

A schematic of our experimental apparatus is presented in Fig. 3.1. The setup consist of an all-glass vacuum cell where we prepare an ensemble of cold trapped atoms with a magneto-optical trap (MOT) followed by polarization gradient cooling. The cell is surrounded by a plexiglass cube supporting several sets of coils that we use to generate constant and rf magnetic fields. Microwave magnetic fields are generated using two horn antennas placed slightly above the plexiglass cube. Measurement is carried out using Stern-Gerlach analysis (SGA) which we implement by letting the atoms fall in a magnetic field gradient provided by the MOT coils and by inferring the magnetic populations from the time-dependent fluorescence excited by a probe beam and detected with a photodiode.

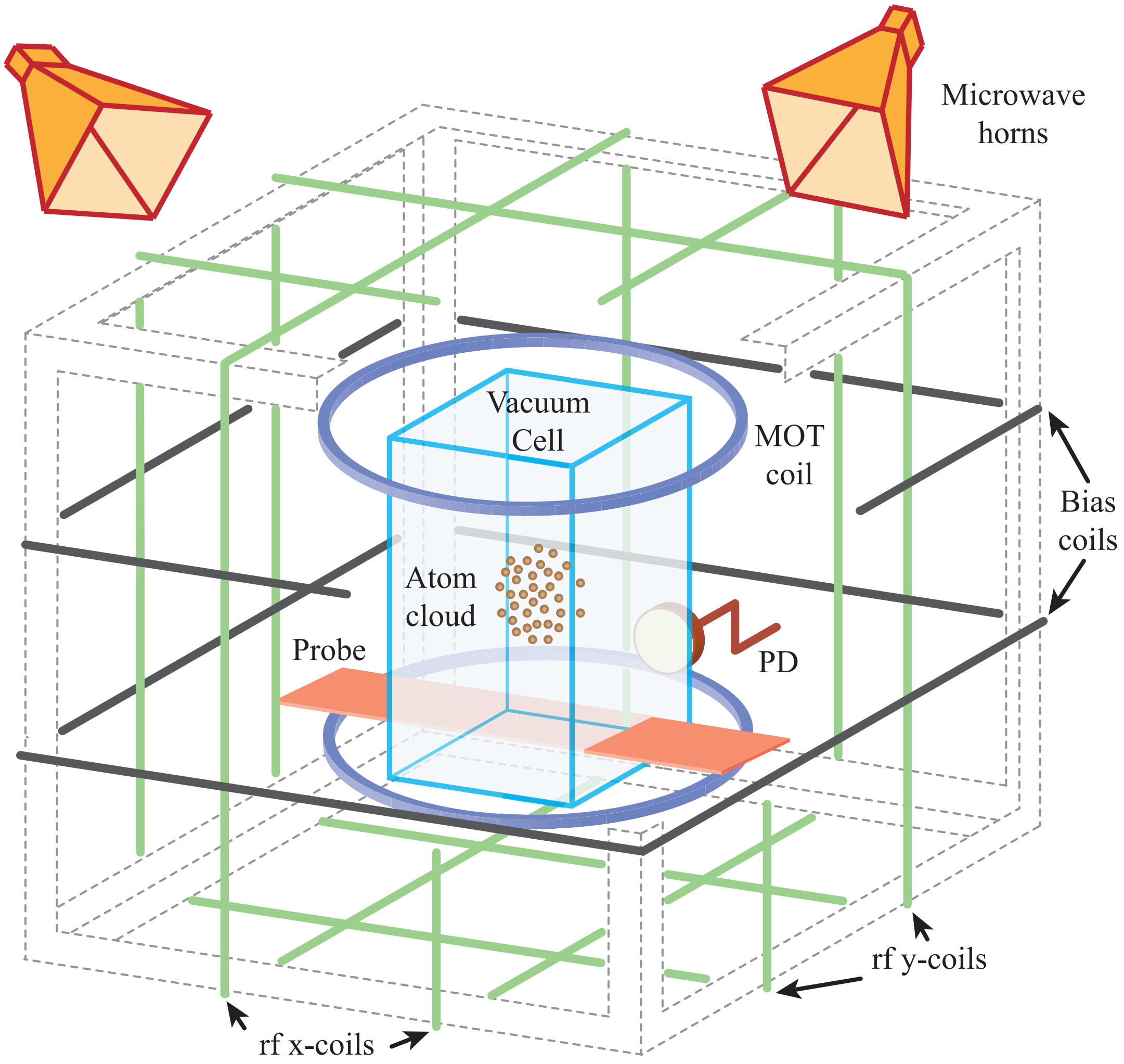

Figure 3.1: Schematic of our experimental apparatus. Laser cooled cesium atoms are prepared in an all-glass vacuum cell centered within a plexiglas cube sustaining the bias and rf coils. Microwave magnetic fields are generated using two horn antennas. Stern-Gerlach analysis is performed by letting the atoms fall in a magnetic field gradient provided by the MOT coils and by inferring the magnetic populations from the time-dependent fluorescence excited by a probe beam and detected with a photodiode (PD).

In order to faithfully generate control magnetic fields in our experiment, transient magnetization effects and eddy currents must be suppressed. To achieve this, we minimize the presence of magnetizable and conductive materials near the vacuum cell. All nearby hardware, including our optical table, is non-magnetic, while sources of DC and AC magnetic fields such as the vacuum pump, power sources and optical isolators are placed at a distance. In addition, we also require a high level of background magnetic field suppression. The use of passive magnetic shielding is not a viable approach, since our experimental setup requires good optical access and application of rapidly time-varying control fields. As an alternative, our experiment is triggered with the power-line cycle such that the constant and spatially inhomogeneous background magnetic fields are reproducible between experimental cycles. We then use our cold atom ensemble as an *in situ* magnetometer to measure the background field and subsequently we cancel it out by applying a "nulling" field. Our background field measurement scheme involves the application of a series of $\mu$w pulses and the full details for the procedure can be found in in Ref. [41]. In order to apply the nulling field we use three orthogonal pairs of compensating coils that surround the entire apparatus (not shown in Fig. 3.1). Each pair of coils is controlled by two independent precision current supplies allowing us to generate constant magnetic field offsets as well as gradients. In our laboratory, we have been able to obtain an average residual background magnetic field which is typically below $\sim 100\ \mu$G [42].

## 3.2 Preparation of the Atomic Ensemble

The first step for all experiments presented in this dissertation is to prepare an ensemble of laser cooled and trapped cesium atoms. Techniques for laser cooling and trapping were extensively developed over the last few decades [43–46] and the details regarding the implementation of these techniques in our experiment can be found in previous dissertations from this group [37, 47]. Here, we review the important features of the setup that are relevant to understand the experiments described in this dissertation.

Cesium atoms are contained as a dilute vapor in a vacuum cell at a pressure of $\sim 10^{-8}$ Torr. A standard 3D MOT is implemented with three pairs of counter-propagating laser beams and a magnetic field gradient produced by driving current through a pair of coils (MOT coils) arranged in anti-Helmholtz configuration (see Fig. 3.2a). The trap loads a sample of a few million atoms in a volume of $\sim 0.5$ mm$^3$ with a temperature of $\sim 100$ $\mu$K in a couple seconds. Then, the magnetic field gradient is turned off and the atoms are released into optical molasses where we apply a position-dependent polarization gradient, generated by the MOT beams. This procedure cools down the atoms further to a temperature of $\sim 3$ $\mu$K after 10 ms. At this point, the laser cooling beams are turned off and the atoms free fall due to gravity for the remainder of the experiment. Typically, the control experiments described in this dissertation take place in a time interval no longer than 15 ms, during which the atomic ensemble have only displaced $\sim 1$ mm from its starting position, meaning that, motional and collisional effects of the atoms can be ignored.

Upon completion of the trapping and cooling phase, the quantum state of the ensemble of atoms is described by a mixed state in the electronic ground $F = 4$ manifold. We proceed to initialize all the atoms into a single spin polarized state via optical pumping. The outcome of this procedure is an atomic ensemble all prepared in a pure quantum state which is easy to verify and offers a convenient starting point for our control experiments.

We begin optical pumping by applying a small bias magnetic field of $\sim 20$ kHz strength along $\mathbf{y}$ using the optical pumping pair of coils shown in Fig. 3.2a, this defines a quantization axis that prevents any magnetic sublevel in the $y$ basis from Larmor precessing. At the same time, we make use of a resonant optical beam driving the $F = 4 \to F' = 4$, $\sigma_+$ electric dipole transitions on the $D_2$ line. Driving this transition causes the atoms to accumulate in the *dark state* $|F = 4, m_{F_y} = 4\rangle$, where there is no longer any resonant transition with $\Delta m_{F_y}$ and photon absorbtion no longer occurs. In addition with the $F = 4 \to F' = 4$ pumping light, we use resonant light driving the $F = 3 \to F' = 4$, $\pi$ and $\sigma_+$ electric dipole transitions on the $D_2$ line to pump atoms out of the $F = 3$ manifold. This is necessary, as scattering of light from the $F = 4 \to F' = 4$, $\sigma_+$ transition can optically pump atoms into $F = 3$. In order to avoid pumping into a dark state in the $F = 3$ manifold, we drive $\pi$ and $\sigma_+$ transitions by aligning the $F = 3 \to F' = 4$ slightly off from the $\mathbf{y}$ axis. Lastly, in our experimental setup, pumping and repumping beams propagate in opposite directions to balance the radiation pressure from each other during the optical pumping process. In our laboratory, this process takes a few milliseconds to be implemented and the resulting population of the atoms in the desired state is $\sim 97\%$. The remaining atoms end up in nearby magnetic sublevels such as $|F = 4, m_{F_y} = 3\rangle$ and $|F = 4, m_{F_y} = 2\rangle$, mostly due to polarization impurity in the beams.

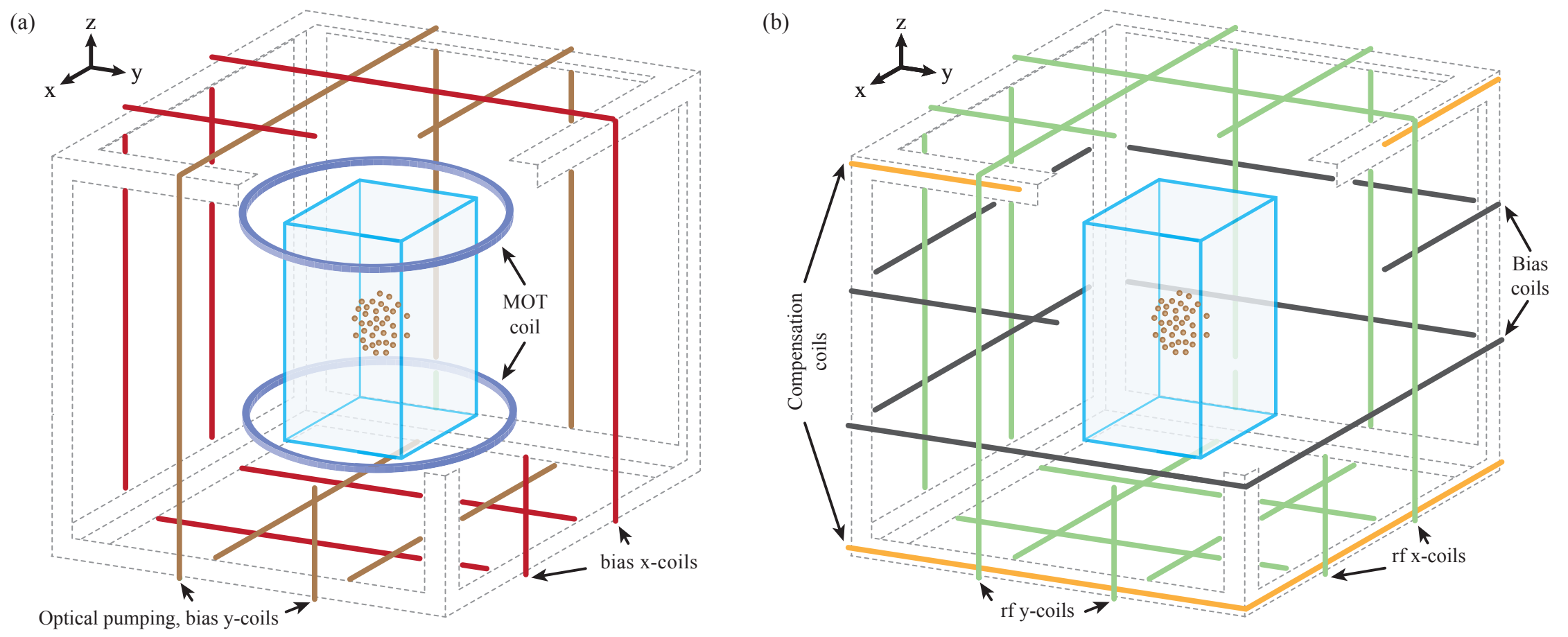

Figure 3.2: Coils arrangement in our experimental setup. MOT coils are connected in anti-Helmholtz configuration to produce a magnetic field which is zero at the center of the plexiglass cube and its magnitude increases linearly in every direction. All other coil pairs are connected in Helmholtz configuration to produce an homogeneous magnetic field at the position of the atoms. (a) Shows the coils used for laser cooling, trapping, and state initialization while (b) shows the coils used for quantum control tasks.

Once the atoms have been optically pumped, the pumping beams and bias magnetic field along **y** are turned off and we apply a short magnetic field along the **x** axis using the coils shown in Fig. 3.2a. The state $|F = 4, m_{F_y} = 4\rangle$ is rotated to the state $|F = 4, m_{F_z} = 4\rangle$ (from now on we will work in the $z$ basis and drop the subscript) via Larmor precession, after which a large bias field of 1 MHz strength is turned on along the **z** axis using the bias coils shown in Fig. 3.2b. This magnetic field will remain on for the duration of our control experiments and ensure that the initial quantum state of the atoms will not evolve until further manipulation is initiated using our rf and $\mu$w control fields.

In order to prepare the atomic ensemble into the purest initial state before our control experiments take place we perform a final preparation step. As men-tioned above, after performing optical pumping we are left with some atoms in $|F = 4, m_F = 3\rangle$ and $|F = 4, m_F = 2\rangle$. To remove these atoms from the ensemble, we apply a $\mu$w $\pi$-pulse to transfer atoms in $|F = 4, m_F = 4\rangle$ to $|F = 3, m_F = 3\rangle$. Then, we briefly turn on an optical field resonant with the $F = 4 \rightarrow F' = 5$ on the $D_2$ line transition such that atoms remaining in the $F = 4$ manifold get pushed out of the ensemble. This leave us with a very pure (>99.5%) atomic ensemble prepared in the single spin polarized state $|F = 3, m_F = 3\rangle$. This state is our starting point for all the control experiments we implement in the laboratory.

## 3.3 Magnetic Fields for Quantum Control

As described in Sec. 2.2, all the experiments presented in this dissertation make use of magnetic fields to manipulate the internal spin state of the atoms. In this section we discuss the necessary hardware employed in the laboratory to generate static, rf, and $\mu$w magnetic fields in order to perform quantum control.

The vacuum cell is surrounded by a plexiglass cube where we have wrapped around multiple sets of square coils designed to approximate Helmholtz coil pairs. This provides spatially homogeneous fields at the center of the cube, where the atomic ensemble is located. Separate, orthogonal pairs of coils are used to generate the bias magnetic field along the $\mathbf{z}$ direction and rf magnetic fields along the $\mathbf{x}$ and $\mathbf{y}$ directions (Fig. 3.2b). The $\mu$w field is produce by two horn antennae, which facilitates the creation of spatial power homogeneity across the ensemble.

The bias magnetic field is produced by connecting the bias coil pair in the $\mathbf{z}$ direction to an Arroyo 4304 constant current laser driver. The Arroyo is a 5A, 9V current source costum modified by the manufacturer to provide fast switching time and to drive the inductive load of the coils ($\sim 40\ \mu$H). In our configuration, the Arroyo is able to produce a static magnetic field of up to 3 G, which generates an atomic energy shift of $\Delta E_m/h = 1$ MHz. However, the field produced by the Arroyo is not completely stable after it turns on, drifting a small amount during the experiment. In order to produce a bias field that is stable to 10 parts per million we make use of an additional pair of compensation coils also oriented along the $\mathbf{z}$ axis, which are driven by a low power amplifier based on an OPA227 op-amp. This extra pair of coils is placed as far as possible from the bias field pair to minimize the mutual inductance effects between them (Fig. 3.2b). The total field produced by the bias current coils and the compensation coils can be then stabilized to the required level. The details of the method used to measure the total bias field and application of the compensation field are discussed in [37].

The rf field currents for the $\mathbf{x}$ and $\mathbf{y}$ directions are each produced by an Amp-Line AL-50-HF-A power amplifier. The amplifier is a 28 $V_{\mathrm{RMS}}$, 100 W, variable gain source with a -3 dB rolloff at 1.2 MHz. Its input is driven by a Tabor 8026 arbitrary waveform generator which can supply arbitrary time-varying voltages with a sampling rate up to 100 MHz. The output of the amplifier is connected in series by coaxial cable to a 30 $\Omega$ Caddock film resistor with an intrinsic inductance of approximately 20 nH. The resistance and inductance for each coil is approximately 3 $\Omega$ and 5 $\mu$H, respectively. Using this configuration we are able to produce rf magnetic fields which generate geometric rotation in each electronic ground hyperfine manifold independently, with an rf Larmor frequency of $\Omega_{\mathrm{rf}} = 25$ kHz.

The $\mu$w magnetic field is produced using a chain of several $\mu$w components shown in Fig. 3.3. The microwave source is an ultra-stable HP 8672A synthesizer running at $f_{\text{HP}} \approx 9.2$ GHz. The signal passes through a digitally controlled switch, allowing us to turn the source on and off during the experiment via computer control. A pre-amp increases the signal amplitude before it is mixed with a $f_{\text{tabor}} \approx 30$MHz signal from a Tabor WW2571A arbitrary waveform generator that provides phase modulation for the $\mu$w control. The mixer is a single sideband mixer whose output is dominated by $f_{\mu\text{w}} = f_{\text{HP}} - f_{\text{tabor}}$. The Tabor WW2571A allows us to arbitrarily modulate the frequency, phase, and amplitude of the lower frequency signal input to the mixer, which correspondingly modulates the output signal of the mixer. The output of the mixer passes through another pre-amp before it is split and goes to two $\mu$w power amplifiers. Splitting the signal allows us to increase the total power radiated and also allows us to empirically adjust the position of the horns to make the resulting intensity pattern more spatially homogeneous at the location of the atoms. Using this configuration we are able to produce $\mu$w fields which generate SU(2) rotations between $|F = 4, m_F = 4\rangle$ and $|F = 3, m_F = 3\rangle$, with a Rabi frequency of $\Omega_{\mu\text{w}} = 27.5$ kHz.

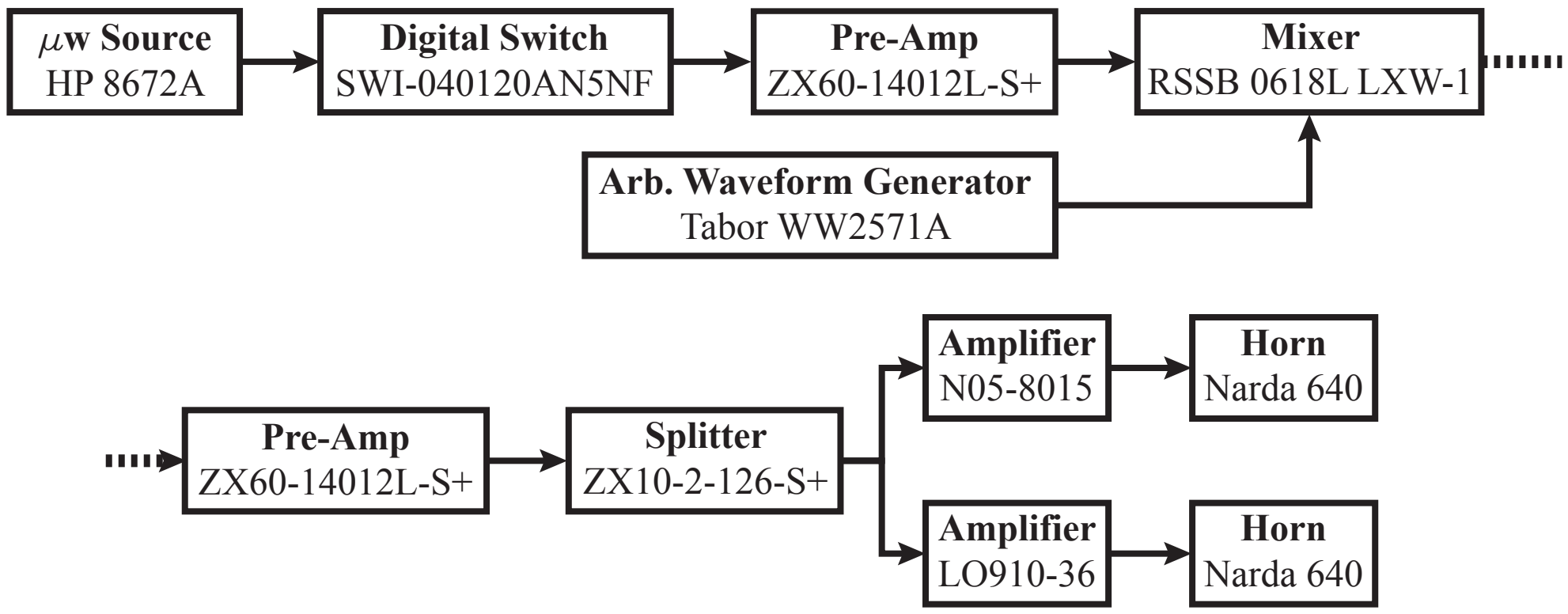

Figure 3.3: Chain of $\mu$w components.

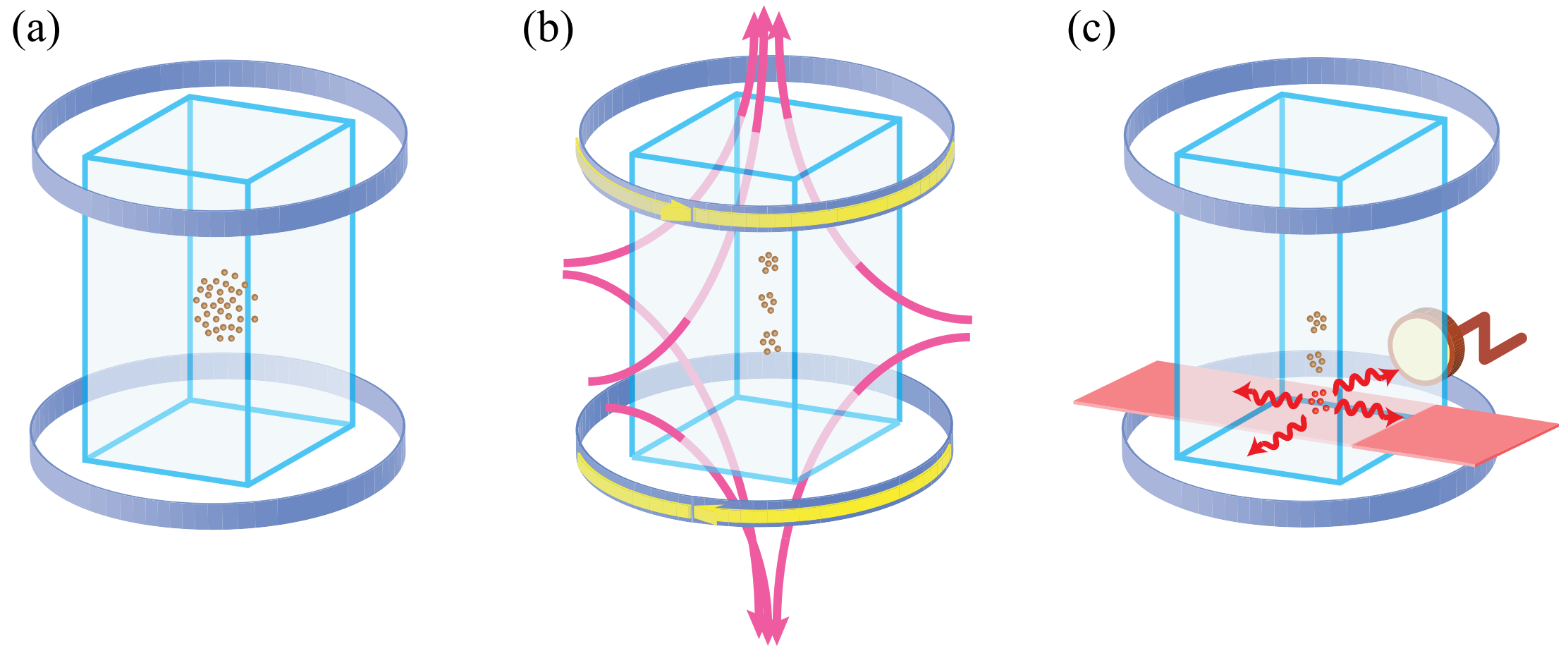

Figure 3.4: Stern-Gerlach analysis. (a) After the control experiment, the cloud of atoms is falling due to gravity. (b) A current running through the MOT coils generates a magnetic field gradient that spatially separates the atom cloud into several groups associated with their $m_F$ spin state. (c) Each group passes through a resonant probe beam that makes the atoms fluoresce. The fluorescence is captured by a photodiode.

## 3.4 Measurement via Stern-Gerlach Analysis

All the measurements performed in our experiments are carried out using Stern-Gerlach analysis (SGA). To perform SGA, the atomic ensemble is released into a magnetic field gradient produced by the coils used for the MOT (Fig. 3.4). As it was described in Sec. 2.4, the interaction between the magnetic field gradient and the atomic spin state of the atoms produces a translation of the momentum degree of freedom of the atoms which is proportional to the $m_F$ value of the magnetic sublevels in which they are. This causes the cloud of atoms to spatially separate as it travels towards the bottom of the vacuum cell. Near the bottom of the cell there are two sheet-shaped laser beams (SGA beams) overlapping with each other, one beam is resonant with the $F = 4 \rightarrow F' = 5$ on the $D_2$ line transition, and the other one is resonant with the $F = 3 \rightarrow F' = 4$ on the $D_2$ line transition. As the different groups of atoms pass through the resonant beams, they fluoresce and part of this fluorescence is captured with a photodiode located near one of the side walls of the vacuum cell.

To measure the fraction of atoms in each magnetic sublevel of the $F = 4$ manifold, we only make use of the $F = 4 \rightarrow F' = 5$ SGA beam. In this case, all the atoms initially in the $F = 3$ manifold are invisible to the beam and because the $F = 4 \rightarrow F' = 5$ is a closed atomic transition, any atom excited to $F' = 5$ from the $F = 4$ manifold fall back to $F = 4$, where it can reabsorb and scatter again producing a Stern-Gerlach analysis signal with multiple scattering events.

To measure the fraction of atoms in each magnetic sublevel of the $F = 3$ manifold, we use both the $F = 3 \rightarrow F' = 4$ and the $F = 4 \rightarrow F' = 5$ SGA beams. In this case, the beam resonant with the $F = 3 \rightarrow F' = 4$ transition is not closed, allowing atoms in the excited $F' = 4$ manifold to decay into both the $F = 3$ or $F = 4$ ground state manifolds. Since multiple scattering events are required to produce sufficient fluorescence signal, the use of the $F = 4 \rightarrow F' = 5$ beam is necessary. In order to avoid measuring atoms initially prepared in the $F = 4$ manifold, we flash on a beam resonant to the $F = 4 \rightarrow F' = 5$ on the atomic cloud before SGA. This beam, which propagates along a single axis, produces radiation pressure which pushes or 'blows away' the $F = 4$ atoms out of the detection region.

An example of experimental signals obtained using SGA is shown in Figs. 3.5a and 3.5d. In these figures we see the raw detected fluorescence signals from a state that has support on all 16 magnetic sublevels. There is a total of 9 and 7 peaks corresponding to each of the magnetic sublevels in the $F = 4$ and $F = 3$ manifolds, respectively. We also see that the distributions of atoms arriving at earlier times are narrower, as these atoms are accelerated towards the SGA beams by the magnetic field and are moving faster as they pass through the beams.

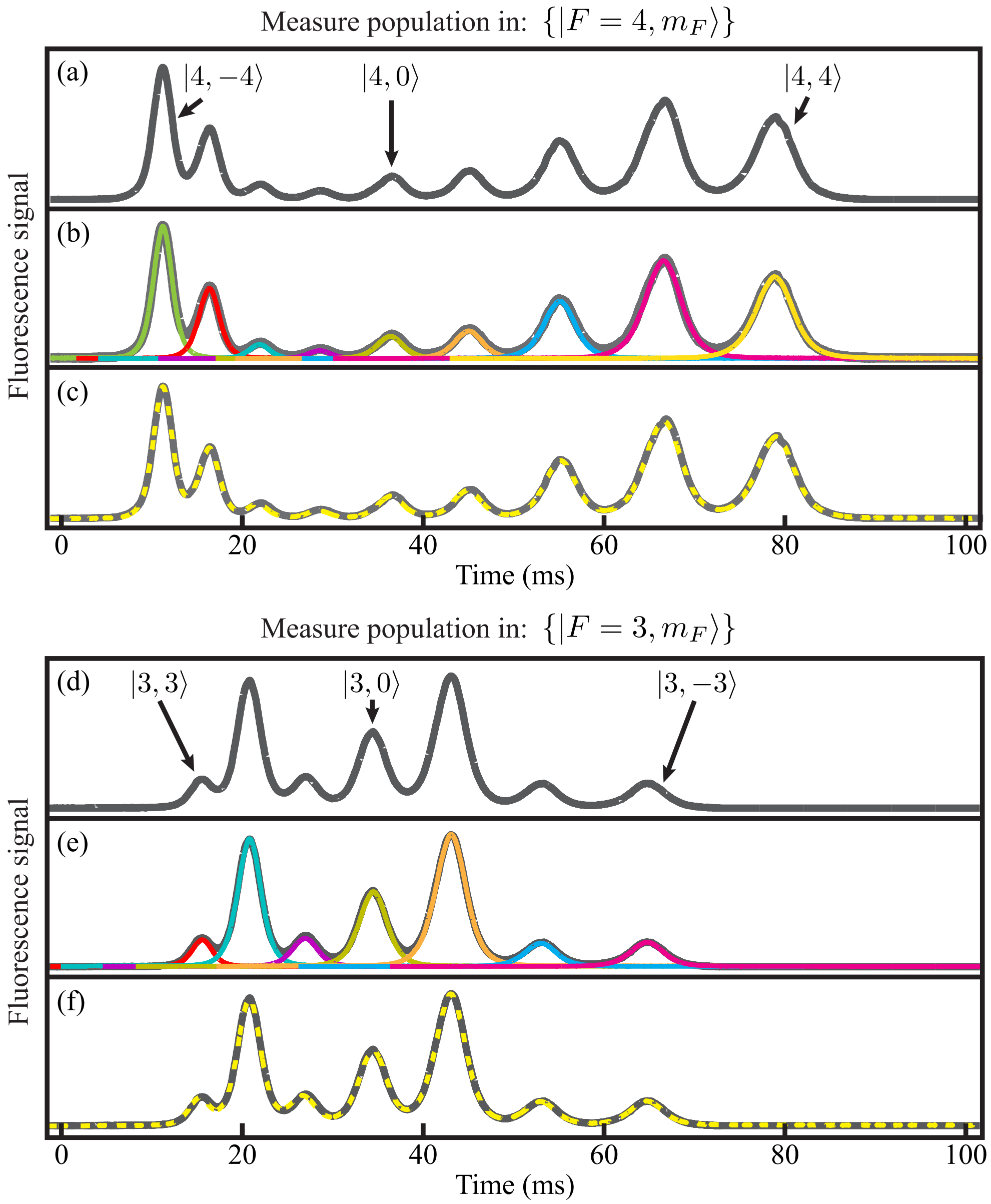

Figure 3.5: Example of Stern-Gerlach signals for the $F = 4$ and $F = 3$ manifolds. (a) and (d) show the raw experimental signals collected from the photo detector. (b) and (e) show the individual fits for each of the peaks in the raw signals, the area under each fitted curve is proportional to the population in each magnetic sublevel. (c) and (f) show the experimental signal in grey and the overall fitted signal in dashed yellow.

It is important to note that unlike the SGA example presented in Ch. 2, the measurement we perform in our laboratory does not correspond to a fully projective measurement since the distributions of atoms for each magnetic sublevel partially overlap with each other. However, in both cases the associated POVM elements that we measure are of the form $E_{F,m_F} = g_{F,m_F}|F, m_F\rangle\langle F, m_F|$, noting that $g_{F,m_F}$ is the distribution of atoms associated with the state $|F, m_F\rangle$, which is known and may be distinct like in the spin-1/2 case or partially overlapping like in our case.

In order to obtain good estimates of the population of atoms in each magnetic sublevel, we fit the signals $S_{\mathrm{SGA}}$ in Fig. 3.5a and 3.5d to a weighted sum of individual distributions

$$S_{\mathrm{SGA}} = \sum_{F,m_F} f_{F,m_F} g_{F,m_F}. \tag{3.1}$$

This yields the set $\{f_{F,m_F}\}$ from which we obtain estimates of the probability of each outcome,

$$f_{F,m_F} \approx p_{F,m_F} = \mathrm{Tr}\left(\rho E_{F,m_F}\right). \tag{3.2}$$

Figs. 3.5b and 3.5e show the individual fits for every peak while Figs. 3.5c and 3.5f show the overall fitted signal in dashed yellow. The previous method is equivalent to estimate $p_{F,m_F}$ from separate, orthogonal measurements on all the individual atoms in the ensemble. The advantage in our approach, however, is that we effectively measure $\sim 10^6$ atoms in parallel, greatly speeding up data acquisition and effectively eliminating measurement statistics as a source of error.

Lastly, because our system allow us to perform a measurement with a total of 16 orthogonal outcomes, we can implement non-orthogonal POVMs with up to 16 outcomes on any chosen subspace using the Neumark extension [48]. The central concept of this extension is to utilize the large $d = 16$ Hilbert space to make a measurement of an orthonormal basis, such that there is a one to one correspondence between the non-orthogonal POVM elements $\{E_\alpha\}$ in the subspace onto the orthogonal POVM elements $\{\tilde{E}_\alpha\}$ in the $d = 16$ space, that is

$$\tilde{E}_\alpha^{(i)} = P E_\alpha^{(i)} P, \tag{3.3}$$

where $P$ is the projector on that subspace.

# CHAPTER 4

# QUANTUM CONTROL EXPERIMENTS

This chapter discusses the experimental results from several quantum control projects implemented in the 16-dimensional Hilbert space associated with the electronic ground state of cesium. We begin with a brief review of the experiment that implements unitary transformations with built-in robustness to static and dynamic perturbations. We then present the results of our exploration of inhomogeneous quantum control. Here, the central idea is based on performing different unitary transformations on qudits that see different light shifts from an optical addressing field. A detailed discussion of the original experiment to implement 16-dimensional unitary transformations can be found in the dissertation of Brian E. Anderson [38].

## 4.1 Unitary Transformations in a Large Hilbert Space

A unitary transformation is the most general input-output map available in a closed quantum system. In a laboratory setup, the primary challenge lies in implementing such transformations with high accuracy in the presence of experimental imperfections and decoherence. For two-level systems (qubits) most aspects of this problem have been extensively studied [3]. Over recent years, the efforts in our laboratory have centered in the implementation of a protocol that can implement any arbitrarily chosen unitary transformations in the 16-dimensional hyperfine ground manifold of cesium atoms. Our control scheme (described in Sec. 2.2) makes use of phase modulated rf and microwave magnetic fields to drive the atomic evolution. The phase modulated control waveforms are found numerically using the tools of optimal control.

In order to implement unitary maps with high accuracy in the presence of experimental imperfections we make use of robust control waveforms designed using Eq. 2.10. In our experimental setup we have found that the dominant source of errors is given by the spatial inhomogeneity of the bias field strength. In this case, it is sufficient to use a search algorithm where the cost function is averaged over two points such that Eq. 2.10 turns into

$$\bar{\mathcal{F}} = \frac{1}{2}\left[\mathcal{F}\left(\Omega_0 + \delta\Omega\right) + \mathcal{F}\left(\Omega_0 - \delta\Omega\right)\right], \tag{4.1}$$

where $\Omega_0 = 1$ MHz and $\delta\Omega = 40$ Hz. Here and elsewhere magnetic field strengths are given in units of Larmor frequency.

To evaluate the performance of the unitary transformations implemented in our laboratory we can, in principle, fully reconstruct the applied quantum map through quantum process tomography (QPT). In practice, process tomography is a very laborious procedure and our most recent studies (see Ch. 6) indicate that our ability to implement QPT is worse than our ability to implement an individual unitary map. As an alternative, we make use of a randomized benchmarking (RB) procedure inspired by the randomized benchmarking technique developed by E. Knill *et al.* [49]. This procedure does not provide the ability to determine the fidelity of a specific unitary transformation due to the fact that it only yields an average fidelity for a given set of transformations. However, it does provide the ability to separate errors present in the unitary transformation from all other experimental error.

Randomized benchmarking is implemented by preparing a randomly chosen initial quantum state, this is referred to as the *preparation*. Preparation is followed by a sequence of $l$ randomly chosen transformations

$$|F=3,m_F=3\rangle \xrightarrow{\text{prep}} |\psi_0\rangle \xrightarrow{\hat{U}_1} |\psi_1\rangle \ldots \xrightarrow{\hat{U}_l} |\psi_l\rangle \xrightarrow{\text{read}} |F=3,m_F=3\rangle. \quad (4.2)$$

The final map from $|\psi_l\rangle$ to $|F=3,m_F=3\rangle$ and the Stern-Gerlach Analysis to measure the population in $|F=3,m_F=3\rangle$ is referred to as the *read out*. The sequence in Eq. 4.2 is repeated many times for different initial states and different unitary transformations. We then fit the decay of the overall fidelity as a function of $l$ using

$$P(l) = \frac{1}{16} + \frac{15}{16}\left(1 - \frac{16}{15}\eta_0\right)\left(1 - \frac{16}{15}\eta\right)^l. \quad (4.3)$$

where $\eta_0$ is the combined error of state preparation and read out, and $\eta$ is the average error per control map. Finally, the *"benchmark"* fidelity is calculated using

$$\mathcal{F}_B = 1 - \eta. \quad (4.4)$$

Fig. 4.1 shows examples of randomized benchmarking data for robust 16-dimensional unitaries (red dots), and nonrobust 16-dimensional unitaries (blue dots) as a function of $l$. The benchmarked fidelities obtained were $\mathcal{F}_B = 0.982(2)$ and $\mathcal{F}_B = 0.971(1)$ for robust and nonrobust control waveforms.

As shown in Sec. 2.2, designing control waveforms to implement unitary maps in our experiment requires us to specify the values for the phase step duration $\Delta t$ and the total control time $T$. Together, this parameters define the number of discrete phase steps given by $N = T/\Delta t$. Because there are three sets of control fields, there

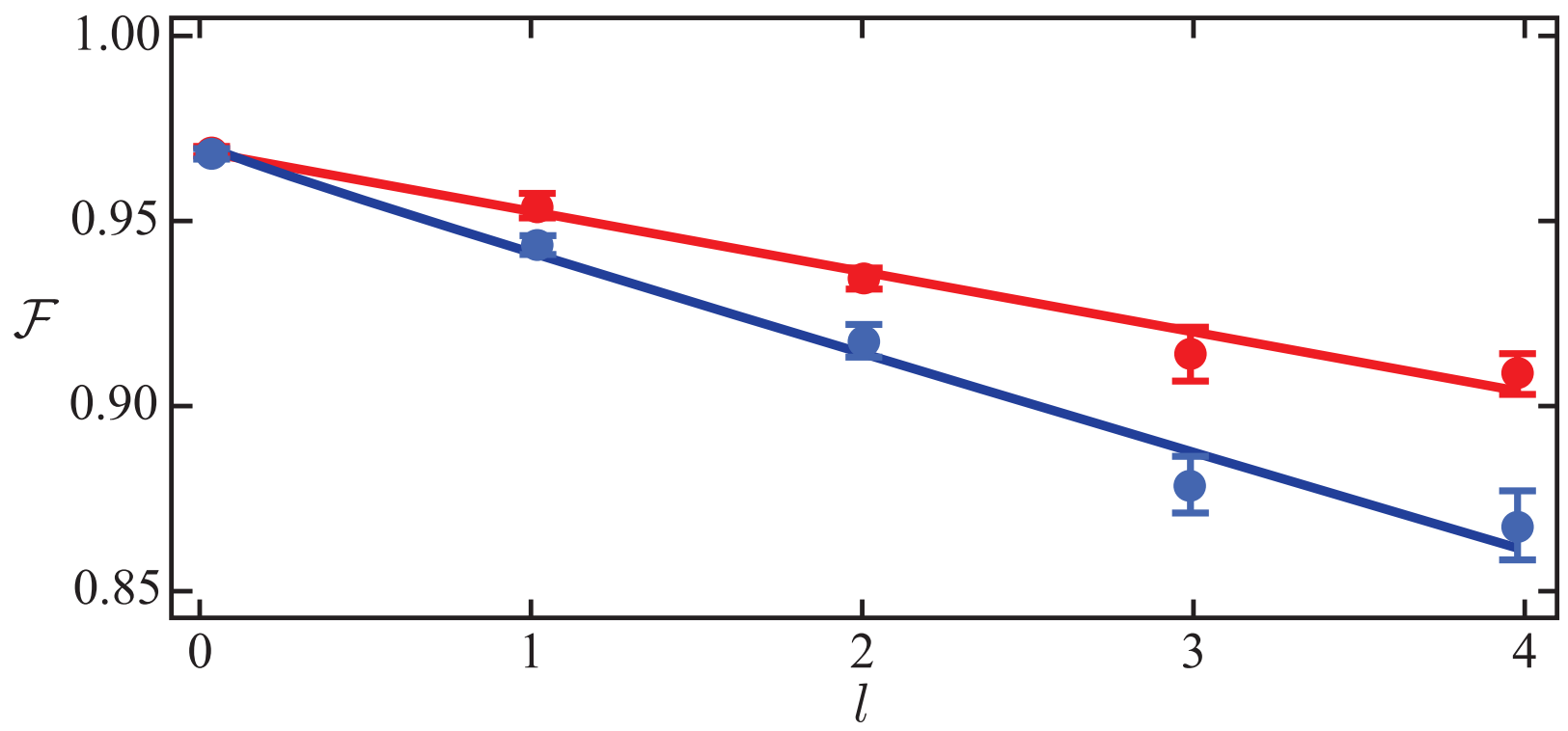

Figure 4.1: Randomized benchmarking data showing the average overall fidelity of robust 16-dimensional unitaries (red dots), and nonrobust 16-dimensional unitaries (blue dots) as a function of $l$. Each point represents an average of 10 sequences; error bars are $\pm$ one standard deviation of the mean. Lines are fits from which the benchmarked fidelity $\mathcal{F}_B$ is determined.

are $3N$ independent control phases in the control waveform. To explore the tradeoff between $T$ and $\Delta t$ for 16-dimensional unitary transformations we implemented a search for control waveforms using several combinations of $(T, \Delta t)$. The search is performed for a set of ten unitary maps chosen randomly according to the Haar measure. Fig. 4.2 shows a calculation of the expected average fidelity for the set of maps as a function of $T$ and $\Delta t$. This calculation is done by simulating the performance of our control waveforms given realistic errors and inhomogeneities in the control parameters (see [38]). It should be noted that the characterization of these imperfections was obtained independently from this project (see [37]). Fig. 4.2 also shows fidelities determined by randomized benchmarking measured at a few discrete points (red numbers). It is worth emphasizing that blue numbers correspond to fidelities calculated using Eq. 2.8 while red numbers are obtained from Eq. 4.4. The relationship between $\mathcal{F}$ and $\mathcal{F}_B$ is studied in detail in [38]. In the figure we see that for sufficiently large values of $T$, the search algorithm consistently finds control

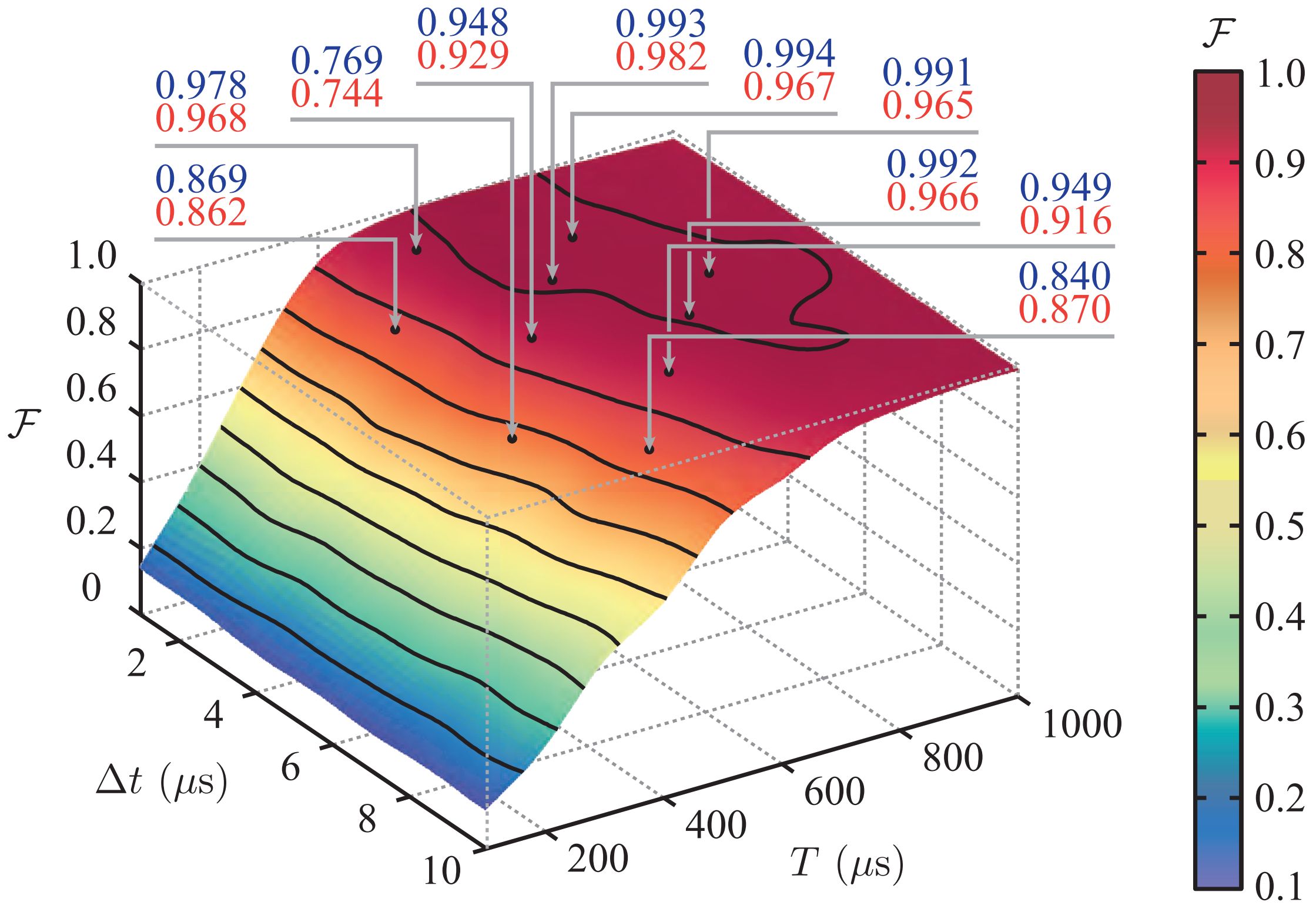

Figure 4.2: From [32]. Expected average fidelity $\mathcal{F}$ reached by a random set of 16-dimensional unitary maps, as a function of the control time $T$ and phase step duration $\Delta t$. Numbers indicate average fidelities $\mathcal{F}$ (blue) and benchmarked fidelities $\mathcal{F}_B$ (red) for a few combinations $T$, $\Delta t$. The top contour line is at $\mathcal{F} = 0.99$. Note that blue numbers are calculated using Eq. 2.8 while red numbers are obtained from Eq. 4.4

waveforms with high fidelity. However, when $T$ is too short there is a rapid drop in fidelity. Lastly, based on the figure, the optimal combination for the parameters are $T = 600$ $\mu$s and $\Delta t = 4$ $\mu$s. Further details and discussion about this study can be found in [38].

Table 4.1 summarizes the control time and step duration combinations used for the control tasks relevant for the experiments presented in this dissertation. These values were obtained for control waveforms which are designed to be robust only against errors in the static bias field strength.

| Control Task | $\Delta t$ ($\mu$s) | $T$ ($\mu$s) |
|---|---|---|
| State-to-state map | 4 | 100 |
| Unitary map in $4d$ | 4 | 300 |
| Unitary map in $7d$ | 4 | 340 |
| Unitary map in $9d$ | 4 | 360 |
| Unitary map in $16d$ | 4 | 600 |

Table 4.1: $T$ and $\Delta t$ combinations used for several control tasks. These values were obtained for control waveforms which are designed to be robust only against errors in the static bias field strength.

### 4.1.1 Unitary Transformations in the Presence of Larger Imperfections

Sec. 4.1 presents results showing that optimal control is an effective tool to implement high accuracy unitary transformations when small imperfections are present in the experimental setup. In order to explore the potential application of optimal control for experimental settings where larger imperfections exist, we now study the performance of robust control waveforms in the presence of deliberately introduced errors. Suppression of these types of errors may prove helpful for quantum control in less than ideal environments such as atoms moving around in the light shift

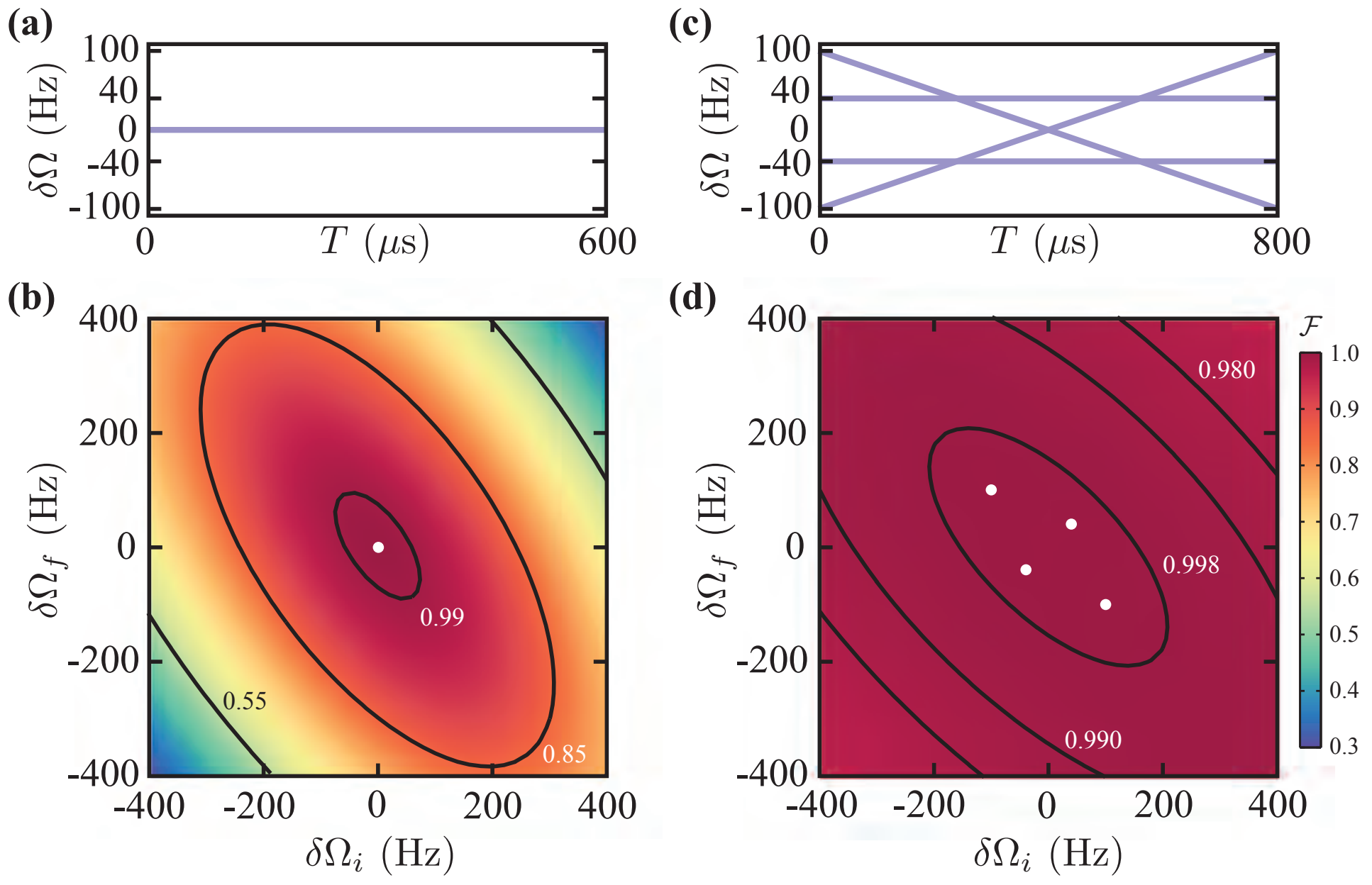

Figure 4.3: From [32]. Fidelity of robust vs nonrobust control waveforms for 16-dimensional unitary maps. (a) Bias field variation $\delta\Omega(t)$ assumed in the design of non robust waveforms. (b) Average fidelity $\mathcal{F}$ predicted for these nonrobust waveforms when the actual $\delta\Omega(t)$ changes linearly from $\delta\Omega_i$ to $\delta\Omega_f$. The central dot corresponds to the variation in (a). (c) Bias field variations $\delta\Omega(t)$ used for the four-point average that goes into the design of robust control waveforms. (d) Average fidelity $\mathcal{F}$ predicted for these robust waveforms as a function of the actual $\delta\Omega_i$, $\delta\Omega_f$. Dots correspond to the variations in (c).

potential of a dipole trap [50].

Our experimental exploration focuses on the application of static and dynamic errors introduced in the bias field strength,

$$\Omega(t) = \Omega_0 + \delta\Omega(t). \tag{4.5}$$

Because $\delta\Omega(t)$ is dominated by the 60 Hz power line cycle and our experiments generally last for $\leq 1$ ms, perturbations during these times will be approximately linear. Thus, $\delta\Omega(t)$ can be fully characterize using the initial and final values of the bias field strength denoted by $\delta\Omega_i$ and $\delta\Omega_f$, respectively. Nonrobust waveforms are designed by maximizing the fidelity only at the nominal bias field strength, $\delta\Omega_i = \delta\Omega_f = 0$ Hz (see Fig. 4.3a). On the other hand, robust waveforms are designed by maximizing the average fidelity for four different settings: two static offsets, $\delta\Omega_i = \delta\Omega_f = \pm 40$ Hz, and two linear variations, $\delta\Omega_i = -\delta\Omega_f = \pm 100$ Hz (see Fig. 4.3c).

Figs. 4.3b and 4.3d show predicted fidelities for unitary maps in the presence of perturbations of the form

$$\delta\Omega(t) = \delta\Omega_i + (\delta\Omega_f - \delta\Omega_i)\, t/T. \tag{4.6}$$

For nonrobust waveforms we see that even small fluctuations yield big reductions in the fidelities of the unitary transformations. In contrast, robust waveforms significantly improve the fidelity of the transformation even for static or dynamic errors 5 times larger than the designed robustness. This increase in robustness comes with the cost of increasing the duration of the control waveforms from $T = 600$ $\mu$s to $T = 800$ $\mu$s.

To verify the performance of robust and nonrobust waveforms in the laboratory, we performed randomized benchmarking at several points along the $\delta\Omega_i = \delta\Omega_f$ and $\delta\Omega_i = -\delta\Omega_f$ diagonals. Fig 4.4 shows the predicted fidelities $\mathcal{F}$ (solid lines) from Fig. 4.3, as well as the observed fidelities $\mathcal{F}_B$ (dots) from randomized benchmarking. Dashed lines are parabolic fits to guide the eye. Our first observation is that in the absence of deliberately applied errors, the fidelities from robust control waveforms are $\sim 98\%$. This indicates that inherent static or time dependent variations in the bias field do not play an important role in limiting the attainable fidelity. In addition, we see that for robust control waveforms even the highest dynamical and static perturbations have modest effect on the performance by decreasing the fidelity to values $\sim 96\%$ in both cases. This is an exceptional result considering that the cost of implementing robust waveform is a modest increase in control time. On the contrary, nonrobust waveforms suffer a large decrease in performance by yielding fidelities $< 80\%$ for the large induced error cases.

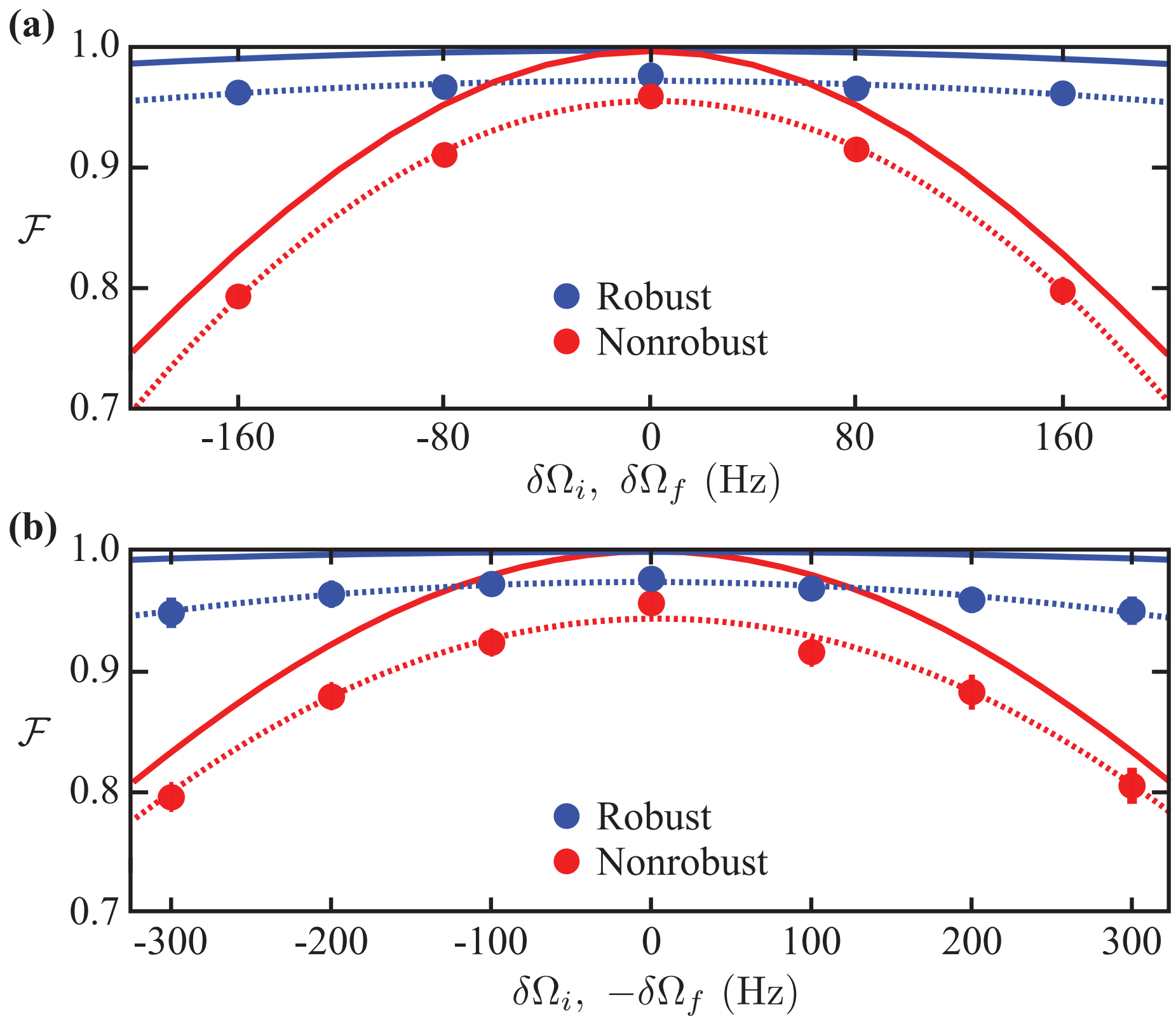

Figure 4.4: From [32]. Fidelity of robust vs nonrobust control waveforms for 16-dimensional unitary maps. (a) Measured and predicted fidelities for robust (blue) and nonrobust (red) control waveforms, along the diagonal $\delta\Omega_i = \delta\Omega_f$. (b) Same along the diagonal $\delta\Omega_i = -\delta\Omega_f$. Data points show the average $\mathcal{F}_B$ for the set of maps; error bars are $\pm$ one standard deviation of the average. Dashed lines are parabolic fits to guide the eye. Solid lines show the predicted $\mathcal{F}$.

## 4.2 Inhomogeneous Quantum Control

Our experimental testbed consists of a large ensemble of atoms which are all controlled using global sets of magnetic fields. Because of the inherent spatial extent of the ensemble, the atoms show variation (inhomogeneity) in some of the parameters that govern the dynamics of the system. So far, the different dynamics generated on different members of our atomic ensemble have been associated with unwanted errors, and we have shown that by using robust control we are able to suppressed their effect. In this section, we present an experiment where an inhomogeneous perturbation is deliberately imposed on the ensemble. In this scenario, the objective is to design a global control waveform to perform different unitary transformations for different members of the ensemble depending on the value of the applied perturbation. This problem is known in the literature as inhomogeneous quantum control. Inhomogeneous control has been extensively studied in the context of Nuclear Magnetic Resonance (NMR) [51–53] and more recently in neutral cold atoms [54, 55].

To test the basic idea of inhomogeneous control in the laboratory, we designed control waveforms to implement two target unitary transformations, $\hat{U}_1$ and $\hat{U}_2$, based on the presence or absence of a light shift generated from an optical addressing field. As introduced in Sec. 2.3, the addressing field is capable of producing an effective rf and/or $\mu$w detuning in the Hamiltonian of our system. Therefore, the total control Hamiltonian governing the dynamics of our system is given by the addition of Eq. 2.4 and Eq. 2.14,

$$\hat{H}_{\mathrm{C}} = \hat{H}_0 + \hat{H}_{\mathrm{rf}} + \hat{H}_{\mu\mathrm{w}} + \hat{H}_{LS}. \tag{4.7}$$

When the optical addressing field is turned off, $\hat{H}_{LS}$ vanishes and the evolution of the system is described by the unitary transformation $\hat{U}_{\mathrm{off}}$. On the other hand, if the addressing field is turned on, $\hat{H}_{LS}$ produces a light shift which modifies the control Hamiltonian, consequently modifying the evolution of the system which is now described by $\hat{U}_{\mathrm{on}}$.

Our search algorithm uses a cost function that takes the form

$$\begin{aligned}\bar{\mathcal{F}} &= \mathcal{F}\left(\hat{U}_1, \hat{U}_{\mathrm{on}}\right) + \mathcal{F}\left(\hat{U}_2, \hat{U}_{\mathrm{off}}\right) \\ &= \frac{1}{d^2}\left[\left|\mathrm{Tr}\left(\hat{U}_1^\dagger \hat{U}_{\mathrm{on}}\right)\right|^2 + \left|\mathrm{Tr}\left(\hat{U}_2^\dagger \hat{U}_{\mathrm{off}}\right)\right|^2\right].\end{aligned} \tag{4.8}$$

In order to facilitate the search of control waveforms to implement two distinct target unitary maps, it is desirable to make Eq. 4.7 as different as possible with and without the light shift. This means that we want to make $\hat{H}_{LS}$ as different as possible for at least some of the states in the ground manifold.

The choice of the addressing field parameters (intensity, detuning, and polarization) is important and depends on several considerations. As stated above, we would like to maximize the differential light shift for some states while minimizing the decoherence induced by the optical field [4, 47]. The time window within which coherent dynamics is possible is set by the scattering time $\tau_s = 1/\gamma_s$, where $\gamma_s$ is the characteristic photon scattering rate. In general we want to choose light parameters such that $\tau_{sc} \gg T$. Fig. 4.5 shows a calculation of the different components of the light shift Hamiltonian produced on the $F = 3$ (red lines) and $F = 4$ (blue lines) ground manifolds from an optical field tuned near the $D_1$ line transition. Values were calculated assuming linearly polarized light and the intensity was allowed to vary to ensure $\tau_{sc} = 1500$ ms at every detuning value. Because the polarization of the light is chosen linear, the vector light shift component is always zero. We also find that the scalar component yields the largest light shift by almost a factor of ∼100 compared to the tensor component. Lastly, wee see that the largest scalar differential light shift $V_{LS}$ (Sec. 2.3) occurs at approximately $\Delta_{F=3\rightarrow F'=4} = -4.86$ GHz (green dashed line in Fig. 4.5a). Here, $V_{LS} \approx 1.4$ kHz and corresponds to the largest effective $\mu$w detuning we can introduce in the control Hamiltonian given the constraint in $\tau_s$.

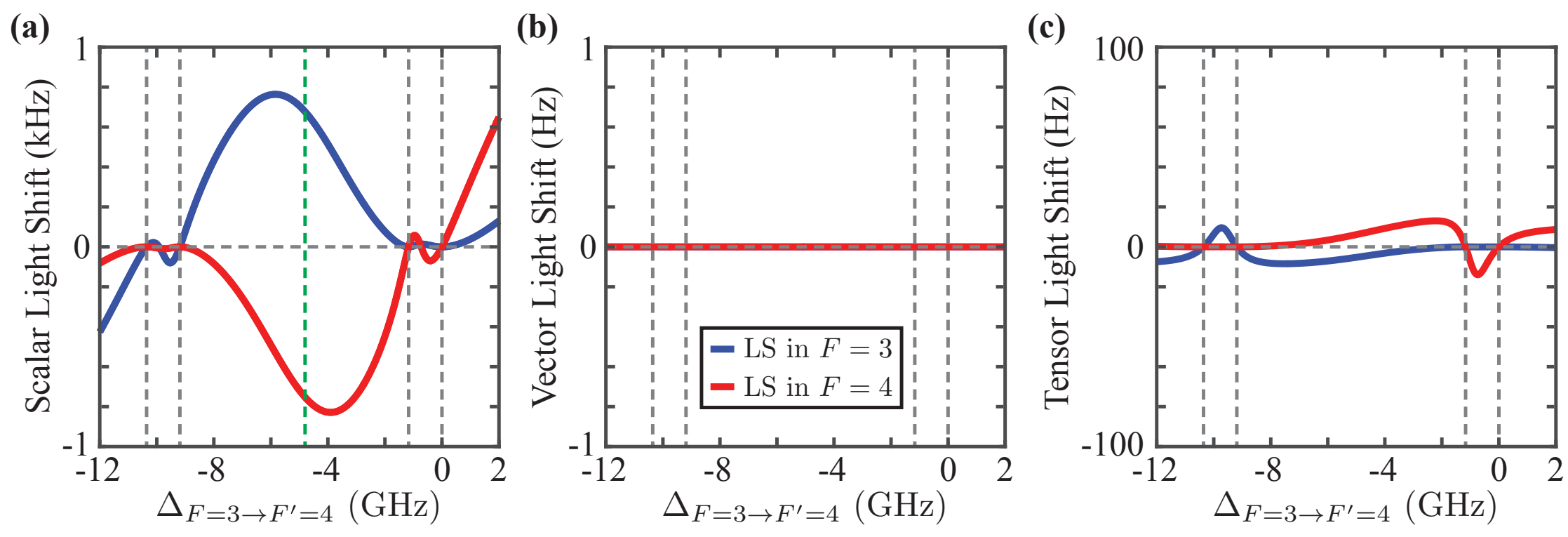

Figure 4.5: Light shifts produced on the $F = 3$ (red lines) and $F = 4$ (blue lines) ground manifolds from an optical field tuned near the $D_1$ line transition. (a) Scalar light shift. (b) Vector light shift. (c) Tensor light shift. Light polarization was assumed linear and intensity was allowed to vary to ensure $\tau_{sc} = 1500$ ms at every detuning value. Green dashed line indicates when $V_{LS}$ is maximum.

The previous calculation shows that, given the $\tau_{sc} = 1500$ ms condition, the best optical addressing field parameters are a nominal intensity of 3.0 mW/cm$^2$, with a frequency of 4.86 GHz red detuned from the $F = 3 \rightarrow F' = 4$ $D_1$ line transition. The combination of intensity and detuning ensures a scattering time which is sufficiently large to ignore decoherence effects. The choice of linearly over circularly polarized light was motivated by a different set of calculations using circular polarized light. There, we found that the contribution from the scalar light shift is still a factor $\sim$10 larger than the vector component. Lastly, from an experimental point of view, high quality linearly polarized light is readily achievable in the laboratory, whereas it is more difficult to achieve a specific circular or elliptical polarization with the required accuracy.

### 4.2.1 Inhomogeneous Control Procedure and Results

As a proof-of-principle demonstration of inhomogeneous quantum control we performed an experiment to address our atomic ensemble with a spatially dependent optical field. In this experiment we start by preparing the entire atomic ensemble in the initial state $|\psi_i\rangle = |F = 4, m_F = 4\rangle$. Then, we apply a global control waveform which makes atoms in the presence of the addressing field undergo a target unitary evolution $\hat{U}_1 = \mathbb{I}$, where $\mathbb{I}$ is the identity operator, such that $|\psi_f\rangle = \hat{U}_1|\psi_i\rangle = |\psi_i\rangle$. At the same time atoms unaffected by the addressing field undergo a different unitary evolution $\hat{U}_2$ that maps the initial state into a coherent superposition state in the $F = 3$ manifold, i.e. $|\psi_f\rangle = \hat{U}_2|\psi_i\rangle = \sum_{j=-3}^{3} \alpha_j |F = 3, m_F = j\rangle$. In the next part of the experiment, we apply a second global control waveform to perform $\hat{U}_1^{-1}$ and $\hat{U}_2^{-1}$ for atoms in the presence and absence of addressing field, respectively. This coher-ently evolve the entire ensemble back into the initial state $|\psi_i\rangle = |F = 4, m_F = 4\rangle$.

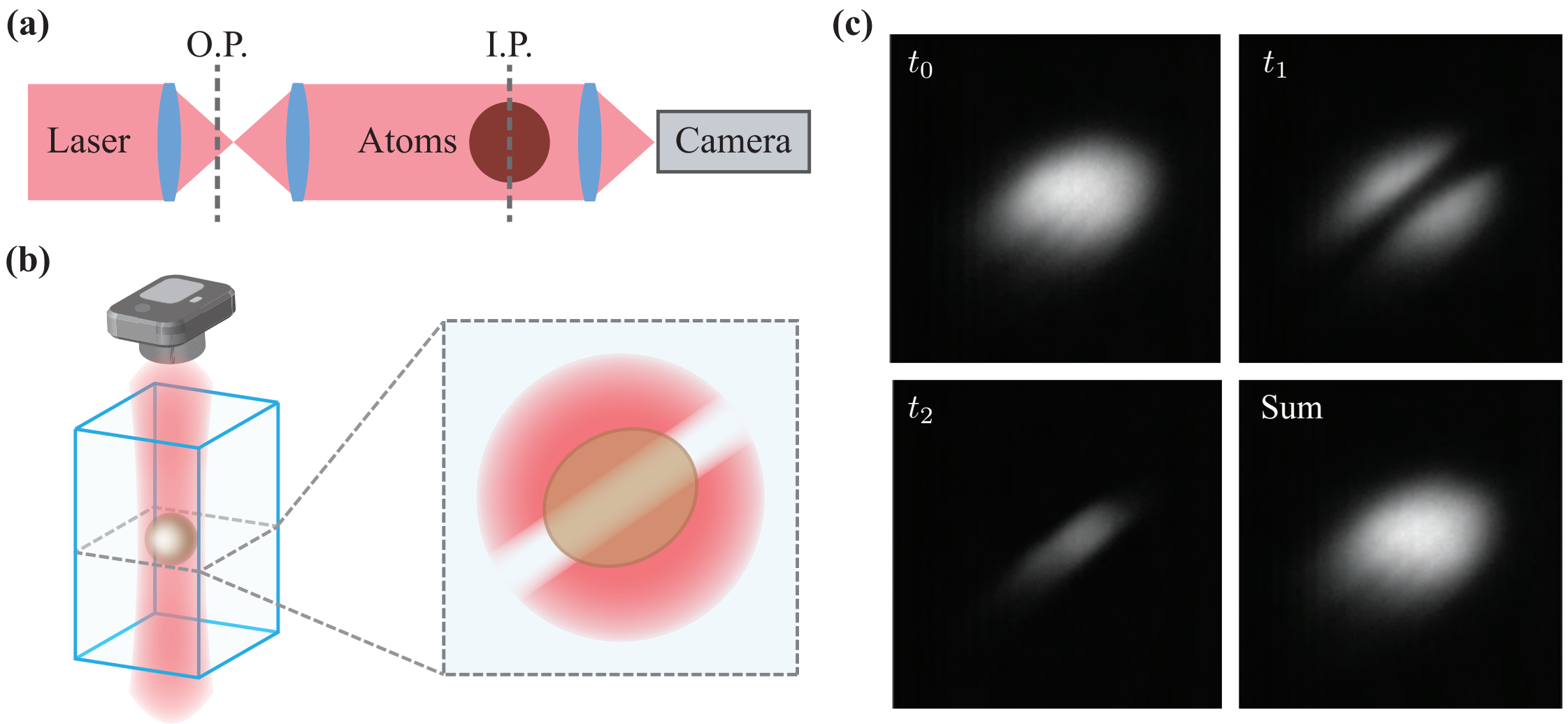

Figure 4.6: Experimental demonstration of inhomogeneous quantum control. (a) Schematic of the setup to image the shadow of a thin opaque string placed at the object plane (O.P.) onto the image plane (I.P.) located at the position of the atomic cloud. (b) Schematic of experimental setup showing the addressing optical field and camera. A transverse cut through the addressing beam at the position of the atoms shows the relative size of the beam, the shadow of the string, and the atomic cloud. (c) Fluorescence images of the atomic cloud at several stages during the experimental sequence. All images are obtained using the MOT beams (not shown here) resonant with the $F = 4 \to F' = 5$ transition in the $D_2$ line.

In order to create the spatial distribution for the optical addressing field, we use a thin opaque string placed on the propagation path of the field such that we image the shadow of the string at the plane where the atoms are located (see Fig. 4.6a). To verify that the ensemble follows the intended unitary evolutions we take fluorescence pictures of atoms in the $F = 4$ manifold at several stages of the experiment (Figs. 4.6b and 4.6c). These pictures are obtained using the MOT beams resonant with the $F = 4 \to F' = 5$ transition in the $D_2$ line. At $t_0$ we take a picture after the ensemble is prepared into $|F = 4, m_F = 4\rangle$. At $t_1$ a picture is taken after the first global control waveform is implemented. Here we see that atoms in the region where no addressing field light is present disappeared from the picture, indicating that they all moved into the $F = 3$ manifold. At $t_2$ a picture is taken after the second global control waveform is implemented. Here we see that atoms originally in the $F = 3$ manifold move back to $|F = 4, m_F = 4\rangle$. The rest of the atoms do not show in the picture because the previous picture at $t_2$ blows them away. The last picture shows the sum of pictures taken at $t_2$ and $t_3$. This image is comparable to the one taken at $t_0$ indicating that our inhomogeneous quantum control waveforms performed as intended.

To obtain a more rigorous estimate for how well our inhomogeneous control scheme works, we performed a second experiment to quantify the fidelity of the control waveforms using randomized benchmarking. In this experiment we designed inhomogeneous control waveforms to implement two distinct unitary transformation acting on states in an 8-dimensional Hilbert space spanned by all the states in the $F = 3$ manifold and the $|F = 4, m_F = 4\rangle$ state in the $F = 4$ manifold. Our exploration was restricted to a $d = 8$ space instead of the available $d = 16$ space because control waveforms designed for the later case require control times where decoherence becomes a serious limitation.

In the laboratory, the control waveforms are evaluated using the randomized benchmarking procedure described in Sec. 4.1. In this case, each and every chain of unitary transformations involved in the RB procedure is implemented for two different experimental settings (Fig. 4.7a). In the first one, all the atoms in the ensemble are illumined by the addressing field inducing the light shift in the control Hamiltonian. Thus, randomized benchmarking yields the average fidelity $\mathcal{F}_B(\hat{U}_{\rm on})$ corresponding to the set of unitary maps $\{\hat{U}_{\text{on}}\}$. In the second setting, the addressing field is completely turned off and thus, randomized benchmarking yields the average fidelity $\mathcal{F}_B(\hat{U}_{\text{off}})$ corresponding to the set of unitary maps $\{\hat{U}_{\text{off}}\}$. Taking the average of $\mathcal{F}_B(\hat{U}_{\text{on}})$ and $\mathcal{F}_B(\hat{U}_{\text{off}})$ yields the overall fidelity of the inhomogeneous control waveforms $\mathcal{F}_B$.

It is important to note that in the case where the atoms are addressed by the optical field, the intensity distribution of the light is inhomogeneous across the atomic ensemble. For this reason it is necessary to modify Eq. 4.8 in order to include robustness against intensity inhomogeneity. This is accomplished following the same approach used to include robustness against inhomogeneities in the bias field strength. Fig. 4.7b shows the fidelities achieved by a single robust (blue line) and nonrobust (red line) control waveform as a function of the addressing beam intensity. Circles indicate the values of the intensity parameter included in the cost function. In the figure we see that robust control waveforms allow us to achieve high fidelities for a wider range of intensities.

To determine what is the minimum control time to successfully find inhomogeneous control waveforms that yield high fidelity, we carried out a numerical exploration. In this study, we searched for control waveforms that are robust against both, inhomogeneities in the bias field strength and addressing optical field intensity. Fig. 4.8 shows results for the maximum achievable fidelity (dashed grey line) as a function of total control time $T$. All control waveforms utilize a phase step duration $\Delta t = 4$ $\mu$s. Here we see that in order to find waveforms with $\mathcal{F} > 0.99$ we must use control times of at least $T = 1.3$ ms. Using waveforms shorter than that, yield control fields which by design will not be able to perform well in the experiment.

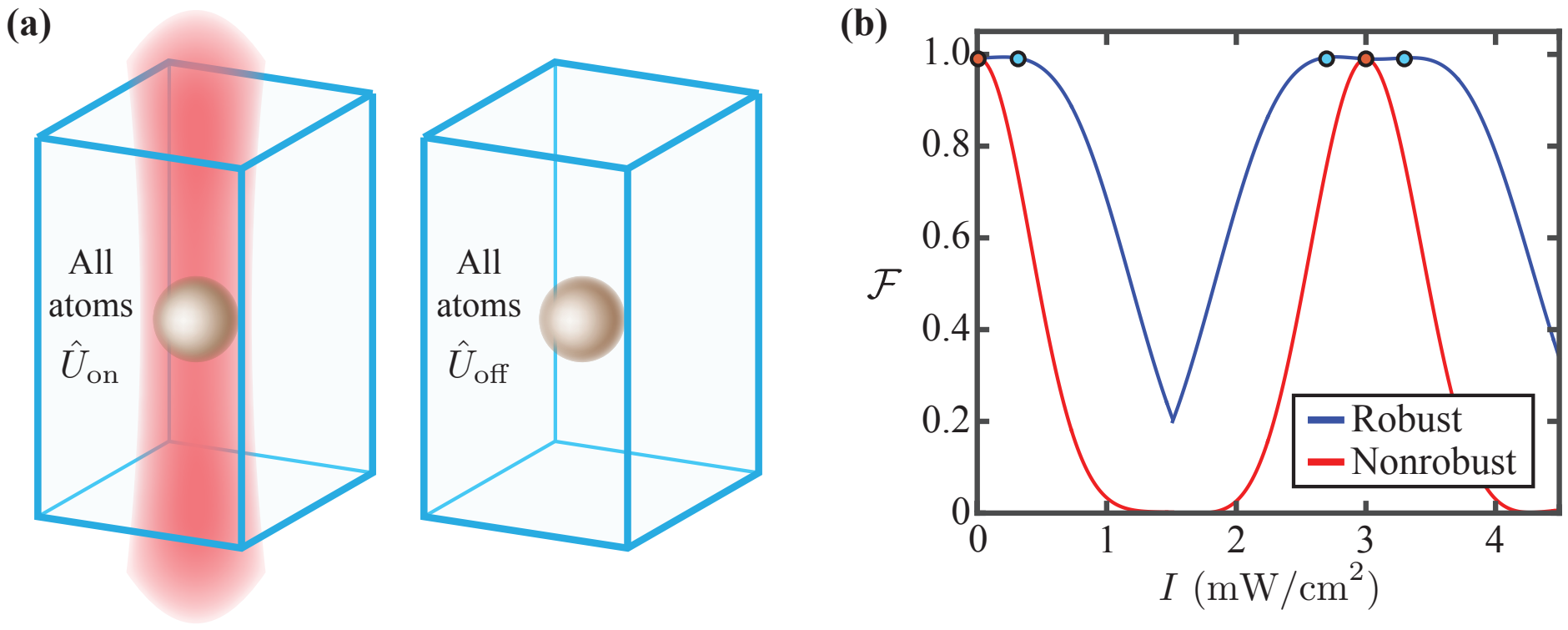

Figure 4.7: (a) Experimental settings to evaluate inhomogeneous quantum control. In the first one, all atoms are illuminated by the addressing field. In the second, the addressing field is completely turned off. Evaluation is carried out in this 2-step procedure to ensure randomized benchmarking only probes the performance of either $\{\hat{U}_{\rm on}\}$ or $\{\hat{U}_{\rm off}\}$. (b) Fidelities achieved by a single robust (blue line) and nonrobust (red line) control waveform as a function of the addressing beam intensity.

However, using an independent experiment (see App. A) we have found that, as a general trend and when everything else is equal, longer control waveforms tend to perform worst in our experiment. This is most likely due to the cumulative effect of experimental imperfections which will gradually reduce the achievable fidelity as the control time increases. Fig. 4.8 shows a solid grey line representing the maximum achievable fidelity for inhomogeneous quantum control multiplied by the function $\mathcal{A}(t)$ (Eq. A.1) which describes the known decay in fidelity due to use of longer control waveforms.

Fig. 4.8 shows the experimental results of inhomogeneous control performed using waveforms of different lengths. Red dots represent the benchmarked fidelities from the set of unitaries $\{\hat{U}_{\rm on}\}$, blue dots represent the benchmarked fidelities from the set of unitaries $\{\hat{U}_{\rm off}\}$, and black dots represent the average benchmarked fidelities $\mathcal{F}_B$. In general, inhomogeneous control is successfully achieved by using waveforms with total control time larger than $T \geq 840\ \mu$s. The best benchmarked fidelity is obtained using control fields with $T = 1.24$ ms with $\mathcal{F}_B = 0.922(18)$.

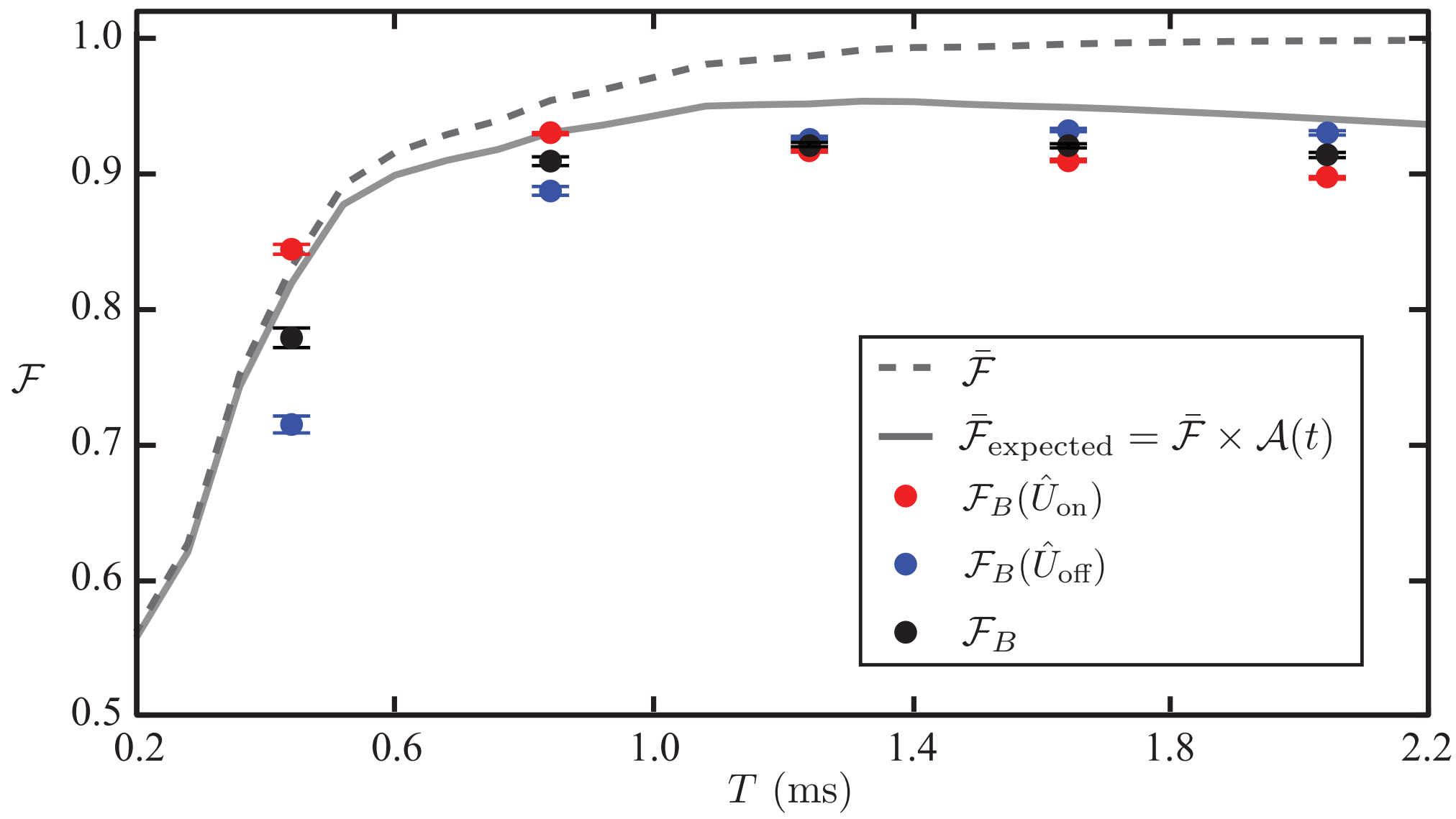

Figure 4.8: Fidelity of inhomogeneous quantum control as a function of total control time $T$. Dashed grey line represents the maximum achievable fidelity $\mathcal{F}$ for different choices of $T$ while keeping $\Delta t = 4\ \mu$s. Solid grey line represent $\mathcal{F}$ modified by the known decrease in performance due to use of longer control waveforms. Red and blue dots are experimental data for the randomized benchmark fidelity for the set of unitaries $\{\hat{U}_{\text{on}}\}$ and $\{\hat{U}_{\text{off}}\}$, respectively. Black dots represent the average benchmarked fidelities $\mathcal{F}_B$.

# CHAPTER 5

# QUANTUM STATE TOMOGRAPHY EXPERIMENTS

This chapter discusses the experimental results for quantum state tomography (QST) implemented in the 16-dimensional Hilbert space associated with the electronic ground state of cesium atoms. We begin with a review of the general procedure to implement QST, followed by a simple example used to introduce the concept of accuracy, efficiency and robustness in the context of QST. We also present the different measurement strategies (known in the literature by the technical term "POVM constructions") which we will use in order to collect the measurement data for QST. We then introduce the different state estimators used for reconstruction. The next sections present and discuss experimental results from several QST experiments. A detailed discussion of the theoretical background for this chapter can be found in the dissertation of Charles Baldwin [56], who along with Ivan Deutsch, contributed greatly to this work.

## 5.1 Quantum State Tomography

As introduced in Sec. 2.5, the general procedure to implement quantum state tomography (block diagram shown in Fig. 5.1) is based on performing a series of measurements (POVMs), each corresponding to a set of POVM elements $\{\hat{\mathcal{E}}_\alpha\}$, on many identically prepared copies of an unknown state $\rho$. The measurements yield a measurement record $\mathcal{M}$ comprising the set of frequencies of outcomes $\{f_\alpha\}$ which correspond to the estimates for the corresponding probabilities of outcomes $\{p_\alpha = \text{Tr}[\rho\hat{\mathcal{E}}_\alpha]\}$. We then use an estimation algorithm to search for an estimated state $\rho_e$ such that the set $\{p_\alpha^{(e)} = \text{Tr}[\rho_e\hat{\mathcal{E}}_\alpha]\}$ provides the best match, according to some chosen metric, to the set $\{f_\alpha\}$ and any prior information about the state.

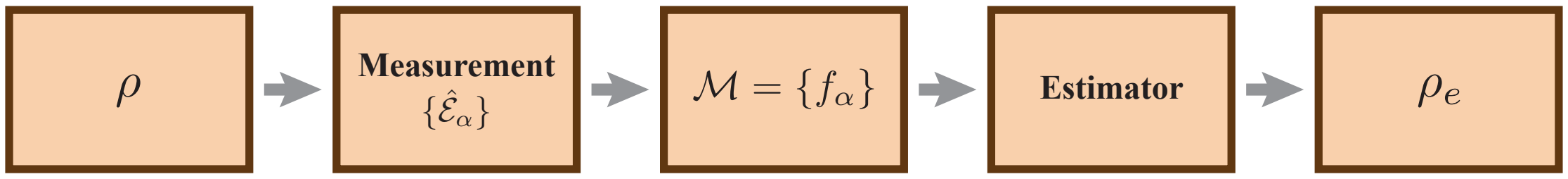

Figure 5.1: General procedure to perform quantum state tomography. Many copies of an unknown state are send to a measurement apparatus which produces a measurement record $\mathcal{M}$. $\mathcal{M}$ is used in an estimation algorithm to obtain an estimate for the unknown state.

In order to develop a better understanding for how each component of the QST procedure looks like, we will consider a simple example where the objective is to reconstruct an unknown state in a 2-dimensional Hilbert space. In this case, the density matrix of the unknown state is given by

$$\rho = \begin{pmatrix} \rho_{1,1} & \rho_{1,2} \\ \rho_{2,1} & \rho_{2,2} \end{pmatrix}. \tag{5.1}$$

Due to Hermiticity and unit trace constrains, there is a total of $d^2 - 1 = 3$ independent, real-value parameters contained in $\rho$. This means that our measurement record must contain frequencies of outcomes for at least three independent POVM elements in order to uniquely reconstruct the unknown state. In our example, one possible choice of measurements consists of the set of Pauli matrices $\{\hat{\sigma}_x, \hat{\sigma}_y, \hat{\sigma}_z\}$, where $\hat{\sigma}_j = P^{(j)}_{|+\rangle} - P^{(j)}_{|-\rangle}$ and $P^{(j)}_{|\pm\rangle} = |\pm\rangle_j\langle\pm|$ are the projectors for each component of the spin in the $j$-direction. In total, our measurement is described by six POVM elements, each producing a frequency of outcome $f^{(j)}_{|\pm\rangle}$ that estimates the probability $p^{(j)}_{|\pm\rangle} = \mathrm{Tr}[\rho P^{(j)}_{|\pm\rangle}]$. In a noiseless measurement, $f^{(j)}_{|+\rangle} + f^{(j)}_{|-\rangle} = 1$, and thus from the set of six frequencies there are only three which are independent.

In order to carry out the measurements of POVM elements we can use Stern-Gerlach analysis (Sec. 2.4). SGA yields the estimates for the probabilities of outcomes such that the measurement record is given by

$$\mathcal{M} = \{f^{(x)}_{|+\rangle}, f^{(x)}_{|-\rangle}, f^{(y)}_{|+\rangle}, f^{(y)}_{|-\rangle}, f^{(z)}_{|+\rangle}, f^{(z)}_{|-\rangle}\}. \tag{5.2}$$

The simplest estimator one can use to perform QST is linear inversion (sometimes called linear state tomography) [57]. In linear inversion, we try to find a state $\rho_e$ that matches the observed set of frequencies $\{f_\alpha\}$, that is,

$$\text{minimize:} \sum_\alpha |\mathrm{Tr}(\rho\hat{\mathcal{E}}_\alpha) - f_\alpha|^2. \tag{5.3}$$

In a 2-dimensional Hilbert space system where the POVM elements set $\{\hat{\mathcal{E}}_\alpha\}$ is given by the Pauli matrices projectors, Eq. 5.3 has the exact solution

$$\rho_e = \frac{1}{2}\begin{pmatrix} 1 + \langle\hat{\sigma}_z\rangle & \langle\hat{\sigma}_x\rangle - i\langle\hat{\sigma}_y\rangle \\ \langle\hat{\sigma}_x\rangle + i\langle\hat{\sigma}_y\rangle & 1 - \langle\hat{\sigma}_z\rangle \end{pmatrix}, \tag{5.4}$$

where $\langle\hat{\sigma}_j\rangle = f^{(j)}_{|+\rangle} - f^{(j)}_{|-\rangle}$. The main advantage of linear inversion is its simplicity, however it also presents some major drawbacks. For example, when the measurements themselves are subject to experimental imperfections, the reconstructed state out of Eq. 5.4 may not be physical, i.e. eigenvalues of $\rho_e$ may be $\leq 0$. To circumvent this issue, one typically makes use of more sophisticated state estimators which search for a good match to the measurement data only from within the set of physical states.

The previous example shows that by measuring the set of Pauli matrices one can collect sufficient information to estimate the density matrix in Eq. 5.1. However, this particular choice of measurements is not a unique option when performing QST. In general, the optimal choice of measurement strategy and state estimator should be motivated by the particular objectives, limitations, and prior knowledge about the experimental setup. For example, if we know the state to be reconstructed is pure or nearly-pure, then there are highly efficient strategies that can yield a good estimate from a much reduced set of POVMs. Additionally, experiments performed on certain physical systems are strongly limited by the sheer amount of work necessary to obtained good estimates for the probabilities of outcomes. In this situation, QST is constraint by measurement statistics, and so it is desirable to use a strategy that yield maximum information about the state from a minimum number of POVMs. Lastly, as we show in what follows, the presence of errors in the measurements themselves may favor yet other strategies, e. g., those that collect redundant information from a larger set of POVMs so that the effect of errors can average out in the final state estimate. Thus, in a real-word setting there is no such thing as an "optimal" protocol for QST; the best choice of measurement strategy and state estimator will depend on the specifics of the scenario at hand and must necessarily reflect some tradeoff between accuracy, efficiency, and robustness to experimental imperfections.

Thus far, proof-of-principle experiments have successfully demonstrated the use of several measurements strategies (known in the literature by the technical term "POVM constructions") to perform QST [58–60]. However, the variety of experi-mental platforms involved in these demonstrations have prevented a direct quantitative comparison of their performance. The objective in our experiment is to put together a comprehensive study of nine different POVM constructions and three different state estimators implemented on a single experimental testbed consisting of the hyperfine manifold in the electronic ground state of cesium. This will allow us to directly compare results from QST implemented on states in a 16-dimensional Hilbert space and highlight the tradeoffs between their accuracy, efficiency, and robustness against experimental errors.

## 5.2 POVM Constructions

A POVM construction is a set of measurements (POVMs) designed to collect the information required to reconstruct an unknown state, taking into consideration the tradeoffs between accuracy, efficiency, and robustness imposed by the limitations of an experiment.

POVM constructions whose outcome probabilities are sufficient to uniquely identify any state from within the set of all physical states (pure and mixed) are said to be *fully informationally complete* (F-IC). In the case of constructions which are informationally complete only for pure states (rank-1 states), Baldwin *et al.* (see Ref. [61]) have shown that, two notions of informationally complete measurements exist: *rank-1 complete measurements* (R1-IC) and *rank-1 strictly complete measurements* (R1S-IC). In the first notion, a pure state is uniquely identified only from within the set of all pure states, in the second notion the same state is uniquely identified from within the set of all physical states. This subtle distinction has important consequences when performing QST and further discussion will be left for Sec. 5.5.

While quantum state tomography can in principle be performed by any F-IC set of measurements, many special POVM constructions have been proposed with particular objectives and considerations in mind. In our experimental exploration we will test a total of nine POVM constructions: three which are fully informationally complete, four which are rank-1 strictly IC, and two which are rank-1 IC. Here, it is important to note that our study is not intended to be exhaustive in the sense that there are more POVM constructions reported in the literature (see for example: [62–66]) and our selection represents only a subset of them.

The following fully informationally complete POVM constructions were used

1. Generalized Gell-Mann Bases (GMB). The GMB are a set of $d^2 - 1$ matrices which form an orthogonal basis for traceless Hermitian operators acting on a $d$-dimensional Hilbert space. As their name suggests, these matrices are the generalization of the $3 \times 3$ Gell-Mann matrices in $d = 3$, as well as the $2 \times 2$ Pauli matrices in $d = 2$. Using an extension of the ideas presented in [67], our theory collaborators at UNM (see Ref. [61]) have shown that for dimensions that are powers of 2, it is possible to obtain an informationally complete measurement record by implementing $2d - 1$ orthogonal bases, each with $d$ outcomes, for a total of $2d^2 - d$ outcomes. GMB were applied to states in Hilbert spaces with $d = 4$ and $d = 16$ dimensions.

2. Symmetric Informationally-Complete (SIC). A *symmetric* POVM is one where all pairwise inner products between the POVM elements are equal. Originally proposed in [68], a SIC POVM is a set of $d^2$ normalized vectors $|\phi_k\rangle$ that satisfy

$$|\langle\phi_j|\phi_k\rangle|^2 = \frac{1}{d+1}, \quad j \neq k. \tag{5.5}$$

SIC POVMs have a total of $d^2$ measurement outcomes and even though there is not a known systematic construction for every dimension, analytic form for a few dimensions exist, e.g., for $d = 2, 3, 4$. In our experiment, SIC was applied to states in a $d = 4$ Hilbert space using the Neumark extension described in Sec. 3.4.

3. Mutually-Unbiased Bases (MUB). Two orthonormal bases $\{|e_i\rangle\}$ and $\{|f_i\rangle\}$ over a $d$-dimensional Hilbert space are defined to be *mutually unbiased* if the inner product between any state of the first basis and any state of the second basis has the same magnitude, i.e.

$$|\langle e_i|f_j\rangle|^2 = \frac{1}{d} \; \forall \; i, j \in \{1, ..., d\}. \tag{5.6}$$

In a measurement, these bases are unbiased in the sense that if a system is prepared in a state belonging to one of the bases, then all outcomes of the measurement with respect to the other bases will occur with equal probability. In the context of quantum state tomography, MUB were originally proposed in [69] and are given by a set of $d+1$ orthonormal bases, each with $d$ measurement outcomes, for a total of $d^2 + d$ outcomes. MUB were applied for both $d = 4$ and $d = 16$ systems in the experiment.

The following rank-1 strictly informationally complete POVM constructions were used

4 Five Gell-Mann Bases (5GMB). This POVM construction was originally proposed in [67] and consist of the first five orthonormal bases of the GMB set. 5GMB produces a total of $5d$ measurement outcomes. 5GMB were applied to states in Hilbert spaces with $d = 4$ and $d = 16$ dimensions.

5 Five Mutually-Unbiased Bases (5MUB). Our theory collaborators at UNM (see Ref. [56]) have produced numerical simulations indicating that the first five bases of the MUB construction correspond to a R1S-IC POVM. In the experiment, we only apply 5MUB to the $d = 16$ case since the 5MUB in $d = 4$ corresponds exactly to the full set of MUB, thus becoming F-IC.

6 Five Polynomial Bases (5PB). This POVM construction was originally proposed in [70] and consist of four orthogonal bases that are constructed based on a set of orthogonal polynomials, plus the logical basis $\{|F, m_F\rangle\}$. This construction applies for any dimension and produces a total of $5d$ measurement outcomes. 5PB were applied to states in Hilbert spaces with $d = 4$ and $d = 16$ dimensions.

7 Pure-State Informationally Complete (PSI). This POVM construction was originally proposed in [62] and consist of $3d - 2$ measurement outcomes.

Since PSI is a set of rank-1 measurement operators, here as well, it is natural to use the Neumark extension to take advantage of our large Hilbert space. In this case we perform QST on $d = 4$, where there is $3d - 2 = 10$ measurement outcomes that can easily be mapped onto our large Hilbert space.

Finally, the following two rank-1 informationally complete POVM constructions were used

8 Four Gell-Mann Bases (4GMB). This POVM construction was originally proposed in [67] and consist of four orthonormal bases of the GMB set. 4GMB produces a total of $4d$ measurement outcomes. 4GMB were tested for both $d = 4$ and $d = 16$ in the experiment.

9 Four Polynomial Bases (4PB). This POVM construction was originally proposed in [71] and consist of four orthogonal bases that are constructed based on a set of orthogonal polynomials for any dimension. 4PB yields a total of $4d$ measurement outcomes. 4PB were tested for both $d = 4$ and $d = 16$ in the experiment.

Table 5.1 shows a summary of all the POVM constructions we use in order to collect measurements to perform QST in 4- and 16-dimensional Hilbert spaces. The total number of measurement bases is directly related to the efficiency of each construction, since every measurement basis requires a different measurement configuration in our experimental setup. Looking at the number of measurement outcomes, it is easy to see that for constructions that are F-IC we require $O(d^2)$ total measurements, while for constructions that are R1S-IC or R1-IC we only require $O(d)$ total measurements. This notable reduction in required information can be understood by recalling that an arbitrary quantum state $\rho$ is specified by $d^2 - 1$ real numbers (since it is a Hermitian operator and satisfies $\mathrm{Tr}(\rho) = 1$), while a pure state is specified by $2d-2$ real numbers (since it has $d$ complex amplitudes which are constrained by one normalization condition and the global phase of a physical state can be set to zero without loss of generality). Because the relations $\{p_\alpha = \mathrm{Tr}[\rho\hat{\mathcal{E}}_\alpha]\}$ are linear, it is easy to show that an informationally complete POVM construction must have $O(d^2)$ and $O(d)$ measurement outcomes for arbitrary and pure states, respectively.

| POVM class | POVM construction | Number of POVMs (measurement bases) | Number of POVMs elements (measurement outcomes) |
|---|---|---|---|
| F-IC | SIC | 1 | $d^2$ |
| | MUB | $d+1$ | $d^2+d$ |
| | GMB | $2d-1$ | $2d^2-d$ |
| R1S-IC | PSI | 1 | $3d-2$ |
| | 5MUB | 5 | $5d$ |
| | 5GMB | 5 | $5d$ |
| | 5PB | 5 | $5d$ |
| R1-IC | 4GMB | 4 | $4d$ |
| | 4PB | 4 | $4d$ |

Table 5.1: Summary of POVM constructions. The first three rows show the constructions that are fully informationally complete, the next four display the rank-1 strictly complete POVMs, and the bottom two show the constructions that are rank-1 informationally complete.

## 5.3 Estimation Algorithms for QST

In quantum state tomography the choice of reconstruction method (also known as estimation algorithm, or *estimator* for short) plays an important role. As with the POVM construction, the choice of estimator should be based on the system under consideration and the application in mind. In this section, we review three well-known estimation algorithms in the context of quantum state tomography.

### 5.3.1 Least-Squares Estimator

In Least-Squares (LS) estimation the objective is to find the state that minimizes the sum of the squares of the difference between the observed frequencies and predicted probabilities of outcomes, under the condition that the estimated state is a physical quantum state, i.e. it is positive and unit trace. To estimate the state we thus solve the optimization program,

$$\begin{aligned} \text{minimize:} \quad & \sum_{\alpha} |\mathrm{Tr}(\rho\hat{\mathcal{E}}_{\alpha}) - f_{\alpha}|^2 \\ \text{subject to:} \quad & \rho \geq 0 \\ & \mathrm{Tr}(\rho) = 1, \end{aligned} \tag{5.7}$$

which can be solved using convex programming since both the objective and constraints are convex functions. In our laboratory, LS and all other estimators presented in this section are implemented using the MATLAB package CVX [72].

### 5.3.2 Maximum-Likelihood Estimator

The objective of Maximum-Likelihood (ML) estimation is to search for the quantum state that is most likely to generate the observed data by maximizing the *likelihood functional* over the state space. During the past decades, ML has found extensive applications in quantum state tomography [73, 74].

The ML strategy consists in maximizing the likelihood functional, which is defined as follows

$$\mathcal{L}(\rho) = \prod_{\mu} \mathrm{Tr}(\hat{\mathcal{E}}_{\mu}\rho)^{mf_{\mu}}, \tag{5.8}$$

for a finite sample of $m$ quantum states. The state that maximizes the likelihood function also minimizes the negative log-likelihood function,

$$-\log[\mathcal{L}(\rho)] = -\sum_{\mu} f_{\mu} \log \mathrm{Tr}(\hat{\mathcal{E}}_{\mu}\rho), \tag{5.9}$$

which is a more convenient function to work with since it is convex. In consequence, the search for $\rho_e$ turns into a convex optimization problem that can be efficiently computed with an algorithm proposed by [75]. To estimate the state we thus solve the optimization program,

$$\begin{aligned} \text{minimize:} \quad & -\log[\mathcal{L}(\rho)] \\ \text{subject to:} \quad & \rho \geq 0 \\ & \mathrm{Tr}(\rho) = 1. \end{aligned} \tag{5.10}$$

ML returns the quantum state $\rho_e$ that maximizes the log-likelihood function and is still within the constrained set of physical states. In the limit that the noise in QST is Gaussian distributed, the likelihood function is well approximated by a Gaussian. In this case, the negative log-likelihood function becomes $-\log[\mathcal{L}(\rho)] = \sum_{\alpha} |\mathrm{Tr}(\rho\hat{\mathcal{E}}_{\alpha}) - f_{\alpha}|^2$, making the LS and ML programs be the same.

### 5.3.3 Trace-Norm Minimization

Trace-norm Minimization (TM) is an estimator originally used in the context of quantum compressed sensing [76, 77]. In compressed sensing one takes advantage of the fact that for states that are pure or nearly-pure, the density matrix often has a number of very small eigenvalues that can be safely ignore. This means that the density matrix is, to a good approximation, low-rank and can be treat it as such. Low-rank matrices are specified by fewer free parameters than an arbitrary matrix. Refs. [78, 79] shown that reconstruction of a low-rank matrix $X$ can be obtained by implementing a convex optimization program of the form,

$$\begin{aligned} \text{minimize: } & \|X\|_* \\ \text{subject to: } & \|M(X) - \boldsymbol{f}\|_2 < \epsilon, \end{aligned} \tag{5.11}$$

where $\|X\|_* = \text{Tr}[\sqrt{X^\dagger X}]$ is the so-called nuclear norm, $\boldsymbol{f} = \{f_\alpha\}$ is the measurement record obtained in the experiment, $M(X) = \{\text{Tr}(X\hat{\mathcal{E}}_1), \ldots, \text{Tr}(X\hat{\mathcal{E}}_m)\}$ represents the set of expected probabilities, and $\epsilon$ is an error threshold that must be chosen before the estimation; this will be discussed further in Sec. 5.5.1. In the context of quantum state tomography the compressed sensing prescription given in Eq. 5.11 can be used by noting that for a physical density matrix, Hermitian and positive semidefinite, the state follows the constraint $\rho \geq 0$ and the nuclear norm is simply the trace,

$$\|\rho\|_* = \text{Tr}(\rho). \tag{5.12}$$

Thus, in order to estimate the state we solve the optimization program,

$$\begin{aligned} \text{minimize: } & \text{Tr}(\rho) \\ \text{subject to: } & \sum_\alpha |\text{Tr}(\rho\hat{\mathcal{E}}_\alpha) - f_\alpha|^2 < \epsilon \\ & \rho \geq 0. \end{aligned} \tag{5.13}$$

TM returns a quantum state $\rho_e$ which will not be properly normalized since the trace was allowed to vary, and so the final estimate must be renormalized to produce a physical quantum state.

## 5.4 Quantum State Tomography in the Laboratory

One of the main goals of this work is to implement and test a broad range of QST protocols on our cesium atom testbed, and to evaluate their performance by applying them to a set of known test states. We here define a *protocol* as a combination of POVM construction and state estimation algorithm. A test run involves the preparation of a test state followed by the measurement of a specified POVM, and then repeating this basic sequence until the entire POVM construction has been implemented. Once an informationally complete data set has been collected, we process it using the LS, ML, and TM algorithms and calculate the fidelity of the resulting state estimates relative to the input test state.

From an experimental standpoint there are two critical steps in the above: (i) accurate preparation of test states (see Sec. 2.2) and (ii) the measurement of generic POVMs. As described in Sec. 3.4, we can use Stern-Gerlach analysis to perform an orthogonal measurement in the basis $\{|F, m_F\rangle\}$ and determine the frequency with which atoms are detected in each of the magnetic sublevels. Furthermore, by preceding SGA with a unitary map

$$\hat{U} = \sum_{\alpha=1}^{16} |(F, m_F)_\alpha\rangle \langle\psi_\alpha| \,, \tag{5.14}$$

we can effectively perform a different orthogonal measurement and determine the frequencies at which atoms are detected in the corresponding basis states $\{|\psi_\alpha\rangle\}$,

$$\begin{aligned} f_\alpha &= \langle (F, m_F)_\alpha | U \rho U^\dagger | (F, m_F)_\alpha \rangle \\ &= \langle (F, m_F)_\alpha | (F, m_F)_\alpha \rangle \langle \psi_\alpha | \rho | \psi_\alpha \rangle \langle (F, m_F)_\alpha | (F, m_F)_\alpha \rangle \\ &= \langle \psi_\alpha | \rho | \psi_\alpha \rangle . \end{aligned} \tag{5.15}$$

This means that with the proper choice of unitary map we can implement any measurement basis required for a particular POVM construction. To ensure good measurement statistics and some degree of averaging over run-to-run variations in our state preparation and unitary maps, we average the Stern-Gerlach signals from 5 successive, identical sequences before fitting as described in Sec. 3.4. The corresponding areas (frequencies of detection) are the raw data for QST.

Table 5.2 summarizes the number of measurement bases each of the POVM constructions evaluated in this work require to obtain an IC measurement record. As an example, when performing QST in the $d = 4$ case using SIC or PSI, we only require 1 measurement basis to obtain an IC measurement record. This is due to the fact that we take advantage of the Neumark extension when designing the unitary map to implement the POVM. In contrast, when performing QST in the $d = 16$ case using GMB, we use 31 different measurement bases in order to obtain an IC measurement record. Lastly, it is worth noting that for the 5GMB, 5PB, 4GMB, and 4PB constructions the number of measurement bases is independent of Hilbert space dimension.

The entire process to obtain an informationally complete measurement record is repeated for a set of 20 pure test states in a 4-dimensional Hilbert space and 20 pure test states in a 16-dimensional Hilbert space. All states were chosen randomly according to the Haar measure.

| POVM class | POVM construction | Number of meas. bases in $d = 4$ | Number of meas. bases in $d = 16$ |
|---|---|---|---|
| F-IC | SIC | 1 | N/A |
| | MUB | 5 | 17 |
| | GMB | 7 | 31 |
| R1S-IC | PSI | 1 | N/A |
| | 5MUB | N/A | 5 |
| | 5GMB | 5 | 5 |
| | 5PB | 5 | 5 |
| R1-IC | 4GMB | 4 | 4 |
| | 4PB | 4 | 4 |

Table 5.2: Number of measurement bases for each POVM construction in $d = 4$ and $d = 16$. In $d = 4$, 5MUB is not available because in this dimension 5MUB corresponds exactly to MUB. In $d = 16$, SIC and PSI are not available because our Hilbert space is not large enough to take advantage of the Neumark extension.

To process the experimental data we import the measurement records to a computer and then perform the state reconstruction by using the estimation algorithms described in Sec. 5.3. The algorithms yield the estimated state $\rho_e$ which is then used to calculate the fidelity of reconstruction with respect to the target test state $\rho$ given by

$$\mathcal{F} = \mathrm{Tr}\left(\sqrt{\sqrt{\rho_e}\rho\sqrt{\rho_e}}\right), \tag{5.16}$$

as well as the infidelity (error) of reconstruction given by $\eta = 1 - \mathcal{F}$.

Figures 5.2 and 5.3 show examples of quantum state tomography for one test state in $d = 4$ and one test state in $d = 16$, respectively. In both cases the reconstruction was done using the least squares (LS) estimator and the entire informationally complete measurement record (i.e. all measurement bases) for each POVM construction. Both figures show the absolute values of the density matrices, with the test state on the left and the estimated states for a selection of POVM constructions on the right. In Fig. 5.2 we see that GMB and MUB performed the best among all POVM constructions yielding infidelities of reconstructions lower than 1%. The worst performing POVM construction was PSI which catastrophically fails yielding an infidelity of reconstruction very close to 30%. In Fig. 5.3 we see that MUB (one of the F-IC constructions) yields the best performance with an infidelity of reconstruction close to 7%. In contrast, 5GMB and 4PB (corresponding to R1S-IC and R1-IC POVM constructions, respectively) yield fairly poor estimates for this test state.

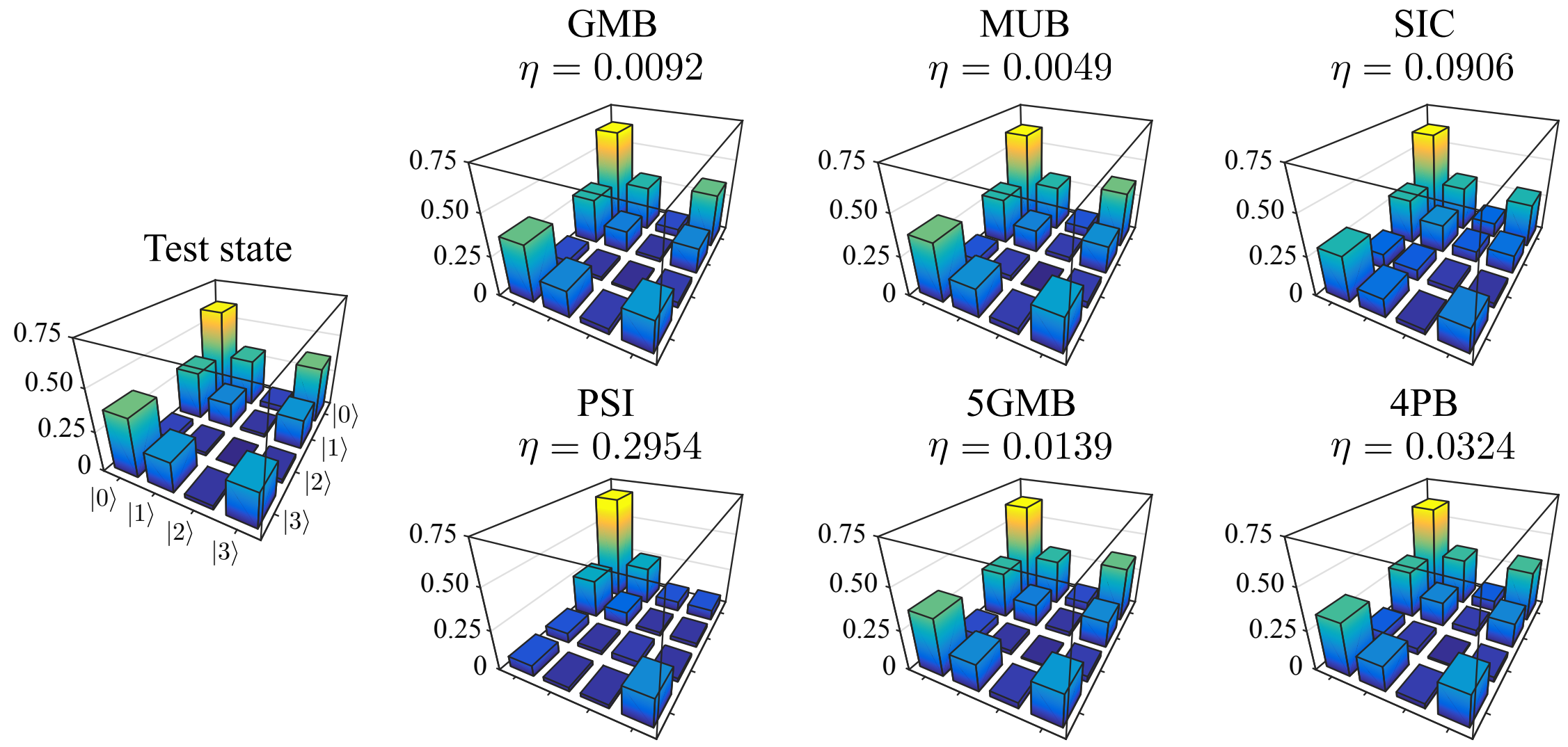

Figure 5.2: Example of quantum state tomography for a test state in $d = 4$. Top row of estimated states shows all F-IC POVM constructions, bottom row shows two R1S-IC and one R1-IC constructions. All reconstructions were done using Least Squares estimator.

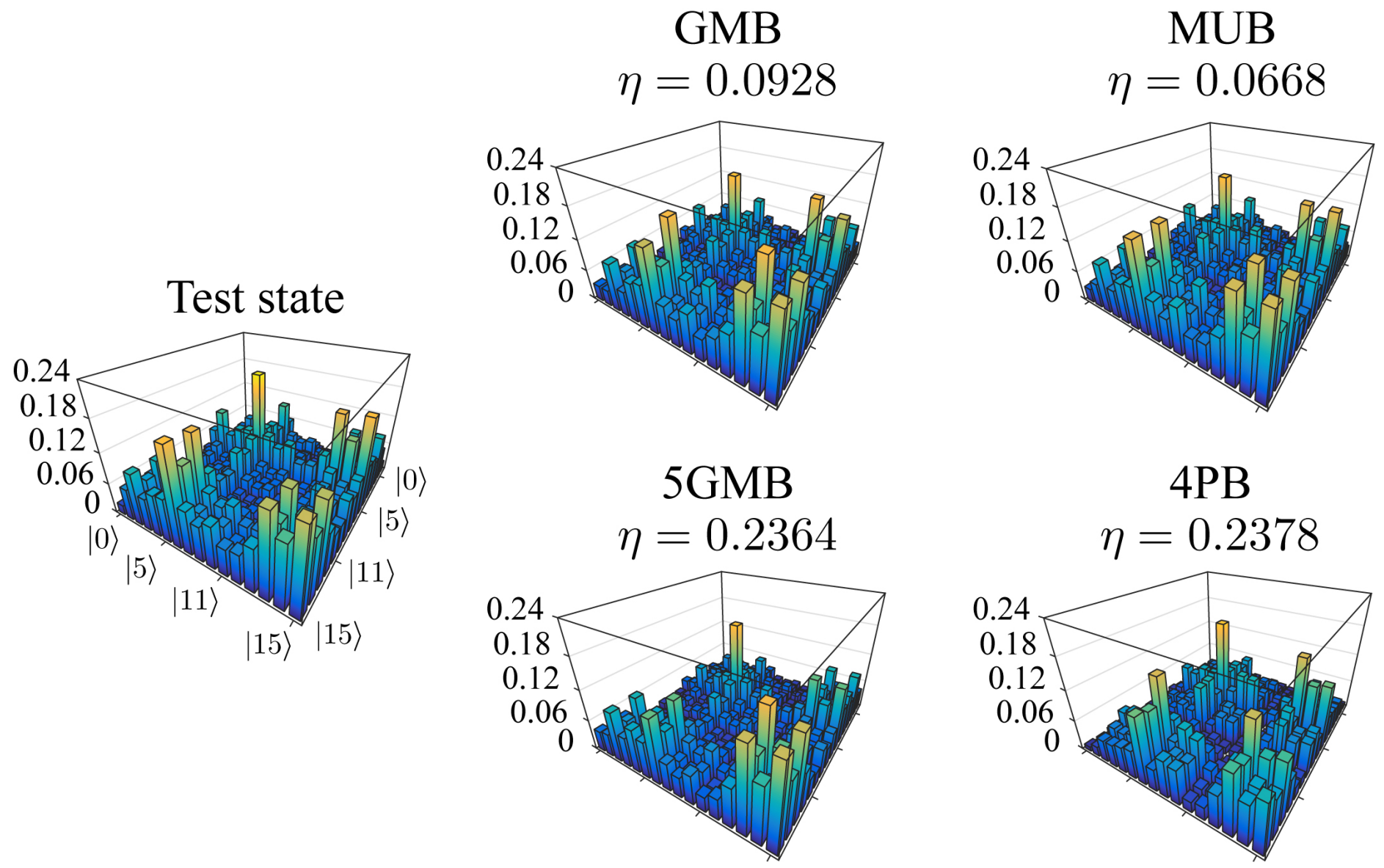

Figure 5.3: Example of quantum state tomography for a test state in $d = 16$. Top row of estimated states shows all F-IC POVM constructions, bottom row shows one R1S-IC and one R1-IC constructions. All reconstructions were done using Least Squares estimator.

## 5.5 State Tomography Results and Discussion

Fig. 5.4 and Table 5.3 show the average infidelity of reconstruction for 20 test states in $d = 4$ and 20 test states in $d = 16$. Overall, reconstructions using the Maximum-Likelihood (ML) estimator consistently yield more accurate results compared to the Least-Squares (LS) estimator. Currently, the superior performance of ML over LS remains an open question. One potential explanation proposed by our theory collaborators at UNM consists in the idea that the positivity constraint build-in in Eq. 5.10 modifies the shape of the likelihood function leading to inconsistent results between ML and LS. In spite of the overall difference in accuracy, results using LS and ML estimators follow similar trends for all POVM constructions. In the following, we focus our analysis on the results produced by the LS estimator noting that all conclusions drawn from the discussion can be equally applied to the ML estimator results.

| POVM construction | $\eta_{\mathrm{LS}}$ $d = 4$ | $\eta_{\mathrm{ML}}$ $d = 4$ | $\eta_{\mathrm{LS}}$ $d = 16$ | $\eta_{\mathrm{ML}}$ $d = 16$ |
|---|---|---|---|---|
| SIC | 0.0661(74) | 0.0625(73) | N/A | N/A |
| MUB | 0.0272(32) | 0.0181(21) | 0.0652(28) | 0.0602(23) |
| GMB | 0.0209(21) | 0.0092(15) | 0.0809(42) | 0.0595(30) |
| PSI | 0.0975(148) | 0.0923(164) | N/A | N/A |
| 5MUB | N/A | N/A | 0.1785(96) | 0.1564(96) |
| 5GMB | 0.0313(41) | 0.0173(28) | 0.2452(132) | 0.2442(173) |
| 5PB | 0.0401(57) | 0.0267(47) | 0.2734(241) | 0.2384(215) |
| 4GMB | 0.0870(213) | 0.0764(221) | 0.2612(151) | 0.2759(217) |
| 4PB | 0.0993(356) | 0.0853(360) | 0.3447(379) | 0.3200(366) |

Table 5.3: Average infidelity of reconstruction for 20 test states in $d = 4$ and 20 test states in $d = 16$. Numbers in parenthesis indicate the uncertainty (standard error of the mean).

Our experimental results show that the average infidelity of reconstruction varies considerably with POVM construction, ranging from 0.0209(21) to 0.0993(356) in $d = 4$, and from 0.0652(28) to 0.3447(379) in $d = 16$. Here and elsewhere, numbers in parenthesis indicate the uncertainty (standard error of the mean). As a general trend, both in $d = 4$ and $d = 16$, we see that the most accurate results are obtained using F-IC POVM constructions, followed by R1S-IC constructions, and last by R1-IC constructions. In addition, we observe that SIC and PSI constructions yield the worst average infidelities out of the F-IC and R1S-IC constructions classes, respectively.

The difference in performance between POVM constructions can be understood by a variety of factors. Most importantly, whereas most theoretical analyses assume that the errors in the measurements arise entirely from the finite number of copies of the state, in our experiment this effect is completely negligible when compared to other sources of noise and errors. In particular, our experiment is dominated by errors in implementation of the POVMs themselves. These errors are a direct consequence of errors in the unitary transformations that precede the measurements in the $\{|F, m_F\rangle\}$ basis. As presented in Sec. 4.1, using randomized benchmarking we have found that the average error per unitary transformation is $\bar{\eta}_U = 0.014(2)$ and $\bar{\eta}_U = 0.018(2)$ in $d = 4$ and $d = 16$, respectively. These errors are predominately systematic, arising from fixed inhomogeneities in some of the control parameters across the atomic ensemble. Because different unitary maps are implemented with very different phase modulation waveforms, the respective unitary control errors tend to be uncorrelated from one to another. As a result, the effect of such errors tend to average out when a POVM construction involves many POVMs, each using their own unitary map.

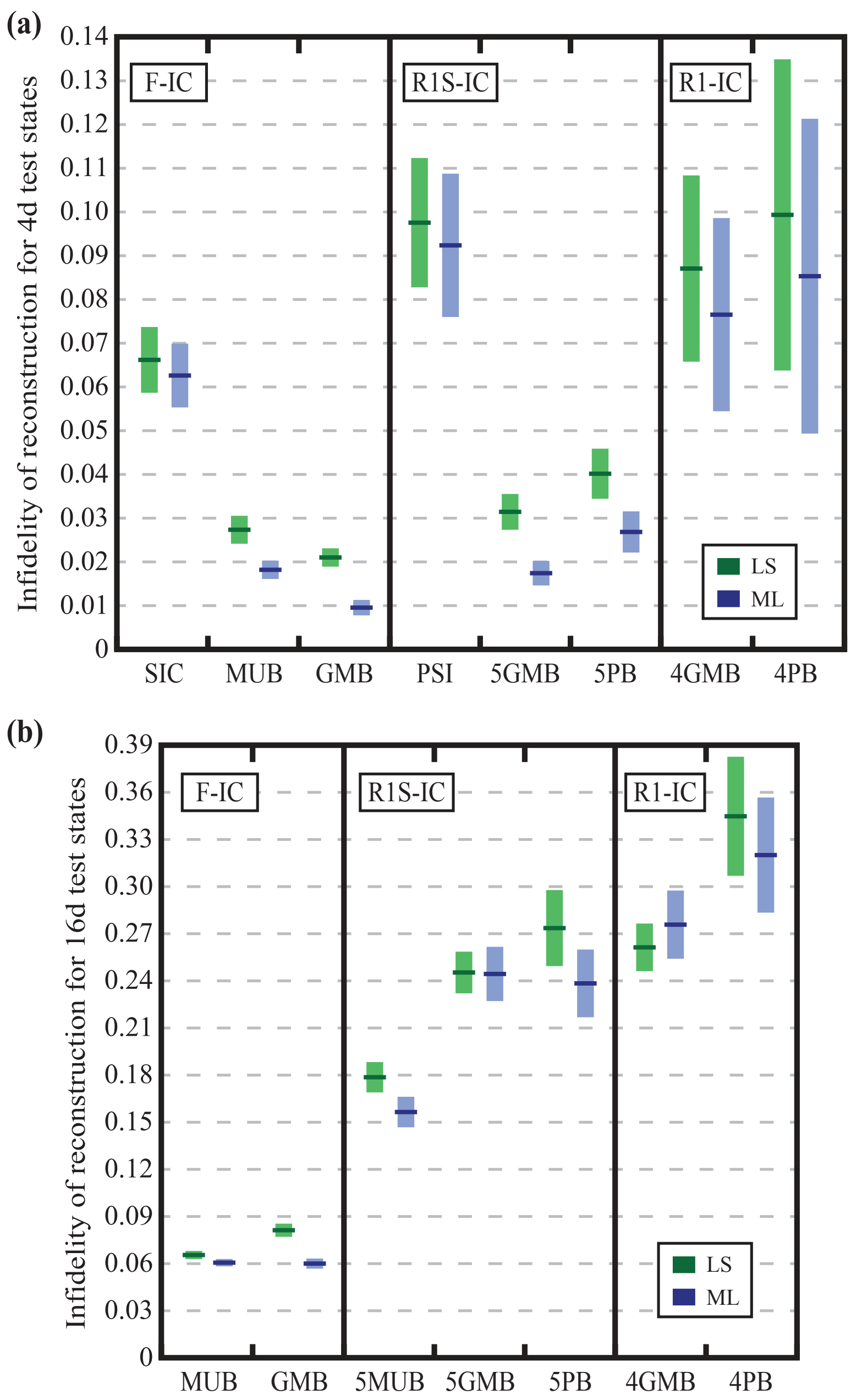

Figure 5.4: Average infidelity of reconstruction for a set of 20 test states in (a) $d = 4$ and (b) $d = 16$. Dark-color horizontal lines represent the average infidelity of reconstruction and light-color bars represent the standard error of the mean. Black vertical lines separate the POVM constructions according to their class (F-IC, R1S-IC, and R1-IC). Green color data corresponds to reconstructions obtained using the Least-Squares (LS) estimator and blue color data corresponds to results obtained using the Maximum-Likelihood (ML) estimator.

The nature of the errors in the measured data largely explains the variations in the performance between POVM constructions. F-IC constructions produce the most accurate results because they implement more than the minimal number of measurements necessary to reconstruct a pure state, hence providing redundancy that helps compensate for systematic errors in the measurements. R1S-IC constructions are next best because they can identify our nearly-pure states from within all physical states. R1S-IC constructions, however, do not provide redundant information resulting in no compensating effect from averaging independent errors in a larger-than-necessary number of unitaries. Finally, R1-IC constructions performed the worst because they do not provide redundancy and they can only identify a state from within the set of pure states. The latter is a critical limitation because when one uses convex optimization algorithms such as LS or ML, the estimators are not searching only among pure states and will often find a mixed state that fits the measured data better, especially when we have noise and errors in the POVMs.

In the particular cases of SIC and PSI constructions we recall that they both are implemented using a single POVM. As a consequence, errors present in the single unitary transformation will have a significant impact on the infidelity of reconstruction. To support this hypothesis we carried out an additional experiment where we implemented QST using SIC and PSI for a subset of 10 out of the 20 initial test states. The objective of this experiment was to reproduce the compensating effect achieved when using more than one measurement basis to obtain the measurement record used in the reconstruction. To achieve this, we take advantage of the fact that control waveforms used to implement a given unitary map are not unique, and each control waveform version of the same map lead to different errors in the measurement. Thus, by using a measurement record consisting of the average of 10 different version of the same POVM, we expect to obtain a compensating effect against the systematic errors in the experiment. In the case of SIC, the average infidelity calculated using only 1 version of the unitary map (original method) is $\bar{\eta}_{\mathrm{SIC}} = 0.0704(133)$. In contrast, using the average measurement record from the 10 versions of the unitary map, the infidelity of reconstruction decreases to $\bar{\eta}_{\mathrm{SIC}} = 0.0284(46)$. For the PSI construction, the results are similar since the infidelity with the original method is $\bar{\eta}_{\mathrm{PSI}} = 0.1461(630)$ compared to $\bar{\eta}_{\mathrm{PSI}} = 0.0532(142)$, obtained using the averaged measurement record. This improvement in fidelity makes the results obtained with SIC and PSI roughly comparable to the ones for MUB and GMB constructions at the cost of implementing more measurements.

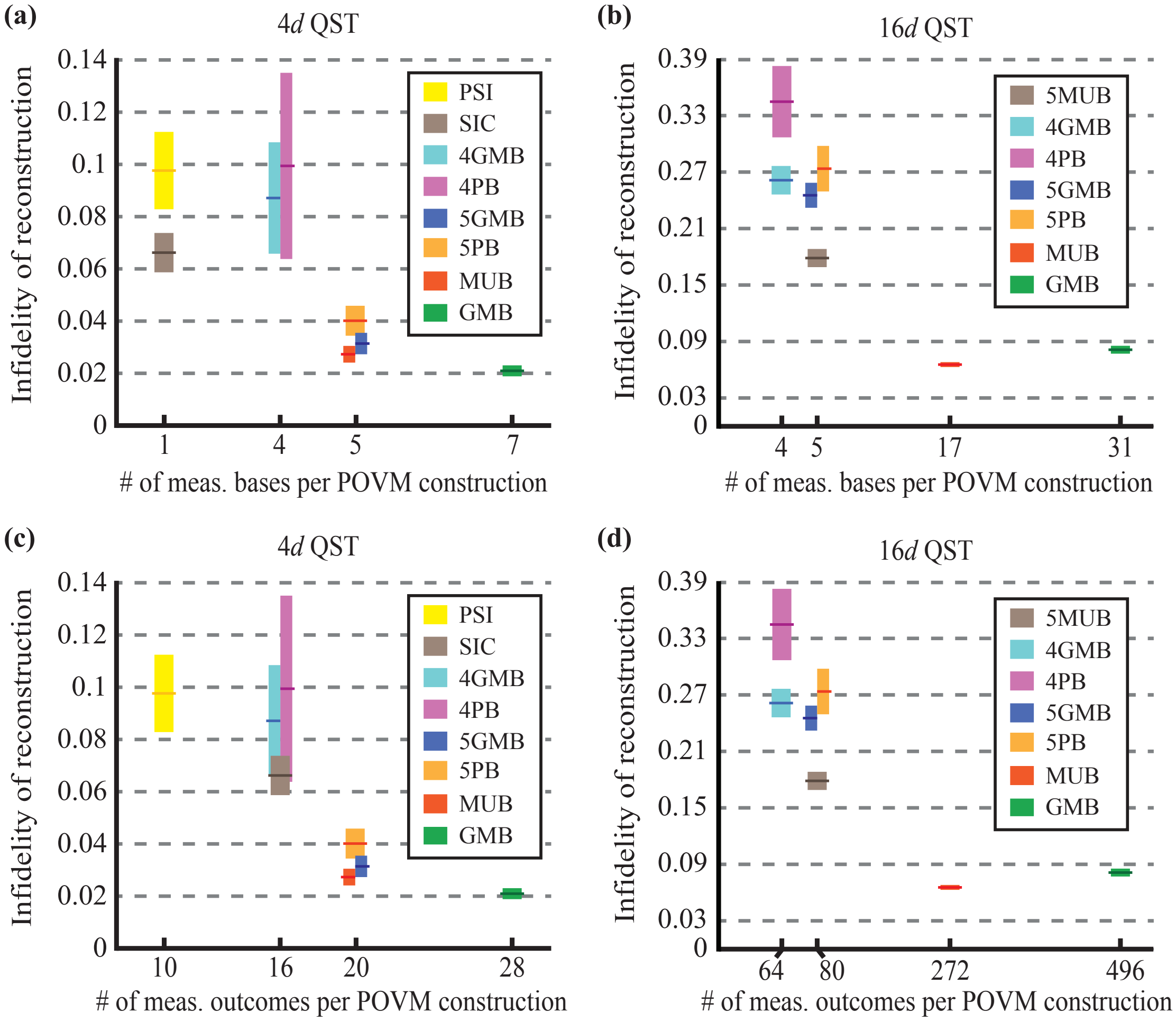

Figure 5.5: Average infidelity of reconstruction as a function of the number of measurement bases in (a) $d = 4$ and (b) $d = 16$. Average infidelity of reconstruction as a function of the number of measurement outcomes in (c) $d = 4$ and (d) $d = 16$. Dark-color horizontal lines represent the average infidelity of reconstruction and light-color bars represent the standard error of the mean. All reconstructions are performed using the Least-Squares (LS) estimator.

The results shown in Fig. 5.4 and Table 5.3 can be rearranged in a different way to motivate a discussion regarding the role of efficiency vs robustness in the reconstruction procedure. Figs. 5.5a and 5.5b show the average infidelity of reconstruction as a function of the number of measurement bases in each POVM construction. Figs. 5.5c and 5.5d show the average infidelity of reconstruction as a function of the number of measurement outcomes in each POVM construction. In this figure we see that regardless of Hilbert space dimension, there is a fairly clear trend showing that POVM constructions with more measurements bases, therefore less efficient, generally performed better in terms of accuracy of reconstruction. This suggests that the efficiency of the POVM constructions is strongly correlated with the robustness against experimental imperfections present in our setup.

Currently, we believe that the robustness effect arises from the fact that when measuring multiple bases, we make use of multiple unitary transformations each implemented using their own control waveform. Each control waveform leads to different, uncorrelated, systematic errors, and thus POVM constructions with larger number of measurement bases have a higher chance to average out the errors in the unitary maps, ultimately increasing the robustness of QST. However, implementing robust POVM constructions such as MUB and GMB comes at the cost of increased overhead in data taking, thus decreasing the efficiency of the reconstruction procedure. Finally, the performance of each POVM construction is potentially shifted relative to the general trend due to some of the other issues discussed in this section.

A final important aspect of the PSI, 4GMB and 5GMB constructions is that they all suffer from failure sets. In the context of QST, a failure set (FS) is a subset of Hilbert space for which the full set of measurements in a POVM construction cannot uniquely identify a quantum state (for details see [56]). Usually QST fails completely on a FS of measure zero, but in the presence of noise and imperfections a FS can be a finite region of state space where reconstruction fidelities are poor. In our experiment, we see that PSI and 4GMB are among the worst performing POVM constructions in the entire study. However, the experiment also shows that 5GMB produces a lower infidelity of reconstruction than, for example, 5PB, which is also a R1S-IC construction but does not have a failure set. Moreover, most of the poor results obtained with the PSI construction can be explained by the lack of robustness inherent to the construction. Therefore, while the failure set might be affecting the accuracy in the estimation, we do not have conclusive indications of its net contribution to the errors in the experimental results.

Based on the previous experimental results, we can now decisively conclude that there is a strong connection between accuracy, efficiency and robustness in the reconstruction procedure. In our experimental setup, the more efficient the POVM constructions is, the less accurate becomes estimating the right test state and the less robust is to the experimental errors present in the experiment.

### 5.5.1 QST Using Non Informationally Complete Measurement Records

In this section we discuss quantum state tomography based on partial measurement records with the LS, ML, and Trace Minimization estimators (Sec. 5.3). Trace Minimization (TM) was originally proposed as a way to harness compressed sensing for tomography, i. e., to obtain good estimates of pure or nearly-pure states based on measurement records that are not F-IC. Before it can be applied in a given experimental setting, however, it is essential to determine the proper value for the parameter $\epsilon$ that appears in the algorithm.

In Sec. 5.3 it was shown that the TM estimator has $\sum_\alpha |\mathrm{Tr}(\rho\hat{\mathcal{E}}_\alpha) - f_\alpha|^2 < \epsilon$ as one of the constrains in the optimization problem. Here we see that the error threshold parameter $\epsilon$ sets the limit above which the mean-square deviation between data and model is considered significant. When the value of $\epsilon$ is chosen too small, the constraint becomes too tight and the TM algorithm fails to find any solution for the problem. On the other hand, if the value of $\epsilon$ is chosen too large, the TM estimator will overemphasize the trace minimization part and underemphasize the matching between data and model part. This typically results in a less accurate estimate for the state.

Fig. 5.6 shows the average infidelity of reconstruction for our set of 20 test states as a function of the value of $\epsilon$ used in the Trace Minimization algorithm. Here we see that for GMB and MUB, both in $d = 4$ and $d = 16$, the final infidelity of reconstruction is highly dependent on the value of $\epsilon$. This is a striking result, because it indicates that for the exact same data one can obtain average infidelities of reconstruction that range from $\bar{\eta} = 0.01$ to almost $\bar{\eta} = 0.90$ by just adjusting the value for $\epsilon$. Moreover, we see that the effect of $\epsilon$ on the fidelity of reconstruction also depends on the choice of POVM construction. MUB yields the same fidelity of reconstruction for a wide range of $\epsilon$, whereas GMB is highly sensitive to its value. This difference in behavior remains an open question.

In order to find the appropriate value for $\epsilon$ we performed a numerical experiment with the objective of generating simulated measurement records in the presence of independently measured imperfections in our experimental setup. These measurement records were then used to calculate $\epsilon$ directly from the mean-squared differ-ence between the probabilities and simulated frequencies for each test state. The simulation starts by assuming our ensemble of atoms is perfectly prepared in the $|F = 3, m_F = 3\rangle$ state. This state serves as the input for a Schrödinger equation integrator which numerically simulates our experimental control sequence. The control sequence consists of a state-to-state map from $|F = 3, m_F = 3\rangle$ to the desired test state, followed by a unitary map according to Eq. 5.14. In order to account for the known errors in our experiment, the Schrödinger equation integrator is run for a set of Hamiltonians that sample statistical distributions around the nominal values for six control parameters as indicated in Table 5.4. These probability distributions are our best estimates for the errors and inhomogeneities of the control magnetic fields used in the experiment, each obtained through independent characterization that predates our QST experiments. Details can be found in [37, 38].

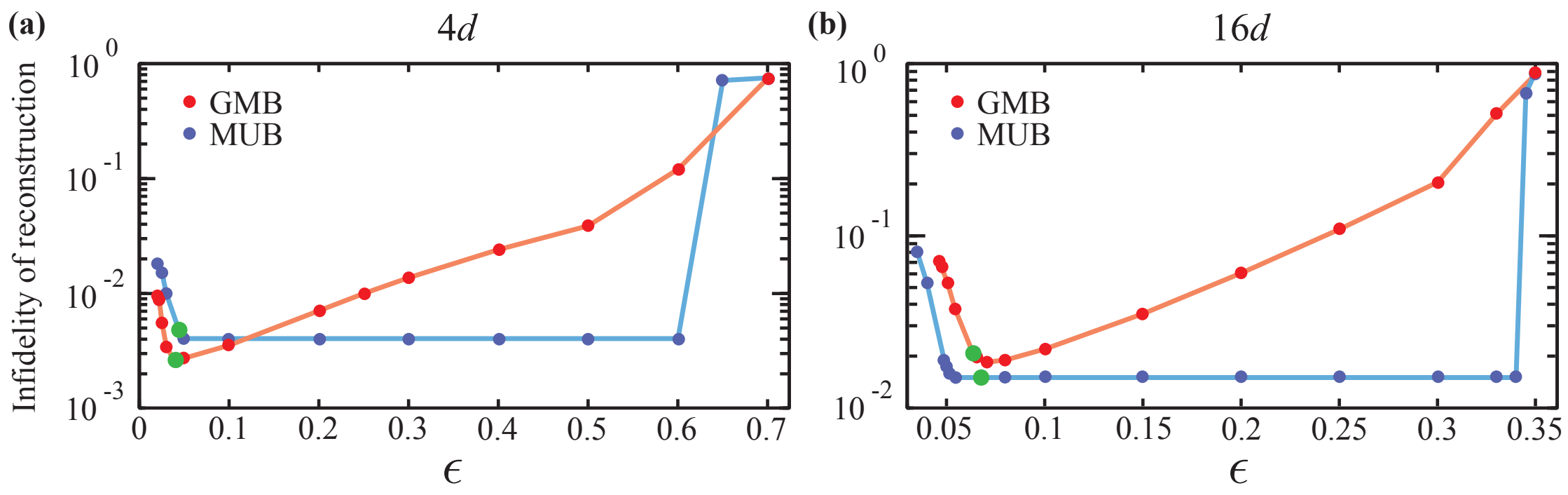

Figure 5.6: Average infidelity of reconstruction as a function of the value for $\epsilon$ used in the Trace Minimization algorithm for the (a) 4- and (b) 16-dimensional case. All reconstructions were performed using the entire informationally complete measurement records.

The output of the Schrödinger equation integrator is a state $\rho^{(\text{sim})}$ from which we can calculate the frequencies $\{\tilde{f}_\alpha^{(\text{sim})}\}$. In addition to errors in the unitary maps, our total measurement error contains a contribution from imperfections and uncertain-ties in our Stern-Gerlach analysis and associated fitting. We model this by adding a random and small number from a normal distribution to each frequency,

$$f_\alpha^{(\text{sim})} = \tilde{f}_\alpha^{(\text{sim})} + \mathcal{N}(\mu = 0, \sigma = 0.01), \tag{5.17}$$

where $\mathcal{N}(\mu, \sigma)$ represents a normal distribution of numbers with mean $\mu$ and standard deviation $\sigma$. The entire simulation is repeated for all the bases belonging to the POVM construction and for the 20 test states originally designed for the experiment. With this data we finally obtain our estimate

$$\epsilon = \frac{1}{\sqrt{nbd}} \left( \sum_{i=1}^{n} \sum_{j=1}^{b} \sum_{k=1}^{d} |\text{Tr}(\rho_i \hat{\mathcal{E}}_{j,k}) - f_{i,j,k}^{(\text{sim})}|^2 \right)^{1/2}, \tag{5.18}$$

where $n$ is the number of test states, $b$ is the number of bases in the POVM construction and $d$ is the number of measurement outcomes per basis. The green dots in Fig. 5.6 indicate the values for $\epsilon$ estimated in this fashion for each POVM construction in dimensions $d = 4$ and $d = 16$. The exact values are: $\epsilon_{4d\ \text{MUB}} = 0.0437$, $\epsilon_{4d\ \text{GMB}} = 0.0408$, $\epsilon_{16d\ \text{MUB}} = 0.0672$, and $\epsilon_{16d\ \text{GMB}} = 0.0638$, and the resulting average infidelities are: $\eta_{4d\ \mathrm{MUB}} = 0.0048(10)$, $\eta_{4d\ \mathrm{GMB}} = 0.0026(5)$, $\eta_{16d\ \mathrm{MUB}} = 0.0150(9)$, and $\eta_{16d\ \mathrm{GMB}} = 0.0217(13)$.

| Hamiltonian Parameter | Values |
|---|---|
| $\Delta_{\text{rf}}$ | 0 Hz + $\mathcal{N}(\mu = 0, \sigma = 15)$ Hz |
| $\Delta_{\mu\text{w}}$ | 0 Hz + $7\Delta_{\text{rf}}$ |
| $\Omega_x$ | 25 kHz + $\mathcal{N}(\mu = 0, \sigma = 25)$ Hz |
| $\Omega_y$ | 25 kHz + $\mathcal{N}(\mu = 0, \sigma = 25)$ Hz |
| $\Omega_{\mu\text{w}}$ | 27.5 kHz + $\mathcal{N}(\mu = 0, \sigma = 27.5)$ Hz |
| $\phi_x - \phi_y$ | $0^\circ + \mathcal{N}(\mu = 0, \sigma = 0.04)^\circ$ |

Table 5.4: Realistic errors and inhomogeneities in the control Hamiltonian used in error simulation. $\mathcal{N}(\mu, \sigma)$ represents a normal distribution of numbers with mean $\mu$ and standard deviation $\sigma$. $\phi_x - \phi_y$ represents the relative phase error between the $x$ and $y$ rf coils.

With these rigorously established estimates for $\epsilon$ it is now possible to use the TM estimator for QST. We limit the discussion to the GMB and MUB F-IC POVM constructions, since that is the situation in which we expect to see a "compressed sensing effect". Fig. 5.7 shows the average infidelity of reconstruction for these POVM constructions as a function of the number of measurement bases in $d = 4$ and $d = 16$.

Because TM is a type of compressed sensing we expect a quick drop in the infidelity well before the number of POVMs reach full informational completeness. Fig. 5.7 shows that this is indeed the case. Looking, for example, at reconstructions using the outcomes of the first five measurement bases, TM yields infidelities of reconstruction $\sim 0.10$ for all cases ($4d$ MUB, $4d$ GMB, $16d$ MUB, and $16d$ GMB). Moreover, we can also see that in the $d = 16$ case, using measurement data from additional POVMs yields only a modest further reduction in infidelity of reconstruction. The clearest example being the one for the GMB, where the difference in infidelity between 5 and 31 measurement bases is 8%, despite the fact that we use six times as much measurement data in the later case.

In regards to the LS and ML estimators, we see that they too show a rapid decrease in the infidelity well before the measurement record reaches full informational completeness. This can be understood from a recent theoretical study by Kalev *et al.* (see Ref. [80]) which proved mathematically that "compressed sensing measurements"(e.g., ones satisfying the Restricted Isometry Property) plus positivity are "strictly complete" measurements for QST. This means that there is only one physical density matrix consistent with the measured data. Their result implies that if one makes use of an optimization algorithm that searches for a physical (positive) quantum state, a quantum tomography estimator such as the Least-Squares and Maximum-Likelihood will exhibit the “compressed sensing effect” build-in in Trace Minimization. In contrast, if one makes use of an optimization algorithm without the positivity constraint one must use Trace Minimization to obtain the “compressed sensing effect”.

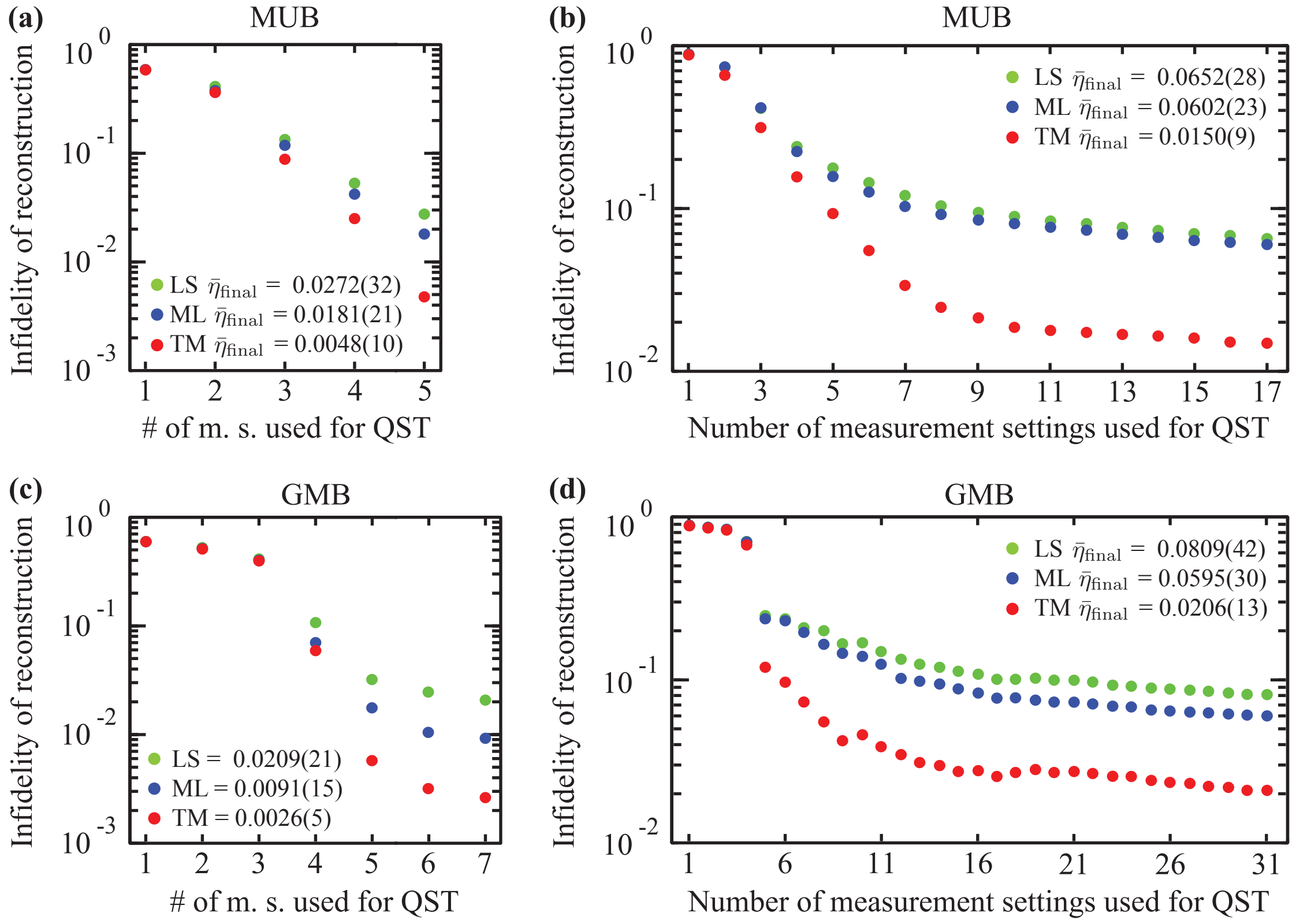

Figure 5.7: Average infidelity of reconstruction for 20 test states as a function of the number of measurement bases used in the reconstruction algorithm. (a) Shows results for MUB in $d = 4$, (b) results for MUB in $d = 16$, (c) results for GMB in $d = 4$, and (d) results for GMB in $d = 16$. Green, blue and red dots correspond to results obtained using the LS, ML, and TM estimators, respectively.

# CHAPTER 6

# QUANTUM PROCESS TOMOGRAPHY EXPERIMENTS

This chapter discusses the experimental results for quantum process tomography (QPT) implemented in the 16-dimensional Hilbert space associated with the electronic ground state of cesium atoms. We begin with a review of the standard procedure to implement quantum process tomography, followed by the introduction of a scheme to implement efficient QPT. We then present the process matrix estimator used for reconstruction. The next sections present and discuss experimental results from several QPT experiments. A detailed discussion of the theoretical background for this chapter can be found in the dissertation of Charles Baldwin [56], who along with Ivan Deutsch, contributed greatly to this work.

## 6.1 Quantum Process Tomography

Quantum process tomography (QPT) is a procedure by which one seeks to estimate a quantum process $\mathcal{W}$ that maps an initial state to a final state,

$$\mathcal{W}[\rho_{\text{in}}] = \rho_{\text{out}}. \tag{6.1}$$

In general, the process satisfies two conditions: *complete positivity* (CP) and *trace preservation* (TP). A CP process is one that when applied to a positive input quantum state $\rho_{\text{in}}$, produces a positive output state, i.e. when $\rho_{\text{in}} \geq 0$ then $\mathcal{W}[\rho_{\text{in}}] \geq 0$. A TP process is one that preserves the trace of the quantum state, i.e. $\mathrm{Tr}(\rho_{\mathrm{in}}) = \mathrm{Tr}(\mathcal{W}[\rho_{\mathrm{in}}])$.

For a $d$-dimensional Hilbert space, such processes can be represented in a basis of orthonormal Hermitian $d \times d$ matrices $\{\Upsilon_\alpha\}$ [21], such that

$$\mathcal{W}[\rho_{\mathrm{in}}] = \sum_{\alpha,\beta}^{d^2} \mathcal{X}_{\alpha,\beta} \Upsilon_\alpha \rho_{\mathrm{in}} \Upsilon_\beta^\dagger, \tag{6.2}$$

where $\mathcal{X}$ is a $d^2 \times d^2$ matrix known as the *process matrix*. A rank-1 process matrix corresponds to a unitary map $\hat{U}$. In quantum process tomography the goal is to estimate the process matrix, which is specified by a total of $d^4 - d^2$ parameters when constrained to be trace preserving.

In the standard procedure, the process matrix is reconstructed by evolving a sequence of $d^2$ linearly independent pure states using the unknown process and performing full quantum state tomography on the resulting output states. Since each instance of state tomography yields $d^2 - 1$ parameters, the entire procedure yields $d^4 - d^2$ parameters and is therefore informationally complete. Thus, in a 16-dimensional Hilbert space, reconstructing a process matrix would require enough measurement data to accurately estimate 65280 parameters. This is well beyond feasible in our current experiment and will likely remain so for the foreseeable future.

The objective of our project is to improve the efficiency of quantum process tomography by taking advantage of a key idea proposed by Baldwin *et al.* in [81]. Their idea for the so-called "intelligent probing" derives from the fact that standard QPT assumes an arbitrary process about which we have no prior information. In fact, QPT can be made substantially more efficient if we know in advance that the process is unitary or near-unitary. For a unitary process Eq. 6.1 takes the form

$$\mathcal{W}[\rho_{\text{in}}] = \rho_{\text{out}} = \hat{U}\rho_{\text{in}}\hat{U}^{\dagger}. \tag{6.3}$$

Here, the unknown process is given by the map $\hat{U}$ which is specified by only $d^2 - 1$ real parameters, and one can fully reconstruct it by intelligently probing it with a particular set of $d$ pure states.

To better understand the idea of intelligent probing, we briefly summarize the relevant part of [81]. A unitary map is a transformation from the orthonormal basis $\{|n\rangle\}$ to its image basis $\{|u_n\rangle\}$,

$$\hat{U} = \sum_{n=0}^{d-1} |u_n\rangle\langle n|. \tag{6.4}$$

In essence, the task in QPT of a unitary map is to fully characterize the bases $\{|u_n\rangle\}$, along with the relative phases of the summands $\{|u_n\rangle\langle n|\}$. Let now the set of $d$ intelligent probe states be:

$$\begin{aligned} |\psi_0\rangle &= |0\rangle \\ |\psi_n\rangle &= \frac{1}{\sqrt{2}}\left(|0\rangle + |n\rangle\right),\ n = 1, \ldots, d-1. \end{aligned} \tag{6.5}$$

The tomographic procedure works as follows. First let the map act on $|\psi_0\rangle$ and make an IC measurement on the output state $\hat{U}|\psi_0\rangle = |u_0\rangle$, from which we can obtain the state $|u_0\rangle$ (up to a global phase that we can set to zero). Next, let the unitary map act on $|\psi_1\rangle$ and perform an IC measurement on the output state $\hat{U}|\psi_1\rangle$. From the relation $\hat{U}|\psi_1\rangle\langle\psi_1|\hat{U}^{\dagger}|u_0\rangle = \frac{1}{2}(|u_0\rangle + |u_1\rangle)$ we obtain the state $|u_1\rangle$, including its phase relative to $|u_0\rangle$. The procedure is repeated for every state $|\psi_n\rangle$ with $n = 1, \ldots, d-1$, thereby obtaining all the information about the basis $\{|u_n\rangle\}$, including the relative phases in the sum of Eq. 6.3, and completing the tomographic procedure for a unitary map.

The previous discussion shows that QPT of a unitary map can be achieved using the intelligent probing scheme. However, in any real-world implementation the process is never exactly unitary due to errors and imperfections in the experimental setup. In our case, most of these errors arise from imperfect quantum control (Sec. 4.1) resulting in implementation of near-unitary maps in the experiment. In order to make the procedure for QPT robust against such errors we take advantage of the results obtained in our project to study quantum state tomography. In Sec. 5.5 we were able to show that for our particular experimental setup, the optimal POVM construction to perform QST, in both the 4- and 16-dimensional cases, was MUB. MUB provided the best tradeoff between accuracy, efficiency and robustness against our experimental imperfections. For that reason, we will make use of the MUB construction to obtain the informationally complete measurement record for each of the output states involved in QPT. By doing this, we expect that the procedure for process tomography will also be robust against our experimental imperfections.

## 6.2 Estimation Algorithm for QPT

In order to introduce the mathematical form for the estimation algorithm used in process tomography we first recall the expression for the probabilities of the outcomes of a measurement as a function of POVM element and state being measured.

Given a POVM construction with POVM elements $\{\hat{\mathcal{E}}_l\}$, the probability of observing an outcome $\mathcal{E}_l$ for a state $\rho_j^{\text{out}}$ is given by

$$p_{j,l} = \text{Tr}(\rho_j^{\text{out}} \hat{\mathcal{E}}_l), \tag{6.6}$$

which can be expressed in terms of the process matrix using Eq. 6.2,

$$\begin{aligned} p_{j,l} &= \text{Tr}\left(\sum_{\alpha,\beta}^{d^2} \mathcal{X}_{\alpha,\beta} \Upsilon_\alpha \rho_j^{\text{in}} \Upsilon_\beta^\dagger \hat{\mathcal{E}}_l\right), \\ &= \text{Tr}\left(D_{j,l}^\dagger \mathcal{X}\right), \end{aligned} \tag{6.7}$$

where $D_{j,l}$ is a $d^2 \times d^2$ matrix given by $(D_{j,l}^\dagger)_{\alpha,\beta} = \text{Tr}(\rho_j^{\text{in}} \Upsilon_\beta^\dagger \hat{\mathcal{E}}_l \Upsilon_\alpha)$. Eq. 6.7 provides a convenient way to relate the process matrix elements to the measurement outcomes from the experiment.

The estimator we employ to perform all the reconstructions in our QPT experiments is the Least-Squares (LS) estimator. As in the case for QST, the LS estimator objective is to find the process matrix that minimizes the distance according to the sum of the squares of the difference between the observed frequencies of outcomes $\{f_{j,l}\}$ and the probabilities of outcomes $\{p_{j,l}\}$, under the condition that the estimated process matrix is a complete positive ($\mathcal{X} \geq 0$) and trace preserving ($\sum_{n,m} \mathcal{X}_{n,m} \Upsilon_m^\dagger \Upsilon_n = 1$) map. Thus, to estimate the process matrix we solve the optimization problem,

$$
\begin{aligned}
\text{minimize:} \quad & \sum_{j,l} |\mathrm{Tr}(D_{j,l}^{\dagger}\mathcal{X}) - f_{j,l}|^2 \\
\text{subject to:} \quad & \sum_{n,m} \mathcal{X}_{n,m}\Upsilon_m^{\dagger}\Upsilon_n = 1, \\
& \mathcal{X} \geq 0,
\end{aligned}
\tag{6.8}
$$

where $f_{j,l}$ is the frequency of outcome for the POVM element $\mathcal{E}_l$ while measuring the state $\rho_j^{\mathrm{out}}$. For QPT performed in low dimensional systems such as $d = 4$ and $d = 7$, this problem is solved using the MATLAB package CVX. However, for large dimensional systems such as $d = 16$, the computation effort required to solve Eq. 6.8 using CVX becomes infeasible on a regular desktop computer. In this case, the optimization problem is solved using a *gradient projection algorithm* inspired by methods presented in [82]. Details about the exact form of the gradient projection algorithm used in this dissertation can be found in [56]. The output of either algorithm is the estimated process matrix $\mathcal{X}_e$. Here, it is important to highlight that when the reconstruction is cast using Eq. 6.8, the estimation algorithm directly produces an estimate for the process matrix and never explicitly reconstructs the output states (although the necessary information to do so is available).

To evaluate the performance of the procedure for QPT, we calculate the fidelity between the reconstructed process matrix $\mathcal{X}_e$ and the process matrix of a target unitary map $\mathcal{X}$

$$
\mathcal{F} = \frac{1}{d^2}\left(\mathrm{Tr}\sqrt{\sqrt{\mathcal{X}}\mathcal{X}_e\sqrt{\mathcal{X}}}\right)^2, \tag{6.9}
$$

as well as the infidelity (error) of reconstruction given by $\eta = 1 - \mathcal{F}$.

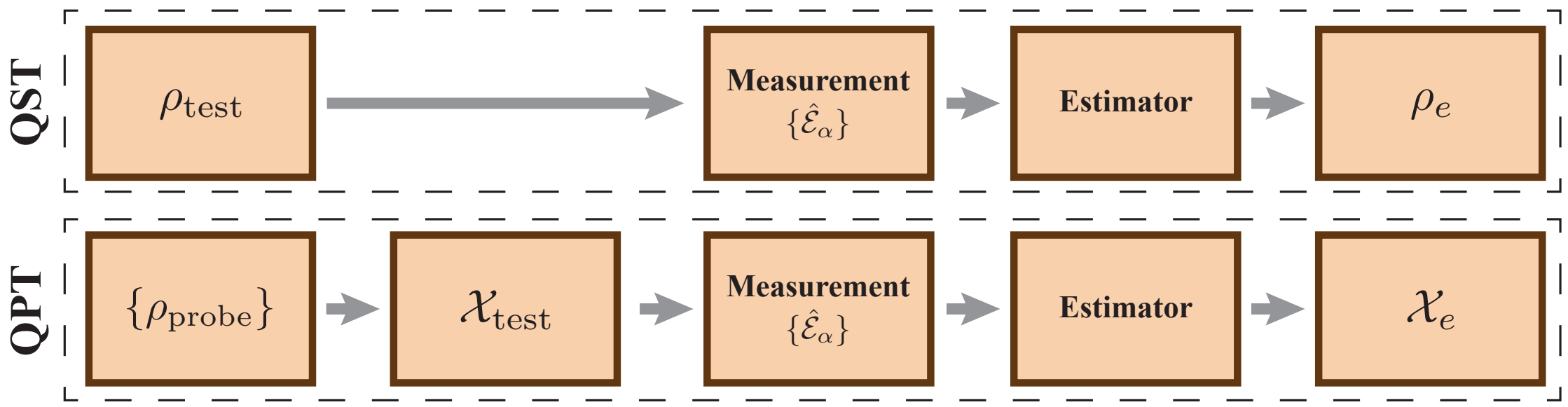

Figure 6.1: Comparison between the procedures for quantum state tomography and quantum process tomography. In QPT the set of input states $\{\rho_{\text{probe}}\}$ consist of $d^2$ states for standard QPT and $d$ states for the intelligent probing scheme. The test unitary process is immediately implemented after state preparation and right before the measurement protocol which is identical to the one for QST. In QPT, the estimator uses the combined measurement records from all output states to generate the estimated process matrix $\mathcal{X}_e$.

## 6.3 Quantum Process Tomography in the Laboratory

Our procedure to implement quantum process tomography consist of a very similar sequence of steps compared to the one for quantum state tomography (see Fig. 6.1). In the procedure for QPT, we evolve a set of input states $\{\rho_{\text{probe}}\}$ using the test unitary process and measure an IC-POVM on each output state. In standard QPT there is a total of $d^2$ input states while for the intelligent probing scheme there is $d$ states. For the data analysis we use the combined measurement records from all output states to reconstruct the process matrix $\mathcal{X}_e$ using the LS estimator (Eq. 6.8).

Our experimental implementation of QPT consists of many separate runs of the experiment. Each of these runs prepares the atomic ensemble in one of the input probe states, applies a known test process, and measures one of the bases in the MUB POVM construction. The procedure is repeated for the set of all input probe states and all MUB bases. The aggregate data is then processed according to Eq.

6.8 to obtain an estimate for the test process.

QPT was performed on a set of ten unitary processes in $d = 4$, ten unitary processes in $d = 7$, and one unitary process in $d = 16$. All unitary maps were chosen randomly according to the Haar measure. For the 4-dimensional case, in addition to the $d = 4$ intelligent probe states, we supplemented the information obtained with this optimal set by acquiring measurement records for a set of $d^2 - d = 12$ additional linearly independent states. The acquisition of the extra information was motivated by the fact that standard QPT requires $d^2$ input states and we wanted to evaluate the tradeoff between accuracy and efficiency among the two QPT schemes. In the 7-dimensional case, attempting to acquire measurement records for $d^2 = 49$ input states was too time consuming to be practical, and thus only the set of $d = 7$ intelligent probe states were used in the experiment. Lastly, in the 16-dimensional case, even the use of $d = 16$ intelligent probe states requires enough measurement that QPT becomes impractical, except as a proof-of-principle demonstration for a single unitary map.

| QPT dimension (procedure) | # of test processes | # of input probe states | # of meas bases per input state | # of meas outcomes per process |
|---|---|---|---|---|
| 4 (Int. Probing) | 10 | 4 | 5 | 80 |
| 4 (Standard) | 10 | 16 | 5 | 320 |
| 7 (Int. Probing) | 10 | 7 | 8 | 392 |
| 16 (Int. Probing) | 1 | 16 | 17 | 4352 |

Table 6.1: Summary of QPT experiments. Measurements on the output states were performed using the MUB POVM construction, which requires $d + 1$ measurement bases, each yielding $d$ outcomes.

Table 6.1 summarizes the information about number of test processes, measure-ment bases, and measurement outcomes required to reconstruct the different process matrices in our experiment. Here, it is interesting to see that even when using the intelligent probing procedure, the resources required for QPT scale badly with Hilbert space dimension. However, intelligent probing does provide a clear increase in efficiency with respect to the standard procedure. For example, looking at the number of measurement outcomes required to perform QPT in $d = 7$ using intelligent probing we see that they are comparable to the number of outcomes required to perform QPT in $d = 4$ using the standard approach.

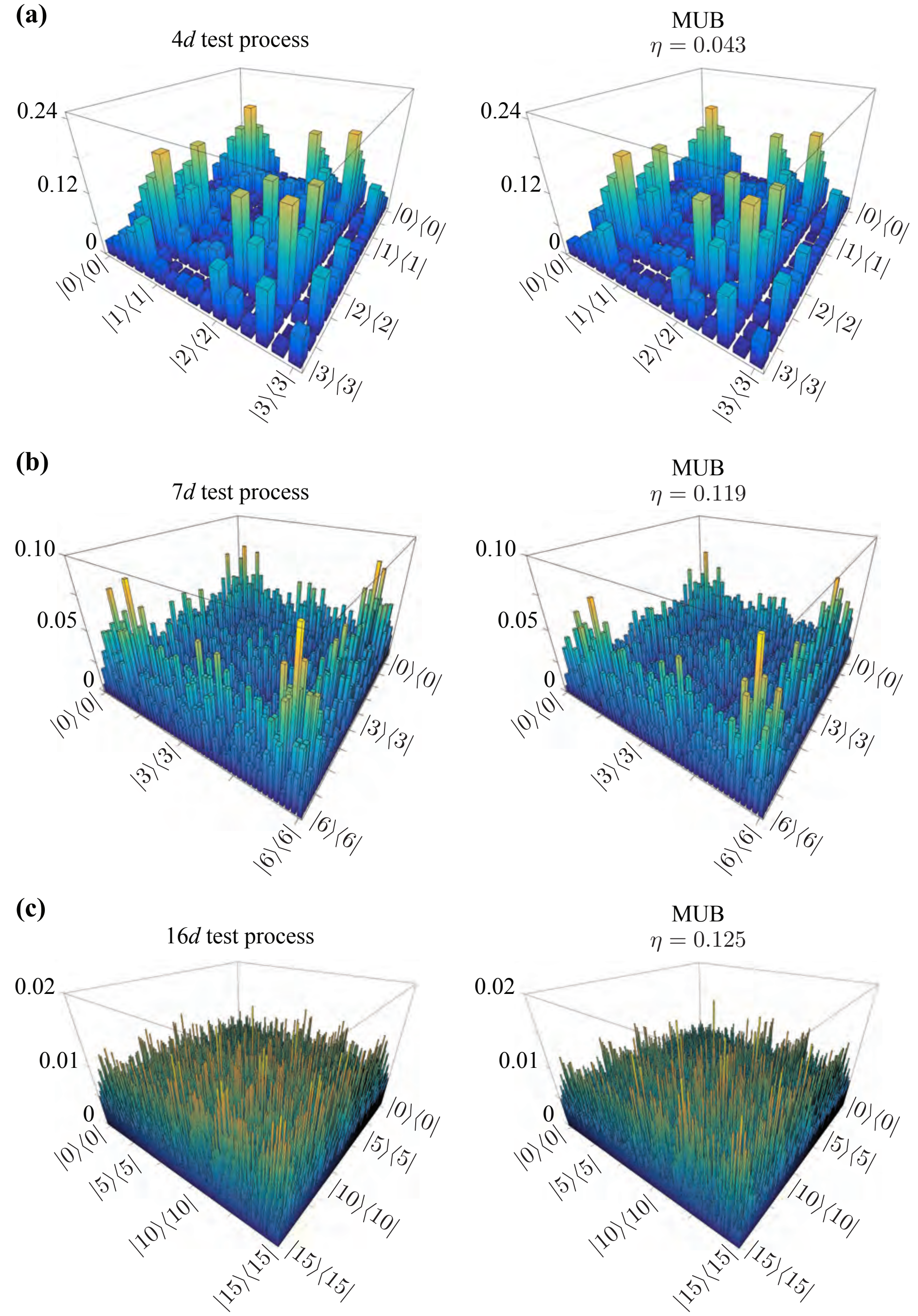

Figure 6.2: Examples of quantum process tomography for unitary processes in (a) $d = 4$, (b) $d = 7$, and (c) $d = 16$. Figures on the left shown the test process matrices and figures on the right show the reconstructed process matrices. All reconstructions were performed using the LS estimator and the measurement records corresponding to $d$ intelligent probe states. In all figures only the absolute values of the matrix elements are shown.

Figure 6.2 shows examples of quantum process tomography for one test process in $d = 4$, one test process in $d = 7$, and the single test process in $d = 16$. In all cases, the reconstructions were done using the Least-Squares (LS) estimator and the measurement record used in the reconstruction algorithm comprised the entire set of frequencies of outcomes for all $d$ input intelligent probe states. We see that for all three cases the infidelity of reconstruction using $d$ input states is close to $\sim 0.10$. This is somewhat worst than the infidelity of QST, which is not surprising given the greater complexity of QPT. In addition, Fig. 6.2 is a direct visualization of the large amount of information contained in a process matrix. In principle, one should be able to use this information to diagnose the errors and imperfections in the underlying experiment. At present, however, it is not known how one may do so, and indeed the use of QPT as a practical diagnostic tool very much remains a challenge for the future. In this dissertation, we focus our discussion on the accuracy of reconstruction (quantified by the process infidelity calculated using Eq. 6.9) and efficiency (quantified by the number of probe states needed to reach IC). A few ideas for using QPT to distinguish between coherent and incoherent errors can be found in [81].

## 6.4 Process Tomography with Intelligent Probing: Results and Discussion

Fig. 6.3a and 6.3b show the results of QPT performed on a set of ten unitary test processes in $d = 4$ and $d = 7$, respectively. Fig. 6.4 shows the results of QPT performed on the single unitary test process in $d = 16$.

In all figures we see that the infidelity of reconstruction quickly drops and reaches a low value when the estimation algorithm employs the set of $d$ intelligent probe states. In the 4-dimensional case the average infidelity using the 4 intelligent probe states is $\bar{\eta}_{(4)} = 0.1002(58)$ while the infidelity using all 16 linearly independent states is $\bar{\eta}_{(16)} = 0.0592(19)$. This shows that reconstructing unitary processes with the standard QPT procedure yields a modest improvement in fidelity compared to the intelligent probing procedure. In the 7-dimensional case the infidelity using the 7 intelligent probe states is $\bar{\eta}_{(7)} = 0.1574(46)$. Lastly, in the 16-dimensional case the infidelity for the single test unitary map using the 16 intelligent probe states is $\eta_{(16)} = 0.1247$, while reconstruction using an additional 4 linearly independent states yields $\eta_{(20)} = 0.1354$. To our knowledge, this is the largest Hilbert space in which process tomography has been successfully implemented in, and it was only possible by taking advantage of the intelligent probing approach.

The previous results allows us to conclude that the intelligent probing scheme proposed in [81] does provide an efficient way to obtain high fidelity results when performing QPT on unitary maps.

In Figs. 6.3 and 6.4 we see that even though our procedure for QPT produces high fidelity results, it was unable to render perfect estimates for the process ma-trices. As with the QST project, we believe the main reason behind this lies in the fact that experimental imperfections in our setup ultimately limit our ability to implement perfect state preparation and unitary maps. This translates into imperfect preparation of the probe states, imperfect implementation of the test unitary processes and imperfect implementation of the measurements (MUB POVMs).

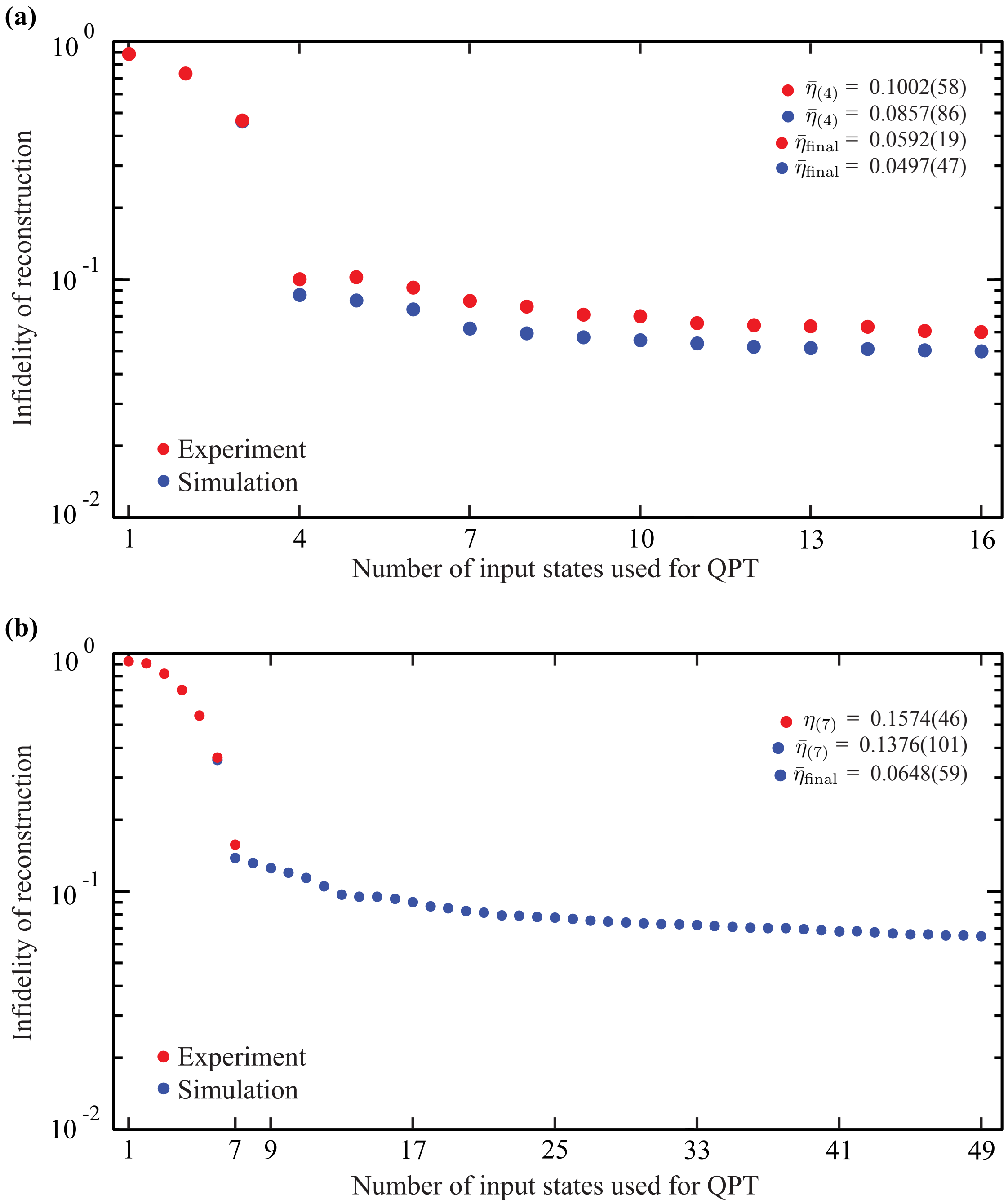

Figure 6.3: Average infidelity of QPT for 10 test unitary processes as a function of the number of input probe states used in the reconstruction algorithm. (a) shows results for $d = 4$ and (b) results for $d = 7$. Reconstruction was performed using the LS estimator, implemented with the CVX package in MATLAB. Red dots correspond to experimental data and blue dots correspond to simulations.

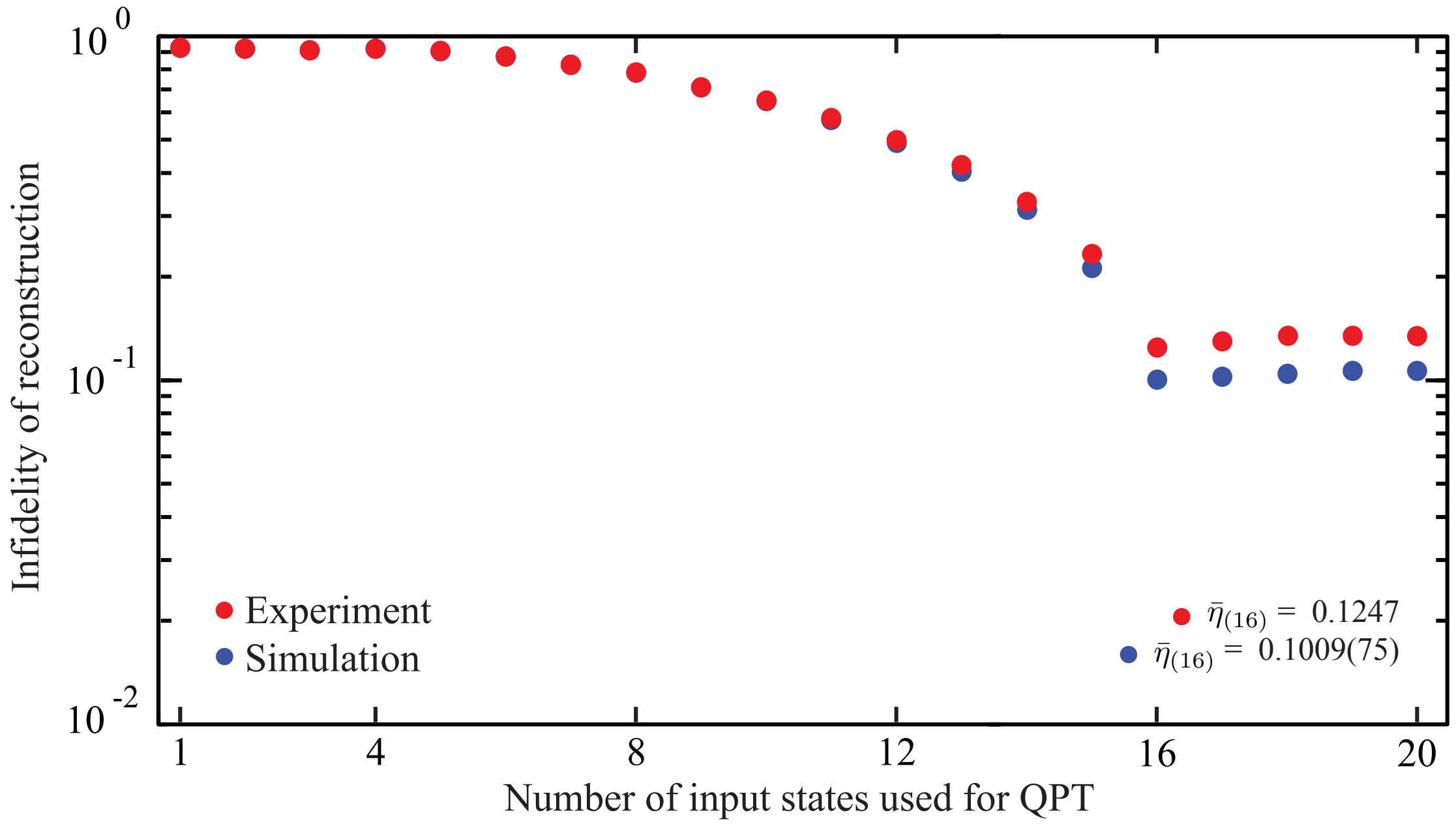

Figure 6.4: Average infidelity of QPT for a single test unitary process in $d = 16$ as a function of the number of input probe states used in the reconstruction algorithm. Reconstruction was performed using the LS estimator implemented with a gradient projection algorithm [56]. Red dots correspond to experimental data and blue dots correspond to simulations.

In order to gain a good idea about the effects of our experimental imperfections and provide a baseline for the performance of our procedure for QPT, we performed a numerical experiment in which we simulated our experimental sequence using a similar approach to the one presented in Sec. 5.5.1. To start the simulation, our ensemble of atoms is assumed to be perfectly prepared in the state $|F = 3, m_F = 3\rangle$. This state serves as the input for a Schrödinger equation integrator which numerically simulates our experimental control sequence. The control sequence consist of three control waveforms: state preparation, test unitary process, and unitary map to implement measurement. In order to account for the known errors in our experiment, the Schrödinger equation integrator is run for a set of Hamiltonians that sample statistical distributions from each of the inhomogeneous control parameters (see Sec. 5.5.1). We then use the output state of the Schrödinger integrator to calculate the frequencies of outcomes, which are then put into Eq. 5.17 to simulate errors present in SGA. Finally, the resulting measurement record $\mathcal{M}_{\text{sim}}$ is feed into our LS estimator to reconstruct the process matrix. The entire procedure to perform QPT was simulated for each test unitary map used in the experiment.

Blue color dots in Fig. 6.3 and Fig. 6.4 represent the simulated average infidelity for the set of 10 test unitary processes in each dimension. As we can see, the results produced by the simulation and the experimental data follow similar trends and correspond fairly well with each other. This allows us to infer that, as expected, most of the errors present in the procedure for QPT are a direct result of errors and imperfections in our implementation of control and measurement.

Lastly, it is worth noting that for the $d = 16$ case (Fig. 6.4) even the simulated experiment was limited to perform reconstruction using only the first 20 probe states. This is because, despite the fact that we can easily produce simulated measurement records for any number of input states, a regular desktop computer simply does not have enough memory to store and efficiently process the amount of data involved, whether using convex optimization or our gradient projection algorithm.

### 6.4.1 QPT Using Non Informationally Complete Measurement Records

Similar to the QST project, we can perform QPT using partial measurement records to attempt reconstruction. In the QPT case, all measurements are obtained by implementing the MUB POVM construction. This means that for a given probe state, an informationally complete measurement record consist of a set of $d^2 + d$ frequencies of outcomes, obtained by measuring $d + 1$ bases, each with $d$ outcomes.

The idea of non informationally complete QPT consist of reconstructing the process matrix from a measurement record obtained by measuring $n_{\mathrm{b}} < d + 1$ MUBs for each of the $n_{\mathrm{p}}$ probe states. Fig. 6.5 shows the average infidelity of reconstruction for our set of 10 unitary test processes in $d = 4$ as a function of the total number of input probe states as well as the number of bases measured for each probe state. Here, we see that when the individual measurement records only include the frequencies of outcomes for the first MUB ($n_{\mathrm{b}} = 1$), the infidelity of reconstruction stays high irregardless of the number of probe states used. However, when the measurement records include, e.g., the first two MUB ($n_\mathrm{b} = 2$) we see that the infidelity of reconstruction using all 16 probe states is comparable to the infidelity using the fully IC measurement record. This is a remarkable results considering that the latter case requires more than twice the number of measurements than the former.

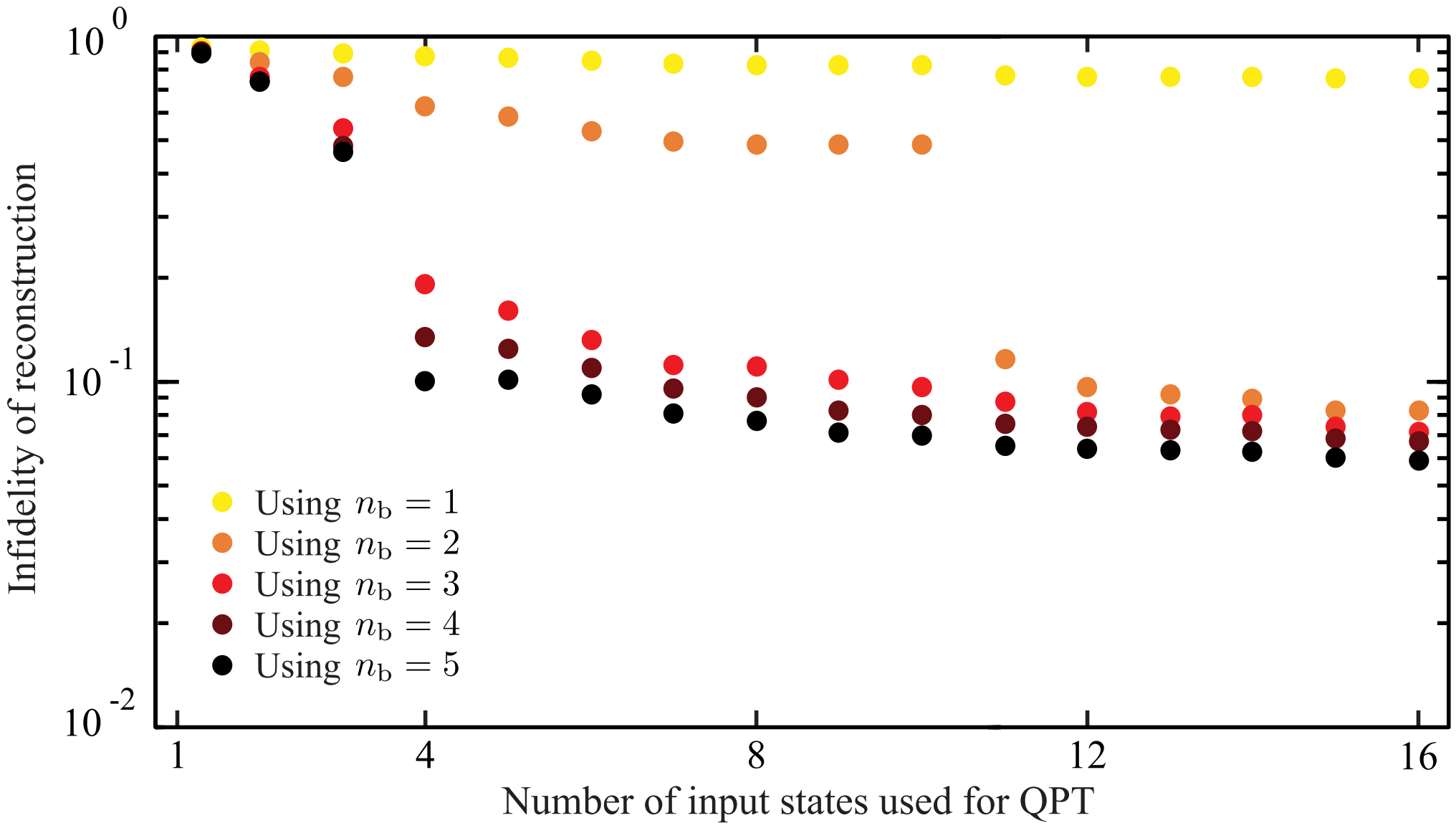

Figure 6.5: Average infidelity of reconstruction for 10 unitary test processes in $d = 4$ as a function of the total number of input probe states, for different number of bases measured per probe state. The case $n_\mathrm{b} = 5$ corresponds to QPT with an IC measurement record.

In order to explore if QPT using non IC measurement records can produce high fidelity results for unitary processes in $d > 4$, we performed a numerical experiment to implement QPT over our set of 10 test processes in $d = 7$ using the simulation protocol discussed in Sec. 6.4. Fig. 6.6 shows the average infidelity of reconstruction as a function of both the total number of input probe states and the number of bases measured for each probe state. In a similar way to the 4-dimensional case implemented in the laboratory, the simulated experiment produces fidelities comparable to fully IC QPT for several scenarios, e.g., if the measurement record consist of the first 2 MUBs and all 49 probe states or if the measurement record consist of frequencies of outcomes produced by the first 5 MUB and the 12 probe states.

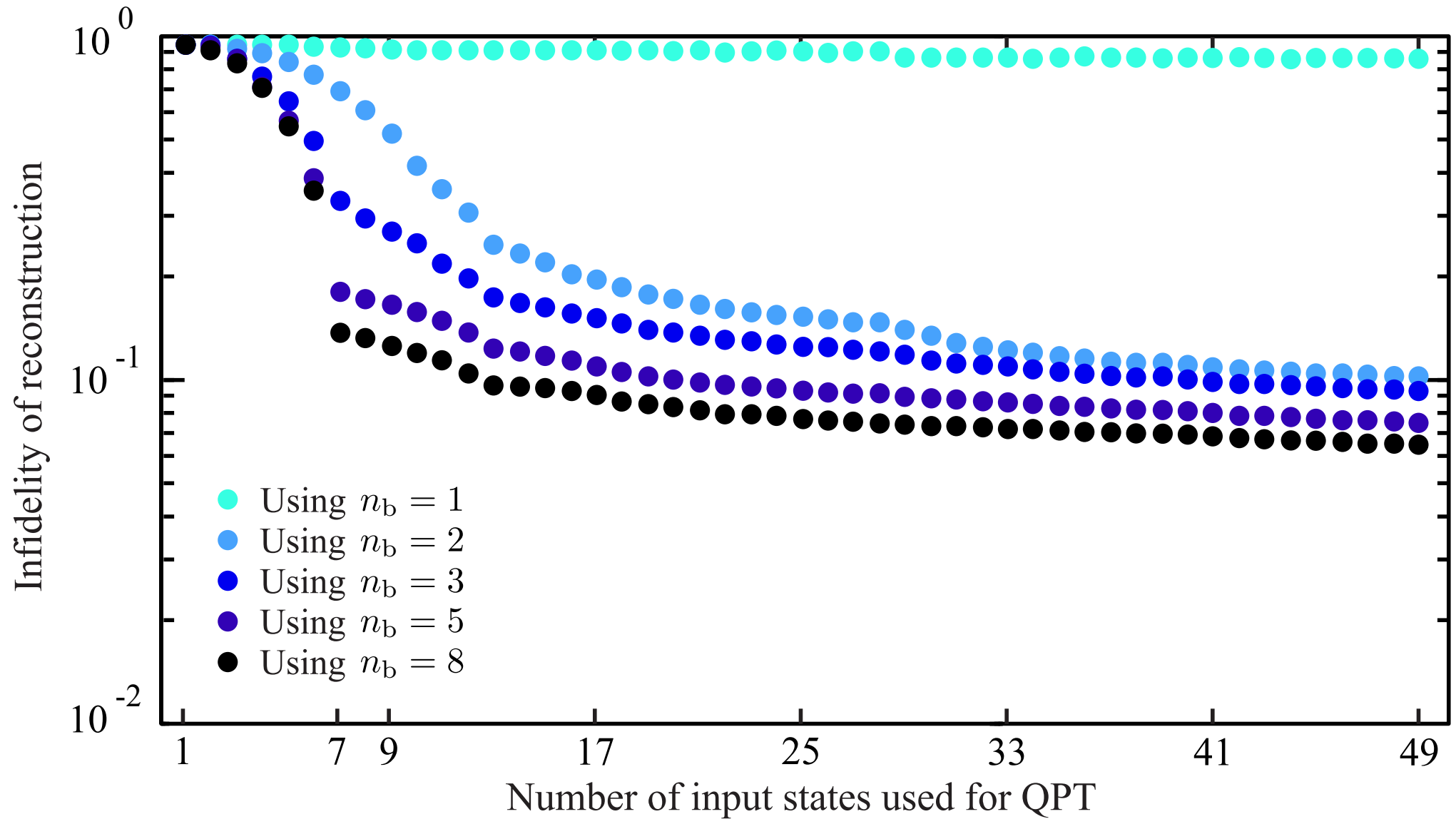

Figure 6.6: Average infidelity of reconstruction for 10 unitary test processes in $d = 4$ as a function of the total number of input probe states, for different number of bases measured per probe state. The case $n_{\rm b} = 8$ corresponds to QPT using an IC measurement record.

In general, both experimental data in $d = 4$ and simulated data in $d = 7$ allows to conclude that performing QPT using non IC measurement records does produce high fidelity estimates. However, performing the reconstruction using the set of $d$ intelligent probe states and an informationally complete measurement record still remains as the most efficient and robust approach to implement QPT.

# CHAPTER 7

# SUMMARY AND OUTLOOK

As the size of the fundamental components in emerging technologies becomes smaller, there will be a point where quantum mechanical effects govern their dominant behavior. Quantum information science provides us with a framework to study the challenges involving the design, construction and control of novel quantum systems. Motivated by the eventual realization of powerful quantum technologies, scientists are now actively developing theoretical and experimental tools to implement quantum control and measurement over qubit and qudit systems. This dissertation presented several quantum control and quantum tomography experiments implemented on our unique testbed consisting of the hyperfine manifold associated with the electronic ground state of $^{133}$Cs atoms. Our system provides long coherence times and can be manipulated with radio-frequency, microwave, and optical fields.

In the first part of chapter 4 we reviewed the experimental implementation of a protocol for arbitrary unitary transformations in our 16-dimensional Hilbert space, using phase modulated rf and microwave magnetic fields to drive the atomic evolution. The fidelity of the resulting transformations is verified experimentally through randomized benchmarking, which indicates an average fidelity better than 98% for robust control waveforms with overall duration (control time) $T = 600$ $\mu$s and phase step duration $\Delta t = 4$ $\mu$s. We also presented results from implementing control fields designed to be robust against static and/or dynamical imperfections and perturbations. In this study we found that robust control waveforms outperformed nonrobust waveform by a significant margin at the modest cost of increasing the control time from $T = 600$ $\mu$s to $T = 800$ $\mu$s. This result shows that optimal control may prove as a good strategy to perform high-accuracy control tasks in qudit systems where less than ideal conditions exist, e.g., atoms moving around in the light shift potential of a dipole trap.

In the second part of chapter 4, we presented a demonstration of inhomogeneous quantum control in an 8-dimensional Hilbert space. In this experiment the objective was to design a global control waveform to perform different unitary transformations for different members of the atomic ensemble, depending on the presence or absence of a light shift generated from an optical addressing field. Experimental results showed that we can successfully implement two distinct unitary transformation by using robust control waveforms with duration $T = 1.24$ ms, reaching an average benchmarked fidelity of $\mathcal{F}_B = 0.922(18)$. Currently, we believe that the main limitation to obtain higher fidelities lies on the fact that the control waveforms used in this experiment are too long and experimental limitations such as the ones discussed in App. A eventually hinder their performance. However, the proof-of-principle experiments presented in this dissertation serve as a good baseline for further exploration of inhomogeneous control for qudits which may ultimately lead to addressable unitary control similar to that demonstrated for qubits in optical lattices.

In the first part of chapter 5, we presented the findings from a comprehensive experimental study in which we performed quantum state tomography comparing nine different POVM constructions and three different state estimators, using our 16-dimensional Hilbert space. We found that, as a general trend, the most accurate reconstruction results are obtained using fully IC (F-IC) POVM constructions, followed by rank-1 strictly IC (R1S-IC) constructions, and last by rank-1 IC (R1-IC) constructions. In addition, we observed that efficient POVM constructions such as SIC and PSI yield the worst infidelities out of the F-IC and R1S-IC constructions classes, respectively. The previous results are understood by the fact that F-IC constructions implement redundant measurements which help compensate for systematic errors in the experiment. R1S-IC constructions are next best because they do not provide redundancy but they can identify our test states from within the set of all physical states. Finally, R1-IC constructions performed the worst because they do not provide redundancy and they can only identify states from within the set of pure states. In the particular cases of SIC and PSI constructions, we believe systematic errors in the unitary maps to implement their respective POVM have a significant impact on their final performance. Using an additional experiment we showed that SIC and PSI constructions can become robust and produce comparable fidelities to the best performing constructions at the cost of including redundancy in the measurements, thus decreasing their efficiency. Overall, our experiment shows that in a real-world setting there is no such thing as an "optimal" protocol for QST; the best choice of measurement strategy and state estimator will depend on the specifics of the scenario at hand and must necessarily reflect some tradeoff between accuracy, efficiency, and robustness to experimental imperfections.

In the second part of chapter 5, we presented results from implementing QST using partial informationally complete measurement records. Our experimental data shows that, as expected, the Trace Minimization estimator exhibits a compressed sensing effect which allows us to obtain good fidelities of reconstructions when using non IC measurements. The same effect was observed when using the Least-Squares and Maximum-Likelihood estimators, which is understood as resulting from imposing a positivity constraint in the reconstruction algorithms. Finally, the experiment also demonstrated that the error parameter $\epsilon$ in the trace minimization estimator plays an important role during reconstruction and choosing its value properly is necessary to obtain reliable reconstruction results.

In chapter 6 we presented an experimental implementation of efficient and robust quantum process tomography which makes use of a set of intelligently chosen probe states to reconstruct unitary processes. The experiment showed that we can successfully implement QPT on test unitary processes in $d = 4$, $d = 7$, and $d = 16$ using a set of $d$ input probe states. We found that in the $d = 4$ case, performing QPT using the set of intelligent probe states produces comparable results to standard QPT, while significantly reducing the amount of information needed for reconstruction. In the $d = 16$ case, successful reconstruction using intelligent probing serves to demonstrate how powerful this QPT approach is. To the best of our knowledge, this is the first time that a unitary process in such a large Hilbert space has been successfully reconstructed. Finally, we presented results from implementing QPT using partially informationally complete measurement records. In this case, we showed that non IC measurement records do produce high fidelities of reconstruction, yet QPT using intelligent probing and an IC measurement record remains as the most efficient and robust scheme.

Altogether, the results presented in chapters 5 and 6 indicate that quantum tomography currently represents a viable tool to verify the performance of state-of-the-art experiments performed on systems with large Hilbert dimension (e.g. 2-4 qubits). However, as new technological advances emerge and the use of larger Hilbert space systems become common, even efficient methods to perform QST and QPT like the ones presented here may not be feasible. In that scenario further theoretical developments will be necessary to keep QT as a useful tool for diagnosing quantum systems.

Looking ahead, there are several control and measurement tools that can be readily explored using our experimental setup. So far, all our experiments have been tested using nearly-pure states prepared by unitary control. In order to expand our control toolbox, it is desirable for us to be able to prepare mixed states with arbitrary purity and rank. The use of mixed states as inputs for the QST procedure could prove useful to complement our experimental exploration regarding the role of informationally completeness. In addition, they can also help us addressing questions related to bias in our estimator algorithms [83]. In order to prepare mixed states in the laboratory there are two approaches we can explore. A first and straightforward approach consist of implementing our standard unitary control scheme to prepare arbitrarily chosen pure states that we can measure using our Stern-Gerlach apparatus. Then, by combining the Stern-Gerlach raw signals from different pure states using an appropriate probability distribution, we can effectively generate a Stern-Gerlach signal corresponding to a chosen mixed state. A second long term approach consist of implementing our standard unitary control, plus one or more optical fields to drive optical pumping between states in the ground hyperfine manifold. Then, by using a master equation to appropriately model the coherent and incoherent parts of the dynamical evolution [84], we can potentially find control waveforms to prepare mixed states in a prescribed fashion.

An important problem in QIS is the ability to simulate quantum mechanical devices [85, 86]. In this regard, our experimental setup can serve as a testbed to simulate the behavior of different quantum systems. In particular, our 16-dimensional Hilbert space associated with the electronic ground state of cesium can be used to simulate a well-studied model of quantum chaos: the Quantum Kicked Top [18]. Given our proven ability to perform high-fidelity unitary transformations and efficient quantum state tomography in our system, we can explore interesting questions such as robust control in the presence of chaos and experimental imperfections.

Finally, as stated in chapters 5 and 6, quantum tomography is, in principle, the ideal set of tools to diagnose errors present in our experimental setup. In this dissertation we have limited our analysis and discussion to quantify the performance of our tomography procedures by using the fidelity of reconstruction as our single figure of merit. Our theory collaborators at the University of New Mexico have started to explore the use of other estimator algorithms which can provide insight as to what the nature of the errors in the procedure for QPT are [81]. However, if quantum tomography is ever going to become a truly useful tool for diagnosis, further theoretical studies are necessary in order to find relevant uses for the large amount of information obtained. For now, a promising alternative to diagnose errors present in our experimental setup is the use of Hamiltonian tomography [87–89]. In this problem one directly attempts to identify the Hamiltonian of the system, which is often specified by fewer parameters. The general idea behind Hamiltonian tomography is the application of intelligently chosen external fields to address specific parts of the complex Hamiltonian describing the system, such that the unknown parameters in the Hamiltonian can be retrieved one by one.

# APPENDIX A

# FIDELITY OF UNITARY TRANSFORMATIONS AS A FUNCTION OF CONTROL TIME

To determine the cumulative effect of experimental imperfections on the performance of unitary transformations, we carried out an experiment to measure the benchmarked fidelity of the unitary maps as a function of their control time $T$.

In this experiment, we performed randomized benchmarking on five sets of control waveforms. In each set, waveforms implemented 16-dimensional arbitrarily chosen unitary transformations. For each of the five sets we make use of a different choice of control time (See Table A.1). Apart from that, all waveforms are designed using the same robustness criteria and the same phase step duration $\Delta t = 4$ $\mu$s. Lastly, choosing $T \geq 600$ $\mu$s for every set ensures that all control waveforms reach designed fidelities $\mathcal{F} > 0.999$.

| $T$ ($\mu$ s) | $\Delta t$ ($\mu$ s) | $N = T/\Delta t$ | $\mathcal{F}_B$ |
|---|---|---|---|
| 600 | 4 | 150 | 0.982(2) |
| 1000 | 4 | 250 | 0.966(2) |
| 1400 | 4 | 350 | 0.944(3) |
| 1800 | 4 | 450 | 0.913(3) |
| 2200 | 4 | 550 | 0.896(4) |

Table A.1: Summary of control parameters

Fig. A.1 shows randomized benchmarking data for the five sets of control waveforms as a function of $l$. Each point represents an average of 10 sequences.

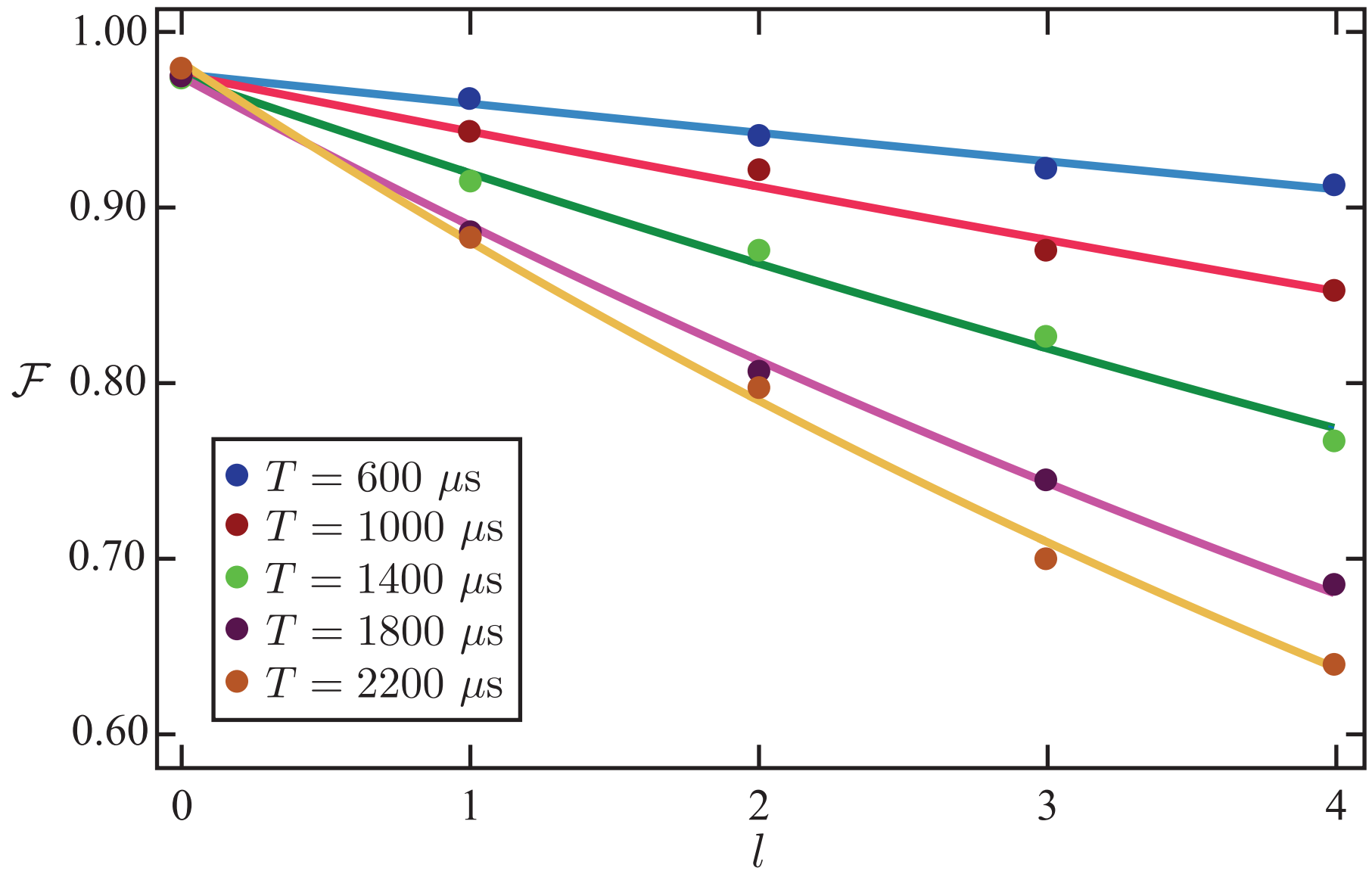

Figure A.1: Randomized benchmarking data for control waveforms with different control time lengths as a function of $l$. Each point represents an average of 10 sequences. Lines are fits from which the benchmarked fidelity $\mathcal{F}_B$ is determined.

The benchmarked fidelities obtained were $\mathcal{F}_B = 0.982(2)$ using $T = 0.6$ ms, $\mathcal{F}_B = 0.966(2)$ using $T = 1$ ms, $\mathcal{F}_B = 0.944(3)$ using $T = 1.4$ ms, $\mathcal{F}_B = 0.913(3)$ using $T = 1.8$ ms, and $\mathcal{F}_B = 0.896(4)$ using $T = 2.2$ ms. The previous data indicates that, as a general trend, longer control waveforms tend to perform worse in our experiment.

Fig. A.2 shows the benchmarked fidelities obtained in the previous experiment displayed as a function of waveform control time $T$. Here, it is easy to see that the decrease in fidelity is well-described by a linear decay that can be fitted using a linear function (blue line). This function is given by

$$\mathcal{A}(t) = 1 - 0.0444t. \tag{A.1}$$

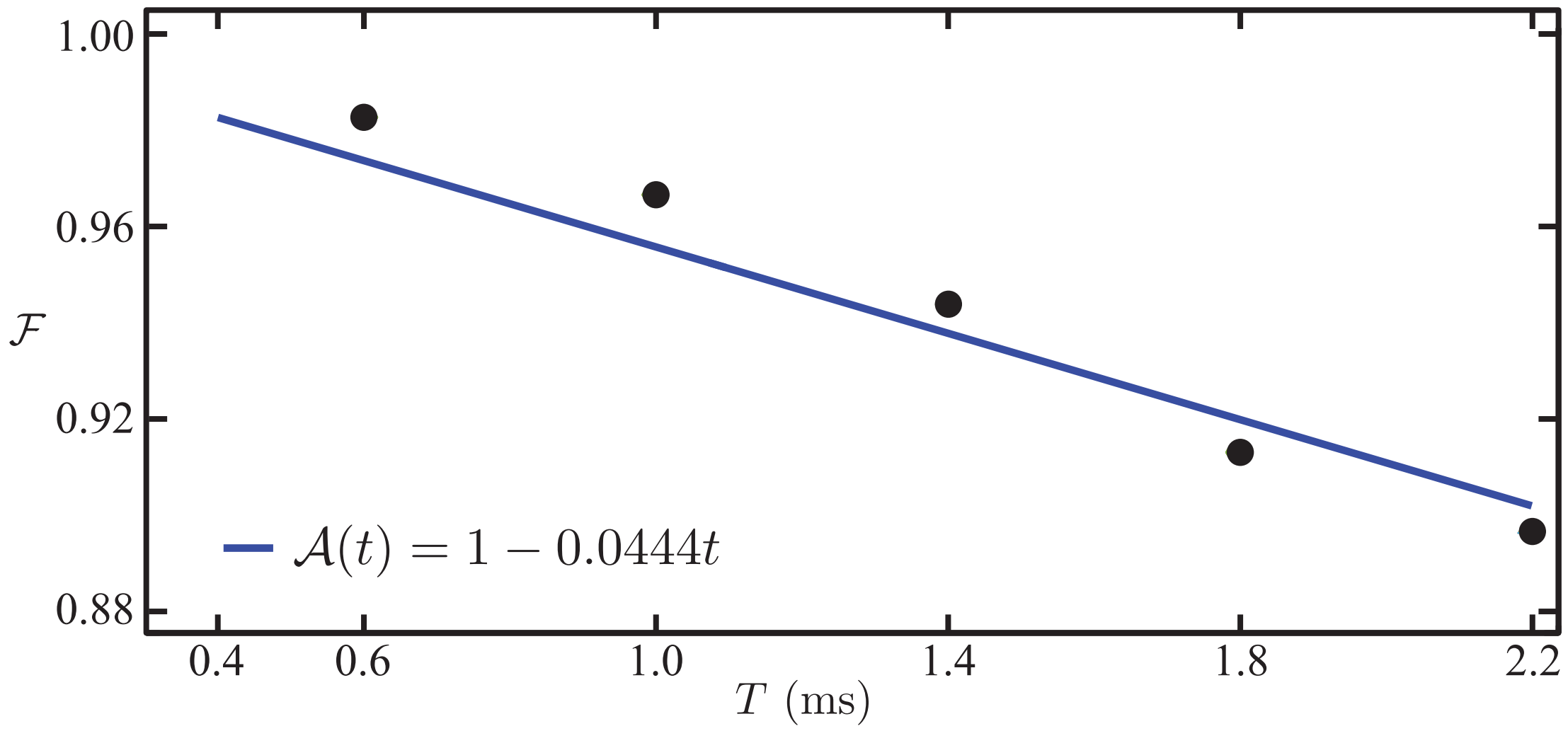

Figure A.2: Benchmarked fidelities as a function of waveform control time $T$. Blue line corresponds to a linear fit of the data.

We believe that the decrease in performance of control waveforms is due to the cumulative effect of experimental imperfections which gradually reduce the achievable fidelity as the control time increases.

One potentially source of such imperfections is the filtering of the rf magnetic fields. As described in Sec. 3.3, the rf circuitry suffers from limited bandwidth. This leads to filtering of the current thought the Helmholtz coils producing the rf magnetic fields. The current at the point of a discontinuous control-phase jump is "smoothed out"and the magnetic fields applied to the atoms no longer correspond to the ideal control fields. This problem can be solved by using circuitry with higher bandwidth, or by accurately characterizing the electronic filter and include it in the numerical design of control waveforms.